\documentclass{aa}  
\usepackage{graphicx}
\usepackage{array}
\usepackage{lscape}                                
\usepackage{natbib}
\usepackage{longtable}
\usepackage{color}
\usepackage[dvipsnames]{xcolor}
\usepackage{mathrsfs} 
\usepackage{subcaption}
\usepackage{siunitx}  
\usepackage{multirow}

\usepackage{txfonts}
\newcommand{\kms} {km\,s$^{-1}$}
\newcommand{\vsini} {$v$\,sin\,$i$}

\newcommand{\Teff} {$T_{\rm eff}$}

\newcommand{\gravc} {log\,{\em g}$_{\rm c}$}

\newcommand{\msun}{M$_{\odot}$}

\newcommand{\RVpp} {$RV_{\rm PP}$}

\begin{document} 

\title{The IACOB project}
   \subtitle{VIII. Searching for empirical signatures of binarity in fast-rotating O-type stars}
   \author{ N.~Britavskiy\inst{1,2,3}, 
            S.~Sim\'on-D{\'\i}az\inst{2,3}, 
            G.~Holgado\inst{4}, 
            S. Burssens\inst{5},
            J. Ma{\'\i}z Apell\'aniz\inst{4},
            J.J.~Eldridge\inst{7},
            Y.~Naz\'e\inst{1}\thanks{FNRS Senior Research Associate},
            M. Pantaleoni Gonz\'alez\inst{4,6},
            A. Herrero\inst{2,3}}

\institute{
Universit\'e de Li\`ege, Quartier Agora (B5c, Institut d’Astrophysique et de Geophysique), All\'ee du 6 Ao\"ut 19c, B-4000 Sart Tilman, Li\`ege, Belgium
\and
Instituto de Astrof\'isica de Canarias. E-\num{38200} La Laguna, Tenerife, Spain.              
\and
Departamento de Astrof\'isica, Universidad de La Laguna. E-\num{38205} La Laguna, Tenerife, Spain.
\and
Centro de Astrobiolog\'ia (CAB), CSIC-INTA, Camino Bajo del Castillo s/n, campus ESAC,  E-\num{28692}, Villanueva de la Cañada, Madrid, Spain
\and 
Institute of Astronomy, KU Leuven, Celestijnenlaan 200D, 3001 Leuven, Belgium.
\and
Departamento de Astrof{\'\i}sica y F{\'\i}sica de la Atm\'osfera, Universidad Complutense de Madrid. E-\num{28040} Madrid, Spain. 
\and
Department of Physics, University of Auckland, Private Bag 92019, Auckland, New Zealand.
}

\offprints{mbritavskiy@uliege.be}

\date{Submitted 5 October 2022 / Accepted 1 February 2023}

\titlerunning{Investigation of the binary nature of Galactic fast-rotating O-type stars}

\authorrunning{N. Britavskiy et al.}
 
  \abstract
   {The empirical distribution of projected rotational velocities (\vsini) in massive O-type stars is characterised by a dominant slow velocity component and a tail of fast rotators. It has been proposed that binary interaction  plays a dominant role in the formation of this tail.}
   {We perform a complete and homogeneous search for empirical signatures of binarity in a 
   sample of 54 fast-rotating stars with the aim of evaluating this hypothesis. This working sample has been extracted from a larger sample of 415 Galactic O-type stars that covers the full range of \vsini\ values.}
   {We used new and archival multi-epoch spectra in order to detect spectroscopic binary systems. We complemented this information with {\it Gaia} proper motions and {\it TESS} photometric data to aid in the identification of runaway stars and eclipsing binaries, respectively. We also benefitted from additional published information to provide a more complete overview of the empirical properties of our working sample of fast-rotating O-type stars. }
  {The identified fraction of single-lined spectroscopic binary (SB1) systems and apparently single  stars among the fast-rotating sample is $\sim$18\% and $\sim$70\%, respectively. The remaining 12\% correspond to four secure double-line spectroscopic binaries (SB2) with at least one of the components having a \vsini\,>\,200~\kms ($\sim$8\%), along with a small sample of 2 stars ($\sim$4\%) for which the SB2 classification is doubtful: these could actually be single stars with a remarkable line-profile variability. When comparing these percentages with those corresponding to the slow-rotating sample, we find that our sample of fast rotators is characterised by a slightly larger percentage of SB1 systems ($\sim$18\% vs. $\sim$13\%) and a considerably smaller fraction of clearly detected SB2 systems (8\% vs. 33\%). Overall, there seems to be a clear deficit of spectroscopic binaries (SB1+SB2) among fast-rotating O-type stars ($\sim$26\% vs. $\sim$46\%).  On the contrary, the fraction of runaway stars is significantly higher in the fast-rotating domain ($\sim$33-50\%) than among those stars with \vsini\,<\,200~\kms. Lastly, almost 65\% of the apparently single fast-rotating stars are runaways. 
  As a by-product, we discovered a new over-contact SB2 system (HD\,165921) and two fast-rotating SB1 systems (HD\,46485 and HD\,152200)
  Also, we propose HD\,94024 and HD\,12323 (both SB1 systems with a \vsini\,<\,200~\kms) as candidates for hosting a quiescent stellar-mass black hole.
  }
  {Our empirical results seem to be in good agreement with the assumption that the tail of fast-rotating O-type stars (with \vsini\,>\,200~\kms) is mostly populated by post-interaction binary products. In particular, we find that the final statistics of identified spectroscopic binaries and apparent single stars are in good agreement with newly computed predictions obtained with the binary population synthesis code BPASS and earlier estimations obtained in previous studies. 
  }

   \maketitle

\section{Introduction}\label{intro}

One decade ago, \citet{deMink_2013, deMink_2014} performed a detailed theoretical evaluation of the impact that binary interaction could have on the spin-rate properties of massive O-type stars (i.e. main sequence stars with masses in the range of $\sim$20\,--\,80~\msun). This study was partly motivated by the necessity to provide an explanation for an empirical result already highlighted by \citet{Conti_1977,Wolff_1982} and subsequently confirmed by some other authors \citep[see further references in][]{Simon-Diaz2014, vfts_2013_otype, Holgado_2022}; namely: the spin-rate distribution of any investigated (large) sample of O-type stars (even in different metallicity environments) is characterised by a main component, including stars spinning with projected rotational velocities (\vsini) below $\sim$150\,--\,200~\kms, and a tail of fast rotators reaching values of \vsini\ up to 400\,--\,600~\kms. Such a tail of fast rotators normally comprises $\sim$20\,--\,25\% of stars in the considered samples.

By performing a specific simulation of a massive binary-star population typical for our Galaxy and assuming continuous star formation, \citet{deMink_2014} found that binary interaction during main sequence evolution could easily explain the existence of the empirically detected tail of fast-rotating O-type stars. In brief, mass (and angular momentum) transfer from the initially more massive star (donor) to the lower mass companion (gainer) is an efficient mechanism to spin up the latter which, under certain circumstances, may become the more massive component of the binary system after Roche lobe overflow \citep[see also e.g.][]{Packet_1981,Pols_1991}.

Indeed, by assuming the empirical distributions of mass ratios, orbital periods, and the corresponding binary fraction obtained by \citet{Sana_2012} as input for their simulations, \citet{deMink_2013} ended up with a fraction of fast rotators (i.e. assuming they are characterised by having a \vsini\,$>$\,200~\kms), which is very similar to the observed one. Furthermore, \citeauthor{deMink_2013} provided some predictions regarding the expected type and percentage of binary products populating the tail of fast rotators in the \vsini\ distribution of O-type stars. This mainly includes mergers and mass gainers orbited by a hot stripped star or a compact degenerate object (neutron star or black hole). 

In this context, we should also note that the merger scenario is considered as one of the mechanisms producing magnetic fields in massive stars \citep[e.g.][]{Ferrario_2009,Schneider_2016,Schneider_2019}. In such magnetic massive stars, rotation is braked fast, so that mergers would lead to slow rotators. Since the debate remains open on the exact outcome of mergers, the merger scenario should still deserve particular attention when investigating the origin of fast rotators.

In addition, some of these fast-rotating O-type stars would be expected to be detected as runaway stars resulting from a disrupted binary after supernova explosion of the initially more massive component of the system \citep[see e.g.][]{Blaa61, Bolton_1986, Walborn_2014}. The latter would imply a complementary or alternative explanation to the dynamical ejection scenario from a stellar cluster to the occurrence of runaway events in the massive star domain \citep{Portegies_Zwart_1999}. 

The empirical confirmation of these theoretical scenarios has important consequences for several topics of modern astrophysics, especially those influenced by our specific knowledge about massive star formation, evolution, and feedback. In these respects, it is important to recall that stellar rotation is known to play an important role in the evolution of high-mass stars \citep[e.g.][]{Maeder_2000, Ekstrom2012}. It not only modifies the evolutionary paths followed by these stars, as well as their lifetimes and final fates, but it has also been proposed to induce the transport of core-processed elements to the stellar surfaces. There is clear evidence that a non-negligible percentage of massive O-type stars rotate at velocities fast enough to be affected by the aforementioned effects. Thus, we need to be careful while using the empirical information compiled for these fast-rotating O-type stars in order to constrain our theories of high-mass star formation and evolution. This is because the theories will be different if the stars have acquired their angular momentum during the star formation process or through any type of binary interaction. For example, if a large percentage O-type stars having a \vsini\ larger than $\sim$200~\kms\ are post-interaction binaries, proposed by \citet{deMink_2013,deMink_2014}, using these stars to investigate the efficiency of rotational mixing in single star evolution models may lead to erroneous results and conclusions \citep[e.g.][]{Hunter_2009, Cazorla2017b}. 

Another point worth mentioning is the exotic nature of the possible companions to fast rotators. 
Indeed, short-period binary systems that are passing through the common-envelope phase may produce binary black holes or neutron star systems that can be progenitors of gravitational wave events \citep[e.g.][]{Langer_2020}. These systems undergo a Roche-lobe overflow resulting in the donor becoming a Helium star while the accretor becomes a fast rotator of the OB-type. Taking into account subsequent evolution of such systems, BH/NS+OB configurations may arise. While some steps towards the detection of BH+OB systems have been taken \citep[e.g.][]{Villase_2021,Mahy_2022,Banyard_2022,Shenar_2022NatAs,Shenar_2022aa,Janssens_2022}, our programme sample is one of the best compilations that can be used in the search of such systems. Finally, we may mention in this context that cooler fast-rotating stars showing Be-type signatures have shown hints of a post-interaction nature \citep[e.g.][]{Bodensteiner_2020,Wang_2021, Klement_2022}.

As a continuation of the efforts devoted by the IACOB project (P.I. Sim\'on-D\'iaz) to provide solid empirical foundations to our knowledge about the physical properties and evolution of massive OB-type stars \citep[see e.g.][]{Simon-Diaz2014, simondiaz2017, Holgado_2020, Holgado_2022}, in this paper, we perform a complete and homogeneous search for empirical signatures of binarity in a statistically meaningful sample of several tens of fast-rotating Galactic O-type stars. Our ultimate goal is to evaluate the scenario proposed by \citet{deMink_2013,deMink_2014} to explain the existence of a tail of fast rotators in this stellar domain. To this aim, we use as starting point the results presented in \citet{Holgado_2022}, where we performed a reassessment of the empirical rotational properties of Galactic massive O-type stars using the results from a
detailed analysis of ground-based multi-epoch optical spectra obtained in the framework of the IACOB \& OWN surveys (see Sect.~\ref{obs-sect2}). The spectroscopic observations considered in \citet{Holgado_2022} can now be complemented with an extended multi-epoch spectroscopic dataset (including a minimum of 3\,--\,5 epochs for all stars in the investigated sample of fast-rotators) and a set of superb quality data about proper motions and photometric variability recently delivered by the {\em Gaia} and {\em TESS} space missions.

The paper is organised as follows. In Sect.~2, we introduce the programme sample and the compiled observational data\,set. Sect.~3 describes the analysis we have performed (including a radial velocity analysis and {\it Gaia} and {\it TESS} data) and the literature overview regarding the programme sample. In Sect. 4, we discuss and interpret the obtained results. Sect. 5 presents the main conclusions of the paper.

\section{Sample definition and observations}\label{obs-sect2}

\begin{figure*}[!t]
\centering
\includegraphics[width=0.495\textwidth]{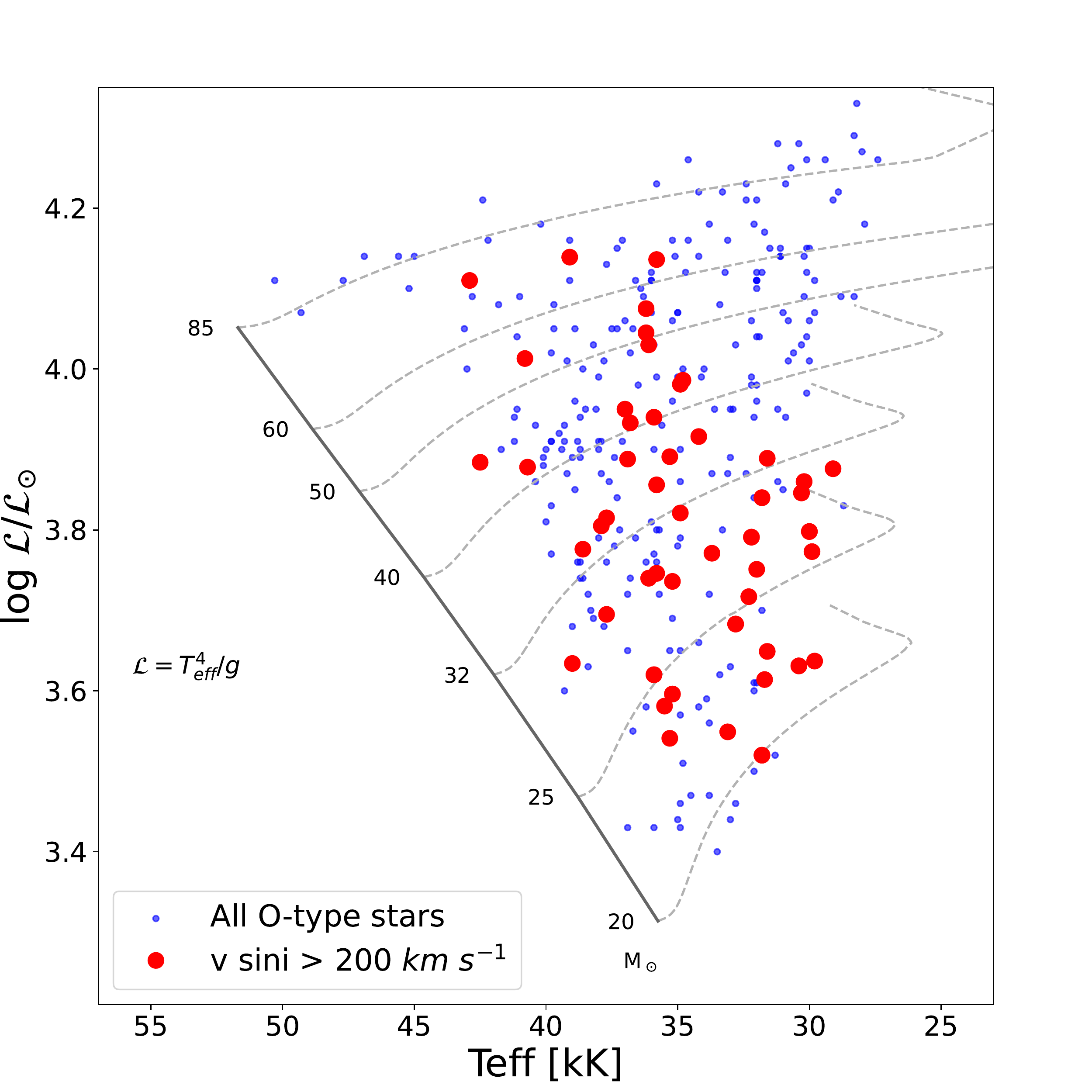}
\includegraphics[width=0.495\textwidth]{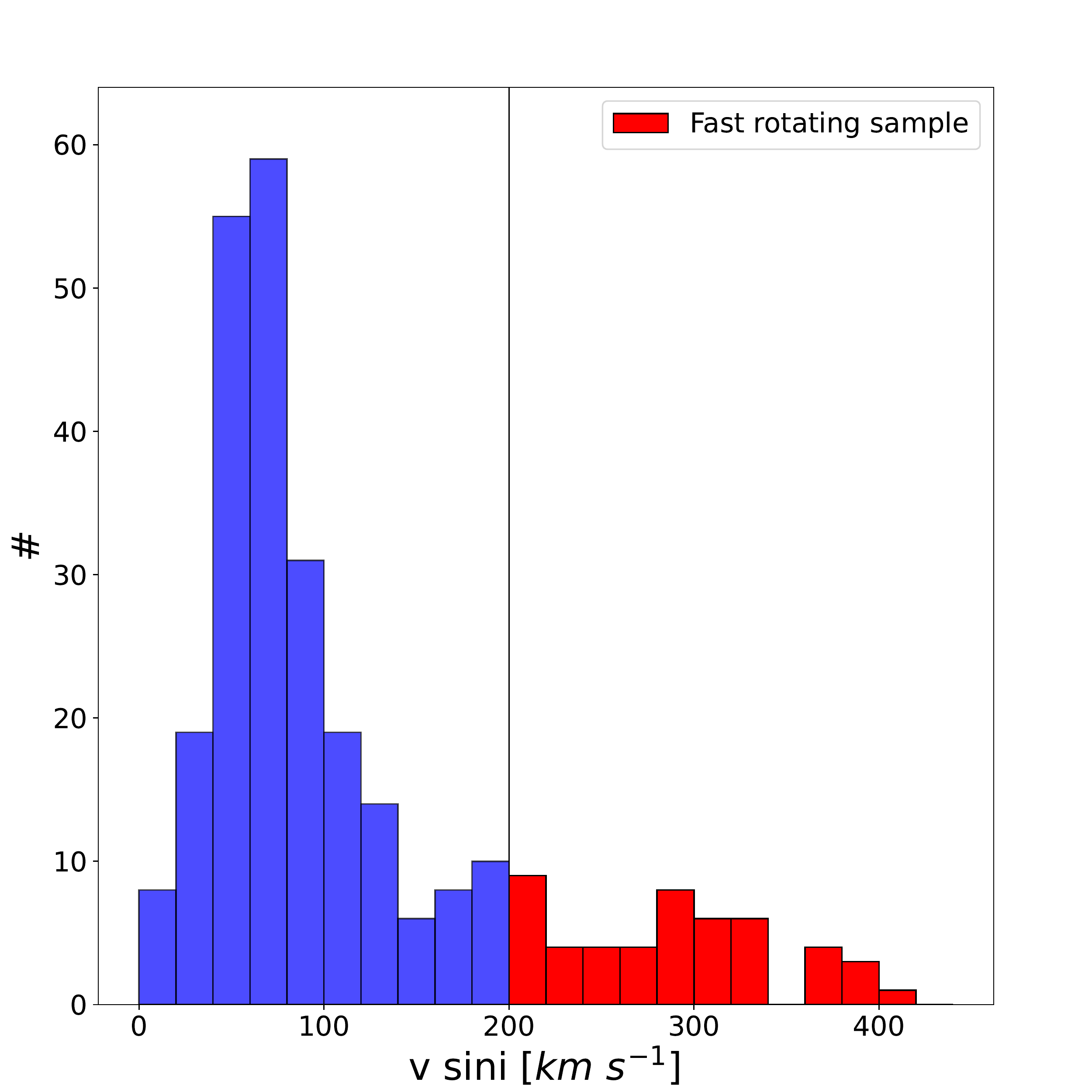}
\caption{ Location in the spectroscopic HR diagram of the 285 Galactic (likely-single and SB1) O-type stars investigated by \citet{Holgado_2022} (left) and associated \vsini\ distribution (right). In both cases, the sample  of likely single and SB1 fast-rotating stars (with \vsini\,$>$\,200~\kms) are highlighted in red. Non-rotating evolutionary tracks from \citet{Ekstrom2012} are also included in left panel for reference.}
\label{figure_hr_brott}
\end{figure*}

\subsection{Programme sample}\label{sample}

To build the sample that forms the basis for the study detailed in the present paper, we benefitted from information presented in \citet{Holgado_2022}. There, a detailed investigation of the spin-rate properties was performed for a sample of 285 Galactic O-type stars identified as apparently single or single-line spectroscopic binaries (SB1), using high-quality optical spectroscopic data gathered by the IACOB and OWN surveys (last described in \citeauthor{SimonDiaz2020} \citeyear{SimonDiaz2020} and \citeauthor{Barba2017} \citeyear{Barba2017}, respectively).

\citet{Holgado_2022} started from an initial sample of 415 Galactic O-type stars and excluded 113 double-line spectroscopic binaries, plus another 17 peculiar stars (i.e. presenting signatures in their spectra that are typically associated with Oe, Wolf-Rayet, or magnetic stars). For the remaining sample of 285 stars, they obtained estimates for the projected rotational velocities and other stellar parameters that can routinely be procured by means of a quantitative spectroscopic analysis. To this aim, they applied for each star the methodology described in \cite{Holgado2018} to the best S/N spectrum \citep[see also][]{Holgado_2020}.

The estimated effective temperature (\Teff), the surface gravity \citep[corrected from centrifugal forces, \gravc, see][]{Herrero_1992,Repolust_2004}, and \vsini\ allowed us to locate the sample in the spectroscopic HR diagram \citep[sHRD,][]{Langer_2014}, as well as to build the corresponding global \vsini\ distribution. As indicated in Sect.~\ref{intro}, and illustrated by the right panel of Fig.~\ref{figure_hr_brott} \citep[see also][]{Holgado_2022}, this \vsini\ distribution is characterised by a dominant low velocity component and a tail of fast rotators extending up to $\sim$\,450~\kms.

Following \cite{deMink_2013}, this tail of fast rotators -- especially above 200~\kms -- is expected to be mostly populated by the evolved post-interaction binaries. To empirically evaluate this hypothesis, we defined as our initial programme sample of stars those targets identified by \citet{Holgado_2022} as having a \vsini\,$>$\,200~\kms\ (see Table~\ref{table:sample}). This corresponds to a total of 50 stars, distributed throughout the O-star domain in the sHR diagram as illustrated in the left panel of Fig.~\ref{figure_hr_brott}. 

While this boundary in rotational velocity is somewhat arbitrary, as predicted by \cite{deMink_2013}, it should allow us to minimise the presence of pre-interacting binary systems in the sample under study. Therefore, except for a small fraction of stars in the \vsini\ range between 200 and $\sim$300~\kms (which could be short-period binary systems with individual components spun up by tides), the large majority of stars in our working sample should be (again in the context of de Mink's scenario) mergers (i.e. genuinely single stars) or mass gainers in which the initially less massive star has now become an O-type star accompanied by a post-mass transfer object (i.e. a compact remnant or a stripped star) or a fast-rotating runaway star. We also note that while we have assumed the same boundary in equatorial velocity ($v_{\rm eq}$) as in \citeauthor{deMink_2013}, we are actually using the quantity \vsini\ and not $v_{\rm eq}$ to build our working sample of fast rotators. Therefore, there could still be fast-rotating stars in Holgado's sample that we are missing because their rotational axis has a low inclination angle. This effect will be taken into account in the discussion and interpretation of our results (see e.g. Fig.~\ref{figure_pp_fin_zoom} and Sects.~\ref{RVppvsini} and \ref{discussion_about_sb1}).

For the sake of completeness, while we were not able to obtain accurate individual \vsini\ measurements for the two components of those stars identified as double-line spectroscopic binaries (SB2), we carried out a careful inspection of all available spectra for the 113 SB2 systems in the initial sample to identify those in which at least one of the components could have a \vsini\ larger than 200~\kms. This information allowed us to have an estimate of the percentage of such systems with at least one of the two components being a fast rotator (see Sect.~\ref{final_statistics} and Table~\ref{table:fast_stat}). The four identified systems fulfilling this criterion are quoted at the bottom of Table~\ref{table:sample}.

\subsection{Multi-epoch spectroscopic observations}\label{spectra-obs}

As was the case of previous papers of the IACOB series dealing with O-type stars \citep[see e.g.][]{Holgado2018, Holgado_2020, Holgado_2022}, the bulk of the spectroscopic observations used in this work comes from three high-resolution spectrographs: the FIES instrument \citep[resolving power, R$\approx$46\,000 or 25\,000,][]{Telting2014} attached to the 2.56-m Nordic Optical Telescope  (NOT), the HERMES spectrograph \citep[R$\approx$85\,000,][]{Raskin2011} attached to the 1.2-m Mercator telescope, and the FEROS spectrograph \citep[R$\approx$48\,000,][]{Kaufer_1997} presently installed at the MPG/ESO2.2-m telescope. The former two instruments are both located in the Roque de los Muchachos observatory (La Palma, Spain), while the latter has been operating in La Silla observatory (Chile) since 2002. Detailed information on the collected spectroscopic data\,set can be found in Table 1 of \citet{Holgado2018}.

Our programme sample comprises Galactic O-type stars in both hemispheres covering a range in B magnitude from 2.5 and down to $\sim$11 mag. Those stars observable from the Canary Island observatory were observed as part of the IACOB survey (P.I. Sim\'on-D\'iaz) during several observing runs allocated between 2008 and 2016. This survey initially included a minimum of three epochs per star; however, we also devoted several new additional campaigns since 2017 (P.I. Holgado) to increase the number of available epochs for the subsample of 23 fast rotators visible from the Canary Islands' observatory. These extra observations, which have allowed us to cover a time-span of more than ten years with more intense time coverage between 2017 and 2018, do not only include FIES and HERMES spectra, but also up to ten additional epochs obtained with the SES high-resolution spectrograph \citep{stella} attached to the STELLA1.2-m robotic telescope operating at Iza\~na observatory (Tenerife, Spain).

For those stars observable from La Silla, the multi-epoch spectroscopic dataset compiled for this work was mostly obtained in the framework of the OWN spectroscopic survey\footnote{We note there are seven stars observable from both the La Silla and the Canary Islands observatories.} (P.I. Barb\'a) between 2007 and 2017. In addition, these observations were complemented with a few spectra downloaded from the ESO-FEROS archive\footnote{\url{http://archive.eso.org/wdb/wdb/adp/phase3_spectral/form}}.

The full list of 54 fast-rotating stars comprising our programme sample\footnote{This includes the four detected SB2 stars with at least one of the components having a \vsini\ larger than 200~\kms.} is quoted in Table~\ref{table:sample}, where we also indicate the corresponding spectral classifications \citep[as provided in version 4.1 of the Galactic O-star catalog, GOSC,][]{MaizApellaniz2013}, the number of FIES, HERMES, FEROS, and STELLA spectra initially available, and the total time-span covered by each set of multi-epoch spectra. Generally speaking, after discarding those spectra with low (less than 5) ${\rm SNR}$  in the 5875~\AA\ region, we have been able to compile a minimum of 4 epochs for almost 90\% of the stars in the sample (see column 8 of Table~\ref{table:sample}), reaching up to 10 epochs in more than 50\% of them. In addition, for some bright Northern targets, we could gather more than 25 epochs. Regarding the covered time-span, except for a few cases, we were able to reach a minimum of 3 years and up to 10\,--\,15 years for $\sim$20 stars (see last column of Table~\ref{table:sample}).

Thanks to {\it Gaia}-EDR3, we also have information about parallaxes for all stars in our working sample (see also Sect.~\ref{extra-obs}). By taking as a first approach the median of the geometric distances provided in \citet{Bailer_Jones_2021}, we see that 51 of the 54 stars are closer than 3 kpc, and only 6 of them have a RUWE value\footnote{We recall that this quantity (the renormalised unit weight error, RUWE) can be used as a quality flag of the {\it Gaia} astrometric solution for each individual target. Following recommendations by the {\it Gaia} team, information for those stars with a RUWE value above 1.4 must be handle with care.} larger than 1.4 (see Table~\ref{table:phys_parameters}). 

As described in Sect.~\ref{parameters}, having access to this information allowed us to obtain estimates for other stellar parameters (radii, luminosities) beyond what can be obtained by spectroscopic means, namely, projected rotational velocities, effective temperatures, surface gravities, or surface abundances of certain elements.

\subsection{Other sources of empirical data considered in this work}\label{extra-obs}

In addition to our main multi-epoch spectroscopic observations described in previous section, we also benefitted from various other sources of empirical data including (a) photometric variability as provided by the light curves obtained by the {\it TESS} mission \citep{tess}, (b) the proper motions delivered by the {\it Gaia} mission \citep{gaia_edr3}, and (c) information about possible close-by visual companions as provided by the {\it Gaia} mission, as well as other on-ground high spatial resolution surveys such as the AstraLux optical survey \citep{astralux_2010}, the Southern Massive Stars at High angular resolution survey \citep[SMASH,][]{Sana_2014}, and the Fine Guidance Sensor resolution survey \citep[FGS,][]{Aldoretta_2015}.

Lastly, we also searched for additional info in the literature about whether any of the programme sample of fast-rotating stars had been previously identified as a spectroscopic, eclipsing, and/or X-ray binary. In this regard, we paid special attention to some specific studies providing information about the orbital parameters of previously detected binaries (see Sect.~\ref{Extra_info}).

\section{Gathering empirical information}
Table~\ref{table:b_fast} summarises all the empirical information of interest compiled for this paper regarding our programme sample (excluding the four clearly detected SB2 systems), ordered by increasing rotational velocity. In addition to the spectral classification (column 2), we quote the projected rotational velocity (column 3), the peak-to-peak amplitude of radial velocity variability, as measured from all available spectra per star (column 4), and the assigned spectroscopic binarity status after a first visual inspection of the variability of the \ion{He}{i}\,$\lambda$5875 line-profile (column 6).

The above-mentioned information -- mainly obtained from the spectroscopic data~set -- is also complemented with other information resulting from the analysis of the additional empirical data mentioned in Sect.~\ref{extra-obs}. The latter includes the type of variability detected in the {\it TESS} light curves (column 7), a determination of whether some close-by visual companions have been detected within 1 and 2 arcmin, respectively (column 9), the runaway status resulting from the analysis of the proper motions provided by {\it Gaia} and other studies in the literature (column 10) and, lastly, the final binary status established for each individual target after revisiting the literature and a more detailed analysis of the available radial velocity curves (column 11).

We list the physical parameters of our sample, gathered from \citet{Holgado_phd} and associated papers, in Table~\ref{table:phys_parameters}. We include estimates for the effective temperature (\Teff, column 2), the $\mathcal{L}$ parameter \citep[defined as \Teff$^4$/$g_{\rm c}$, see][column 3, where $g_{\rm c}$ is the surface gravity corrected from centrifugal forces]{Langer_2014}, as well as the stellar luminosity and radius (columns 4 and 5, respectively), based on the individual {\it Gaia}-EDR3 distances (column 9) provided by \citet{Bailer_Jones_2021}. Below, we describe how each of these pieces of empirical information was obtained.

\begin{table*}
{\footnotesize
\caption{Basic information about working sample of fast-rotating O-type stars extracted from the multi-epoch spectroscopy, {\em TESS} and {\em Gaia} data, as well as some other specific studies from the literature. The list of targets is ordered by increasing \vsini.}
\label{table:b_fast}
\resizebox{!}{7.5cm}{\begin{tabular}{l l l l  c l l l c l l } 
\hline
\hline
Name  & SpC   & \vsini   & RV$_{\rm PP}$ & $\sigma_{\rm RV}$ & SB tag  & Phot. var.    & X-ray & Contamination   &  Runaway?           & Final binary \\ 
      &       &  [\kms]             & [\kms]        & [\kms]    &     &  ({\it TESS}) & $\log(L_{\rm X}/L_{\rm BOL})$ & (1$'$ and 1$'$-2$'$)       & ({\it Gaia}+lit.)    & status \\
\hline
BD+36$^{\circ}$4145   & O8.5\,V(n)        & 200 $\pm$  3 &    3.5 $\pm$ 1.1   & 1.2 & LPV        &  SLF    &       &  0+0    & no      &  LS              \\
HD\,216532           & O8.5\,V(n)        & 200 $\pm$  3 &   15.0 $\pm$ 3.8   & 3.8  & LPV        &  PQ (+fr>5)    &  & 0+1    & no      &  LS            \\
HD\,163892            & O9.5\,IV(n)       & 201 $\pm$ 12 &  83.50 $\pm$ 9.1  &  26.3 & SB1        &  $\_$          & & 0+0    & no      &  \bf{SB1 ($\dagger$)} \\
HD\,210839            & O6.5\,I(n)fp      & 201 $\pm$  8 &   28.5 $\pm$ 5.5   & 6.3 & LPV/SB1?   &  SLF   & $-7.10$$^a$        &  0+0    & yes    &  LS (*)        \\
HD\,308813            & O9.7\,IV(n)       & 205 $\pm$  5 &   41.5 $\pm$ 6.7   & 15.4 & SB1        &  \bf{?}         &   $-7.36$$^l$ &  0+1    & no      &  \bf{SB1}      \\
HD\,36879             & O7\,V(n)((f))     & 205 $\pm$  6 &   10.0 $\pm$ 2.5   & 3.2 & LPV        &  SLF           & & 0+0    & yes     &  LS              \\
HD\,37737             & O9.5\,II-III(n)   & 209 $\pm$ 11 &  156.5 $\pm$ 9.9   & 48.3 & SB1        &  \bf{EB}       &  & $\_$   & no      &  \bf{SB1}      \\
HD\,152200            & O9.7\,IV(n)       & 210 $\pm$ 32 &   32.0 $\pm$ 6.5   & 12.8 & SB1        &  \bf{RM}   & $-6.99$$^b$  & 0+1    & no      &  \bf{SB1}      \\
HD\,97434             & O7.5\,III(n)((f)) & 215 $\pm$ 22 &   14.5 $\pm$ 3.5   & 6.1 & LPV        &  SLF   & $-6.69$$^b$        &  0+2    & no      &  LS           \\
HD\,24912             & O7.5\,III(n)((f)) & 224 $\pm$  8 &   29.0 $\pm$ 4.7   & 6.2 & LPV/SB2?   &  $\_$  & $-7.09$$^c$        &  0+0    & yes     &  LS (*)        \\
BD+60$^{\circ}$2522   & O6.5\,(n)fp       & 231 $\pm$ 23 &   24.5 $\pm$ 5.8   & 8.4 &  LPV/SB2?   &  SPB           & $\sim-7.0$$^k$  & 0+0    & yes     &  LS (*) \\
HD\,89137             & ON9.7\,II(n)      & 233 $\pm$  3 &    3.0 $\pm$ 1.3  & 1.3  &  LPV        &  SLF (+fr>5)   & & $\_$   & yes     &  LS             \\
BD+60$^{\circ}$134    & O5.5\,V(n)((f))   & 234 $\pm$  9 (*) &    7.5 $\pm$ 3.4   & 3.0 & LPV        &  SLF           & & $\_$   & yes     &  LS          \\
HD\,172175            & O6.5\,I(n)fp      & 243 $\pm$ 20 &   19.5 $\pm$ 7.7   & 9.7 & LPV        &  $\_$          & & $\_$   & yes     &  LS              \\
HD\,165246            & O8\,V(n)          & 254 $\pm$  8 &  126.0 $\pm$ 14.9  & 36.3 & SB1        &  \bf{EB      } & & 1+0    & no      &  \bf{SB1}      \\
HD\,5689              & O7\,Vn((f))       & 256 $\pm$ 40 &   12.0 $\pm$ 3.4   & 4.2 &  LPV        &  SLF           & & 0+0    & yes     &  LS             \\
HD\,124314            & O6\,IV(n)((f))    & 256 $\pm$ 10   &   33.0 $\pm$ 10.2  & 9.9 &  LPV/SB2?   &  SLF         &  &  2+0    & no      &  \bf{LPV/SB2?} \\ 
HD\,192281            & O4.5\,IV(n)(f)    & 261 $\pm$  5 (*) &   19.5 $\pm$ 8.6   & 5.6 & LPV        &  SLF + rot?    & & 0+0    & yes     &  LS          \\ 
HD\,76556             & O6\,IV(n)((f))p   & 264 $\pm$ 11 &   10.5 $\pm$ 7.3   & 4.4 & LPV        &  SLF + rot? & $-7.09$$^i$   &  1+1    & no      &  LS           \\
HD\,41997             & O7.5\,Vn((f))     & 272 $\pm$ 12 &   23.5 $\pm$ 4.8   & 6.0 & LPV        &  SLF           &  & 0+0    & yes     &  LS             \\
HD\,124979            & O7.5\,IV(n)((f))  & 273 $\pm$  6 &   12.0 $\pm$ 3.3   & 3.8 & LPV        &  SLF           &  & $\_$   & yes     &  LS             \\
HD\,155913            & O4.5\,Vn((f))     & 282 $\pm$ 10 (*) &    8.5 $\pm$ 8.0   & 3.1 & LPV        &  SLF + rot?    &  & 0+0    & yes     &  LS         \\
HD\,175876            & O6.5\,III(n)(f)   & 282 $\pm$ 16 &   35.5 $\pm$ 6.9   & 10.2 & LPV/SB2?   &  $\_$          & & 0+0    & yes     &  LS (*) \\
HD\,15137             & O9.5\,II-IIIn     & 283 $\pm$  7 &   44.0 $\pm$ 6.3   & 10.8 & LPV/SB1?   &  SLF          & &  0+0    & yes     &  \bf{SB1} (*)  \\
HD\,28446A            & O9.7\,IIn         & 291 $\pm$ 10 &   29.0 $\pm$ 7.5   & 7.0 & LPV        &  SLF (+fr>5)   &  & 1+0    & no      &  LS             \\
HD\,15642             & O9.5\,II-IIIn(*)     & 293 $\pm$ 10 &   23.5 $\pm$ 6.6   & 7.6 & LPV        &  SLF           &  & 0+1    & yes      &  LS         \\
HD\,90087             & O9.2\,III(n)      & 295 $\pm$  2 &   11.5 $\pm$ 3.4   & 4.3 & LPV        &  $\_$          & & $\_$   & no      &  LS              \\
HD\,165174            & O9.7\,IIn         & 299 $\pm$11 &   59.5 $\pm$ 7.4   & 18.3 & SB1        &  $\_$ & $-6.9$$^d$         &  0+0    & no      &  \bf{SB1}      \\
HD\,52266             & O9.5\,IIIn        & 299 $\pm$  7 &   35.0 $\pm$ 6.0   & 10.1 & LPV        &  SLF          &  &  0+0    & no      &  \bf{LPV/SB1? (*)}      \\
HD\,91651             & ON9.5\,IIIn       & 304 $\pm$ 16 &   21.5 $\pm$ 5.8   & 6.2 & LPV/SB2?   & $\beta$ Cep    &  &  0+0    & no      &  \bf{LPV/SB2?} \\
HD\,228841            & O6.5\,Vn((f))     & 311 $\pm$  8 (*) &   18.0 $\pm$ 7.3   & 5.5 & LPV        & SLF    & $-7.25$$^i$        &  $\_$   & yes     &  LS       \\
HD\,52533             & O8.5\,IVn(*)         & 312 $\pm$ 14 &  166.0 $\pm$ 27.7  & 54.4 &  SB1        & \bf{EB}      &  &  3+0    & no      &  \bf{SB1 ($\dagger$)} \\
BD+60$^{\circ}$513    & O7\,Vn            & 313 $\pm$ 11 (*) &   27.5 $\pm$ 7.4   & 12.2 & LPV        & SLF + rot?     & $-7.56$$^e$ &  1+0    & no      &  LS             \\
HD\,229232            & O4\,Vn((f))       & 313 $\pm$ 11 (*) &   36.0 $\pm$ 17.7  & 13.4 & LPV        & SLF + rot?     & &  1+0    & yes     &  LS        \\
HD\,13268             & ON8.5\,IIIn       & 316 $\pm$ 10 &   21.0 $\pm$ 5.4   & 5.2 & LPV        & $\_$  &         &  0+0    & yes      &  LS              \\
HD\,14442             & O5\,n(f)p         & 320 $\pm$ 14 (*) &   15.0 $\pm$ 10.7  &  7.0 &  LPV        & SPB?   &        &  0+1    & no      &  LS        \\
HD\,41161             & O8\,Vn            & 322 $\pm$  8 &   10.5 $\pm$ 3.8   & 3.0 & LPV        & SLF  &          &  0+0    & yes      &  LS              \\
HD\,149452            & O9\,IVn           & 323 $\pm$ 14 &    1.5 $\pm$ 1.2   & 0.8 & LPV        & SLF  &          &  1+0    & yes      &  LS              \\
HD\,203064            & O7.5\,IIIn((f))   & 323 $\pm$  8 &   42.0 $\pm$ 8.7   & 10.6 & LPV/SB1?   & SLF           & &  0+0    & yes     &  LS (*)        \\
HD\,326331            & O8\,IVn((f))      & 323 $\pm$  7 &   17.5 $\pm$ 5.2   & 4.4 & LPV        & SLF  & $-6.80$$^f$          &  2+2    & no      &  LS           \\
HD\,46485             & O7\,V((f))nvar?   & 334 $\pm$ 16 &   29.5 $\pm$ 13.6 & 9.1 &  LPV        & \bf{EB+RM}      &  &  0+0    & no      &  \bf{SB1} (*)  \\
HD\,46056A            & O8\,Vn            & 365 $\pm$ 26 (*) &   29.0 $\pm$ 11.6 & 8.6 &  LPV        & PQ (+fr>5)     &  & 1+0    & no      &  LS         \\
HD\,117490            & ON9.5\,IIInn      & 369 $\pm$ 10 &   17.0 $\pm$ 6.6   & 4.8 & LPV        & SLF (+fr>5)    & & 0+2    & yes     &  LS              \\
HD\,102415            & ON9\,IV:nn        & 376 $\pm$  4 &   35.0 $\pm$ 10.8  & 12.0 &  LPV        & SLF (+fr>5)    & &  0+1    & no      &  LS            \\
HD\,93521             & O9.5\,IIInn       & 379 $\pm$ 14 (*) &   48.5 $\pm$ 8.0  & 11.2 & LPV        & SLF (+fr>5)    & $-8.7$...$9.4$$^g$ & 0+0     & yes     &  LS       \\
HD\,217086            & O7\,Vnn((f))z     & 394 $\pm$  9 &   15.5 $\pm$ 8.3  & 4.9 &  LPV        & SLF     &  $\sim-7.0$$^j$     &  0+0    & no      &  LS         \\
HD\,14434             & O5.5\,IVnn(f)p    & 395 $\pm$ 12 (*) &   21.0 $\pm$ 11.1  & 8.5 & LPV        & SLF  & $-6.77$$^h$          &  0+0    & yes      &  LS               \\
HD\,191423            & ON9\,II-IIInn(*)     & 397 $\pm$ 18 &   37.5 $\pm$ 9.6   & 9.6 & LPV        & SLF    &        &  0+0    & yes     &  LS          \\
HD\,149757            & O9.2\,IVnn        & 400 $\pm$  8 &   32.0 $\pm$ 9.5   & 4.7 & LPV        & $\_$ & $-7.13$$^c$          &  0+0    & yes     &  LS           \\
ALS\,12370            & O6.\,5Vnn((f))    & 444 $\pm$ 13 (*) &   15.0 $\pm$ 11.8 & 5.6 &  LPV        & SLF + rot/EV?  & &  $\_$   & no       &  LS        \\
\hline
\end{tabular}}
In this table we list the measurements of peak-to-peak amplitude of variation (RV$_{\rm PP}$) and standard deviation ($\sigma_{\rm RV}$) of all radial velocity measurements for each star. Apart from this, we indicate the spectroscopic binary and runaway status ('SB tag' and 'runaway?' columns, respectively), the type of detected photometric variability ('phot. var.' column), the relative X-ray flux ('X-ray' column), and the number of detected visual companions within 1 arcmin and between 1 and 2 arcmin, respectively ('contamination' column). In the column 'final binary status' we indicate our final decision regarding spectroscopic binary status. Eclipsing and spectroscopic binaries in the sample, as well as those stars identified to show ellipsoidal variability in the {\em TESS} light curves are highlighted in bold.\\ 
{\bf \vsini:} In those stars marked with (*) the use of the \ion{He}{ii}\,5411 line was necessary to estimate the \vsini; \\
{\bf SB tag:} LPV -- line profile variable; SB1, SB2 -- one or two spectroscopic binary respectively; LPV/SB1?, LPV/SB2? -- uncertain spectroscopic binaries; \\
{\bf Phot. var.:} SPB/$\beta$~Cep -- coherent low/high frequency variability, respectively; SLF -- stochastic low frequency variability; EB -- eclipsing binary; EV -- ellipsoidal variable; rot? -- possible rotational modulation; RM -- reflection modulation; PQ -- poor quality of the {\it TESS} light curve; fr$>$5 -- existence of prominent peaks at frequencies larger than 5~d$^{-1}$; $\_$ -- no {\it TESS} data or no data regarding visual components; "?" -- unknown periodic photometric modulation. \\
{\bf X-ray:}  $^a$ -- \citet{Rauw2015},  $b$ -- \citet{Naze2009, Bhatt_2010}, $^d$ -- \citet{Naze_2020}, $^e$ -- \citet{Rauw_2016},  $^f$ -- \citet{Sana_2006, Naze2009}, $^g$ -- \citet{Rauw_2012}, $^h$ -- \citet{Naze2009}, $^c$ -- \citet{Naze2018, Cohen_2021}, $^i$ -- this work, $^j$ 
 -- \citet{Getman_2006},  $^k$ -- \citet{Toala_2020}, $^l$ -- \citet{Naze_2013}.\\
{\bf Final binary status:}  LS -- likely/apparently single star.  Those stars marked with a (*) symbol have changed their SB status (compared to column 5) after taking into account all available empirical information (see Sect~\ref{spec_binaries} for details). In addition, although we keep them as SB1 for the purposes of this paper, \citet{Mahy_2022} have identified faint lines of a secondary component using a disentangling technique in a much larger spectroscopic data\,set in those stars marked with a ($\dagger$).\\
}
\end{table*}
\subsection{Empirical information extracted from the spectra}

\subsubsection{Spectroscopic and fundamental parameters}\label{parameters}

Throughout this paper, we consistently use the same set of spectroscopic parameters (basically \Teff\ and log\,$\mathcal{L}$) determined by \cite{Holgado_2020} and later utilised by \citet{Holgado_2022}. The corresponding estimates and associated uncertainties are indicated in columns 2 and 3 of Table~\ref{table:phys_parameters}. As commented in Sect.~\ref{sample}, these data allow us to locate our programme sample of stars in the sHRD (see red dots in the left panel of Fig.~\ref{figure_hr_brott})

In addition, the above-mentioned (spectroscopic) parameters were also complemented with other fundamental parameters, such as the radii, luminosities, and spectroscopic masses, which could be derived by considering the extinction corrected V magnitudes provided by \citet{Maiz_Barba_2018} and the {\it Gaia}-EDR3 distances quoted in \citet{Bailer_Jones_2021}. All these additional information can be found in Table~\ref{table:phys_parameters}. 

\subsubsection{Projected rotational velocities}\label{rotation}

Regarding the projected rotational velocities (\vsini), instead of directly using  the values obtained by \citet{Holgado_2022}, we decided to repeat the line-broadening analysis but this time using all available FIES, HERMES, and FEROS spectra per star. This way we wanted to investigate what is the associated uncertainty in the derived \vsini\ resulting from any potential source of spectroscopic variability affecting the line profiles. In addition, those cases in which an important scatter in the time-dependent \vsini\ measurements is present could  also be an indication that the star is actually a double-line spectroscopic binary.

To this aim, we applied again the {\sc iacob-broad} tool \citep{Simon-Diaz2014} to the same diagnostic line considered in \citet{Holgado_2022}. As commented in that paper, for the sake of homogeneity, we tried to use the \ion{O}{iiii}~$\lambda$5591 line in all cases. However, it was only possible for $\sim$40\% of the stars of the sample. For the rest of the stars, in which the \ion{O}{iii} line appear too weak and shallow due to the high \vsini, we needed to use either the \ion{He}{i}~$\lambda$5015 line ($\sim$37\%) or the \ion{He}{ii}~$\lambda$5411 line ($\sim$23\%).

\begin{figure}[!ht]
\centering
\includegraphics[width=0.49\textwidth]{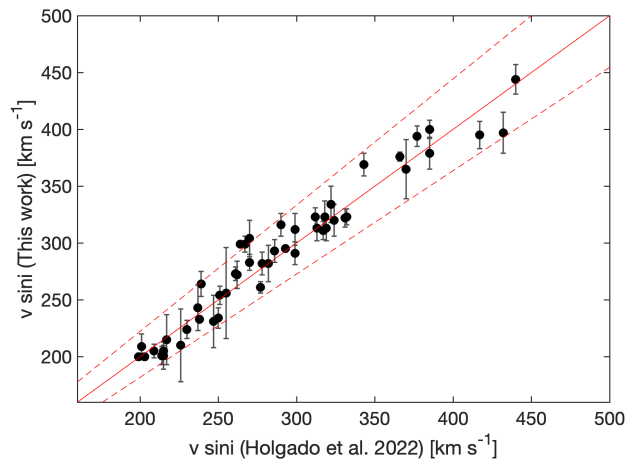}
\caption{Comparison of \vsini\ estimates obtained by \citet{Holgado_2022} and this work. In the case of our measurements, we include as the error bars the standard deviation associated with the multi-epoch line-broadening analysis. Diagonal
lines represent the 1:1 relation and the 10\% deviation,
respectively.
}
\label{figure_vsin_comparison}
\end{figure}

The results of this multi-epoch analysis are presented in column 3 of Table~\ref{table:b_fast}, where we indicate the mean values and associated standard deviations computed from the goodness-of-fit solutions provided by {\sc iacob-broad}. These new estimates are then compared with those obtained by \citet{Holgado_2022} in Fig.~\ref{figure_vsin_comparison}. Generally speaking, there is a very good agreement between both determinations and, except for a few (6) stars, the standard deviation resulting from the multi-epoch analysis is not larger than 5\,--\,6\%. 

In fact, the obtained standard deviation of the \vsini\ measurements is better than 10\% in all stars but three (HD\,152200, BD+60$^{\circ}$2522, and HD\,5689), where it ranges between 15 and 20\%. Interestingly, the {\it TESS} light curve of HD~152200 shows reflection modulation variability (see Sect.\ref{spec_binaries}), likely produced by a deformation of the star, hence affecting the \vsini~ measurements depending on the considered orbital phase. From visual inspection of the variability of some of the line-profiles of BD+60$^{\circ}$2522, we were not completely sure at first whether this star was a two-component spectroscopic binary (see Sect. \ref{visual_inspection}), something which could explain the measured scatter in the \vsini~ estimates. Regarding HD\,5689, we did not notice any peculiarities; however, we suggest using high deviation in \vsini\ as a hint for the possible identification of a faint companion or of peculiar photometric variability.

\subsubsection{Radial velocities}\label{radial}

Deriving radial velocities ($RV$) in stars with broad line profiles is a challenging task. Significant spectral broadening caused by fast stellar rotation affects the shape of the spectral line and simple Gaussian-Lorentzian fitting of such line profiles is not efficient. In addition, other effects such as line blending, the fact that rotational broadening makes the lines to become much shallower, or the occurrence of line profile variability within the broad line-profiles, hinder the application of such fitting technique. In this case, the use of a cross-correlation technique -- being aware that it has some limitations as well, especially when applied to spectra with a limited signal-to-noise ratio -- becomes a viable solution. While this technique allows the possibility to consider a spectral window including several lines, its application to a well isolated line is also a valid option.

For this work, we use the cross-correlation technique described in \citet{Zucker_2003}. We refer the reader to that paper for a detailed description of how a $RV$ estimate and its associated uncertainty can be obtained from the cross-correlation function (CCF) resulting from a specific observed spectrum and an adequate reference spectrum.

\begin{figure}[!t]
\centering
\includegraphics[width=0.49\textwidth]{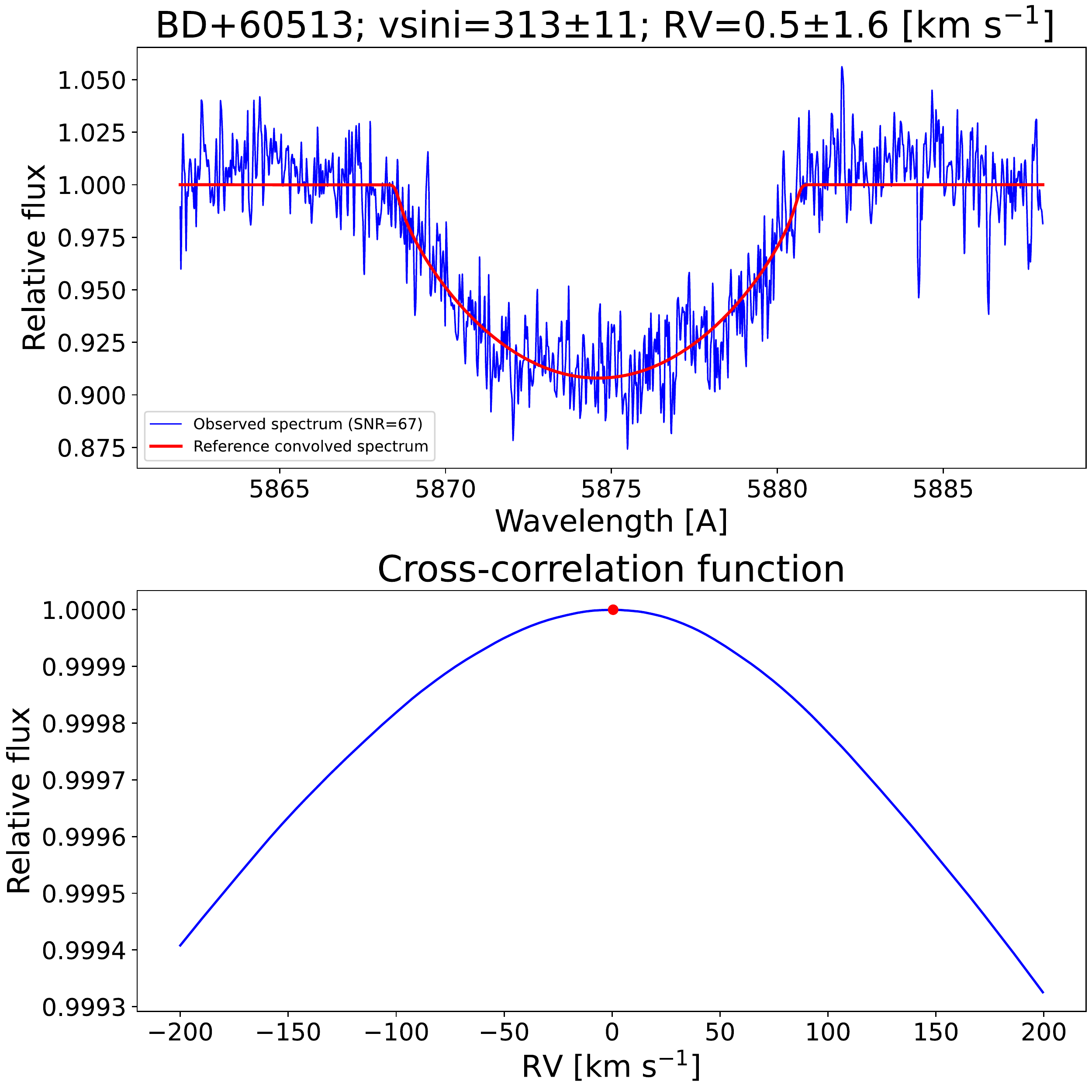}
\caption{Example of an observed He{\sc i}~$\lambda$5875 line-profile ({\it top panel)} and the best-fit synthetic line used as template for the computation of the cross correlation function (CCF, {\it bottom panel}). Following \cite{Zucker_2003}, the radial velocity and its associated uncertainty can be computed from the maximum and the sharpness of the CCF.}
\label{example_spectrum}
\end{figure}

In particular, we decided to apply this technique to just one prominent and unblended diagnostic line which is present in all stars of our programme sample: \ion{He}{i}~$\lambda$5875. This decision was taken after evaluating the possibility of using a larger set of diagnostic lines present in the optical spectra of O-type stars (6 in total). As illustrated in Fig.~\ref{figure_lines_example}, the main outcome of this exercise was that the attempt of using other lines beyond \ion{He}{i}~$\lambda$5875 was neither improving the accuracy in the \RVpp\ estimations (see below for definition of this quantity) or importantly modifying the results obtained using just the most prominent (and always present) \ion{He}{i} line.

Lastly, we are aware that also \ion{He}{i}~$\lambda$5876 line can be affected by stellar wind in some specific cases, this effect should not significantly affect the shape of the CCF if the wind emission is not very pronounced as expected for our sample, which is mostly composed of late O-type stars. Indeed, one of the advantages of the cross-correlation approach is that the behavior of the CCF does not depend on the relative distribution of the input data, that is, if we cross-correlate spectral lines with different widths, it does not affect the resulting function. This is why we can securely cross-correlate one convoluted line profile template with the rest of the spectra even if the lines change width or shape. 

Technically speaking, we started by fitting a rotationally convolved (synthetic) profile to the \ion{He}{i}~$\lambda$5875 line of the first spectrum in the multi-epoch data~set available for a given star (see e.g. the top panel in Fig.~\ref{example_spectrum}). Then, by taking this synthetic line profile as our reference, we cross-correlated the rest of available spectra in the same wavelength range, thus obtaining a CCF as depicted in bottom panel of Fig.~\ref{example_spectrum}. 


\begin{figure*}[!ht]
\centering
\includegraphics[width=0.24\textwidth]{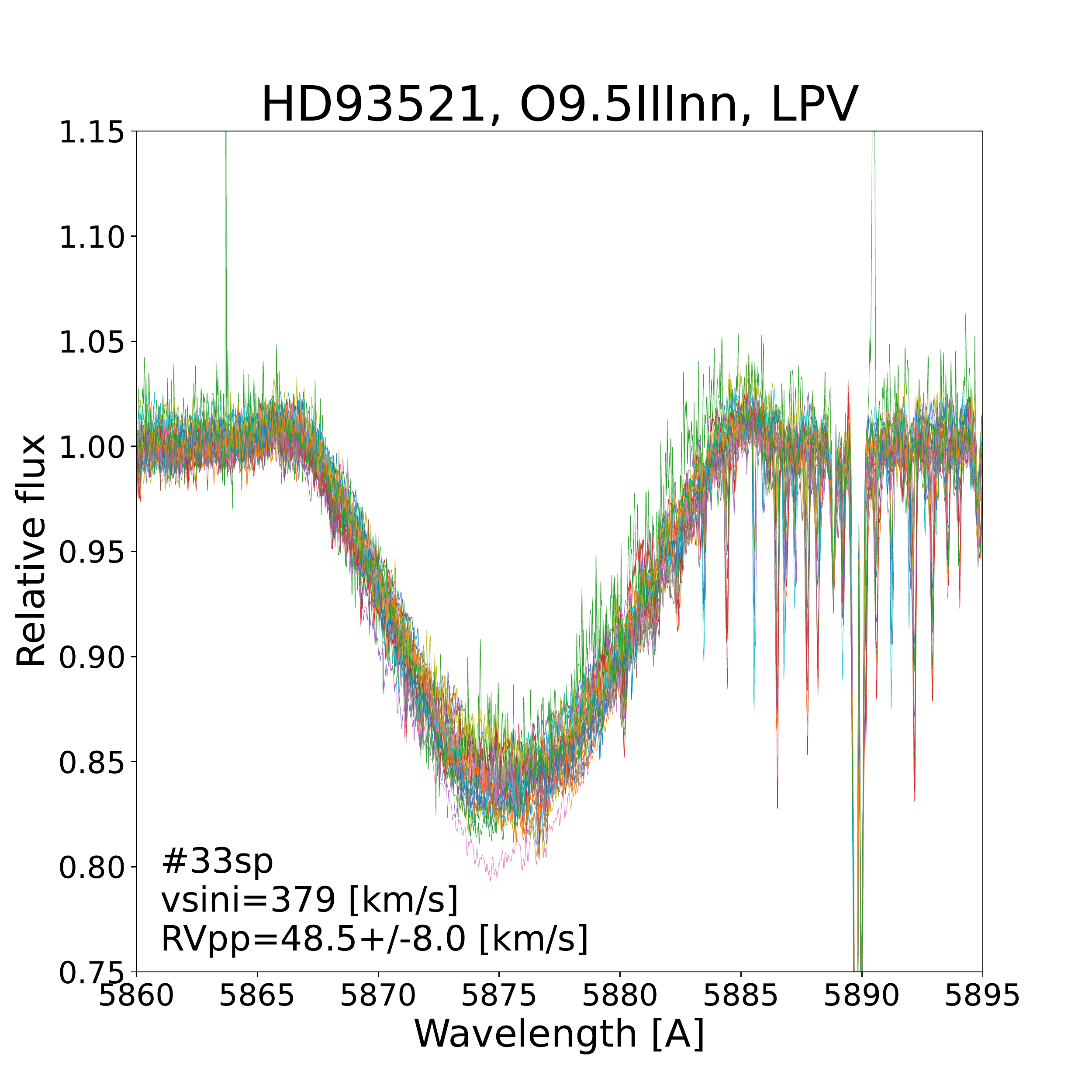}
\includegraphics[width=0.24\textwidth]{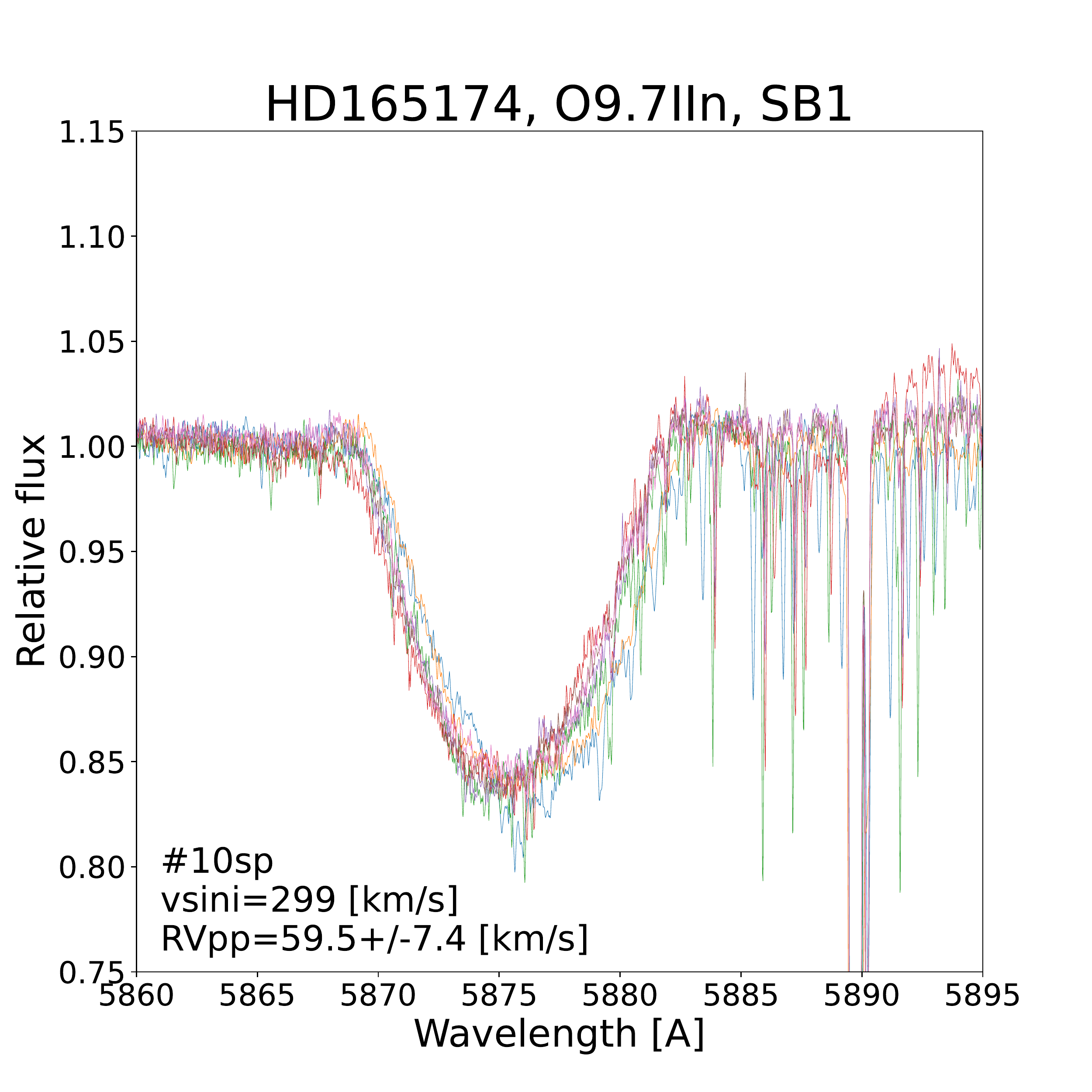}
\includegraphics[width=0.24\textwidth]{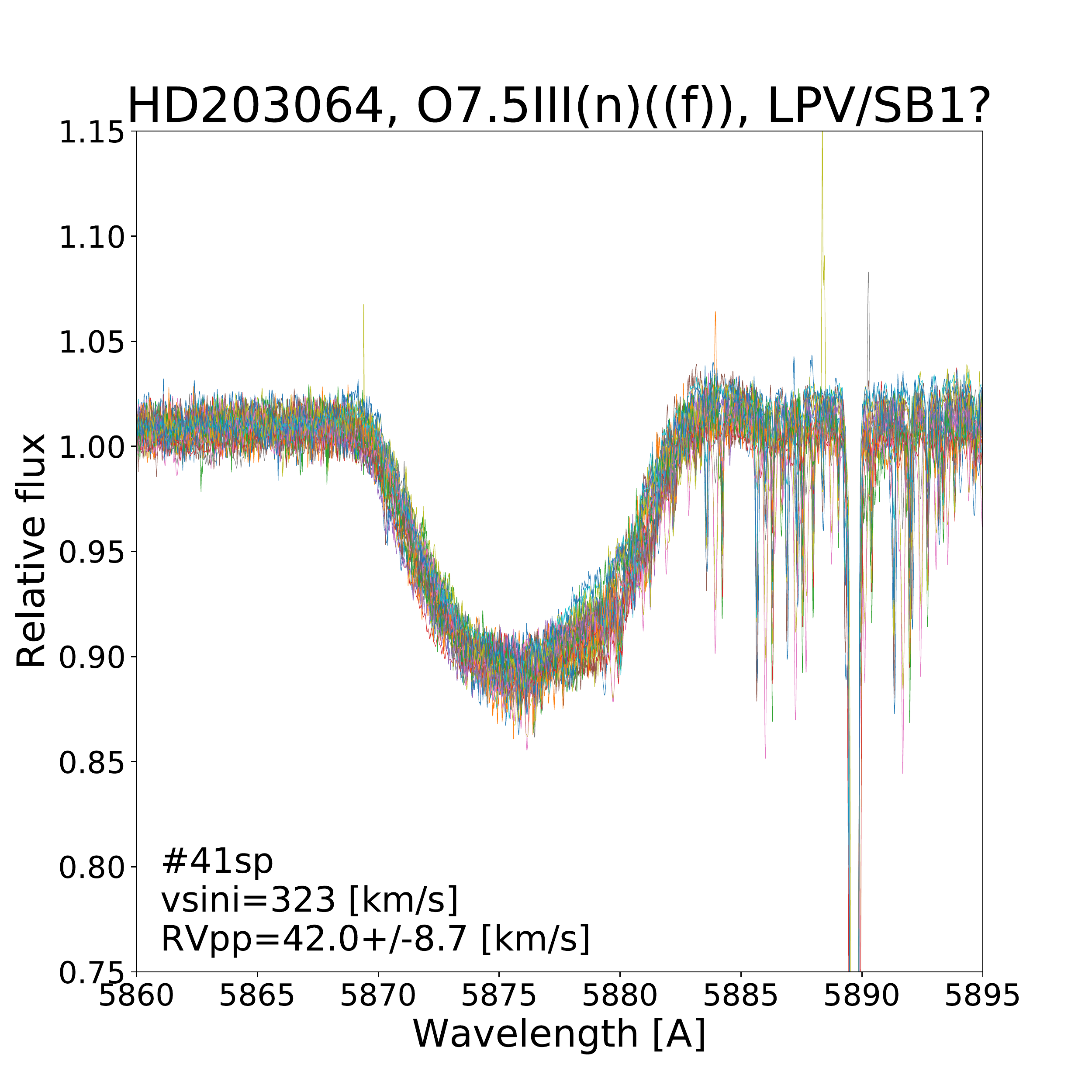}
\includegraphics[width=0.24\textwidth]{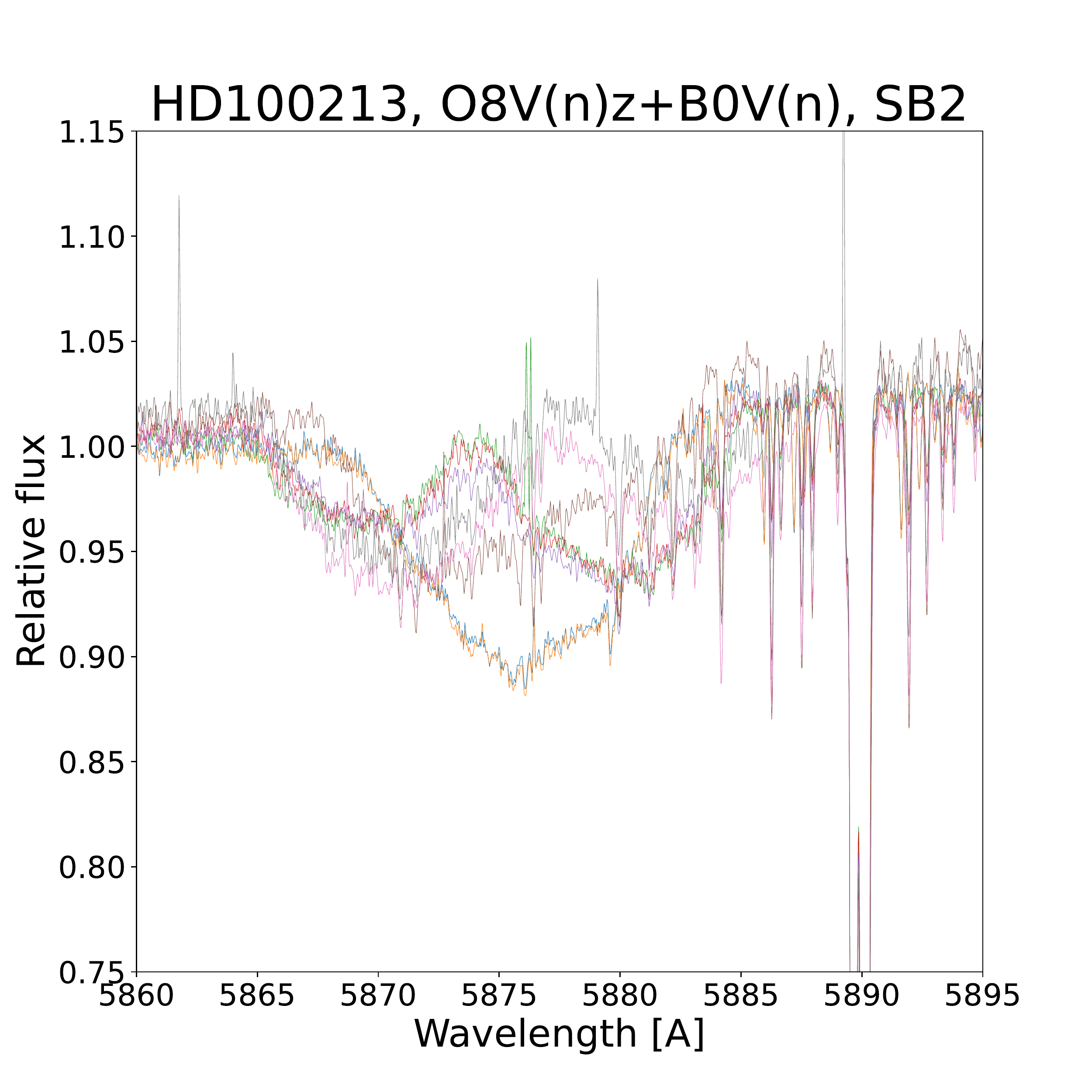}
\includegraphics[width=0.24\textwidth]{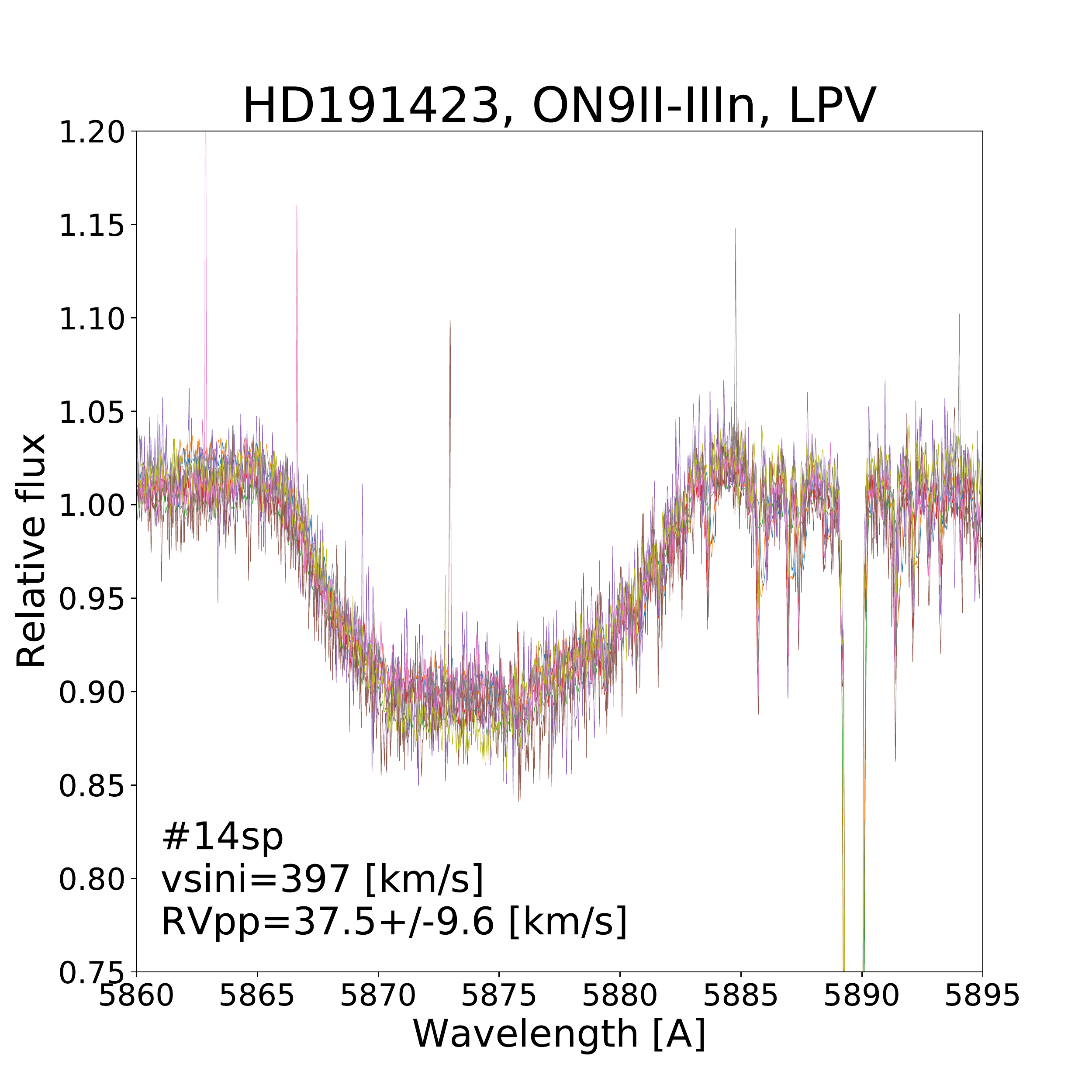}
\includegraphics[width=0.24\textwidth]{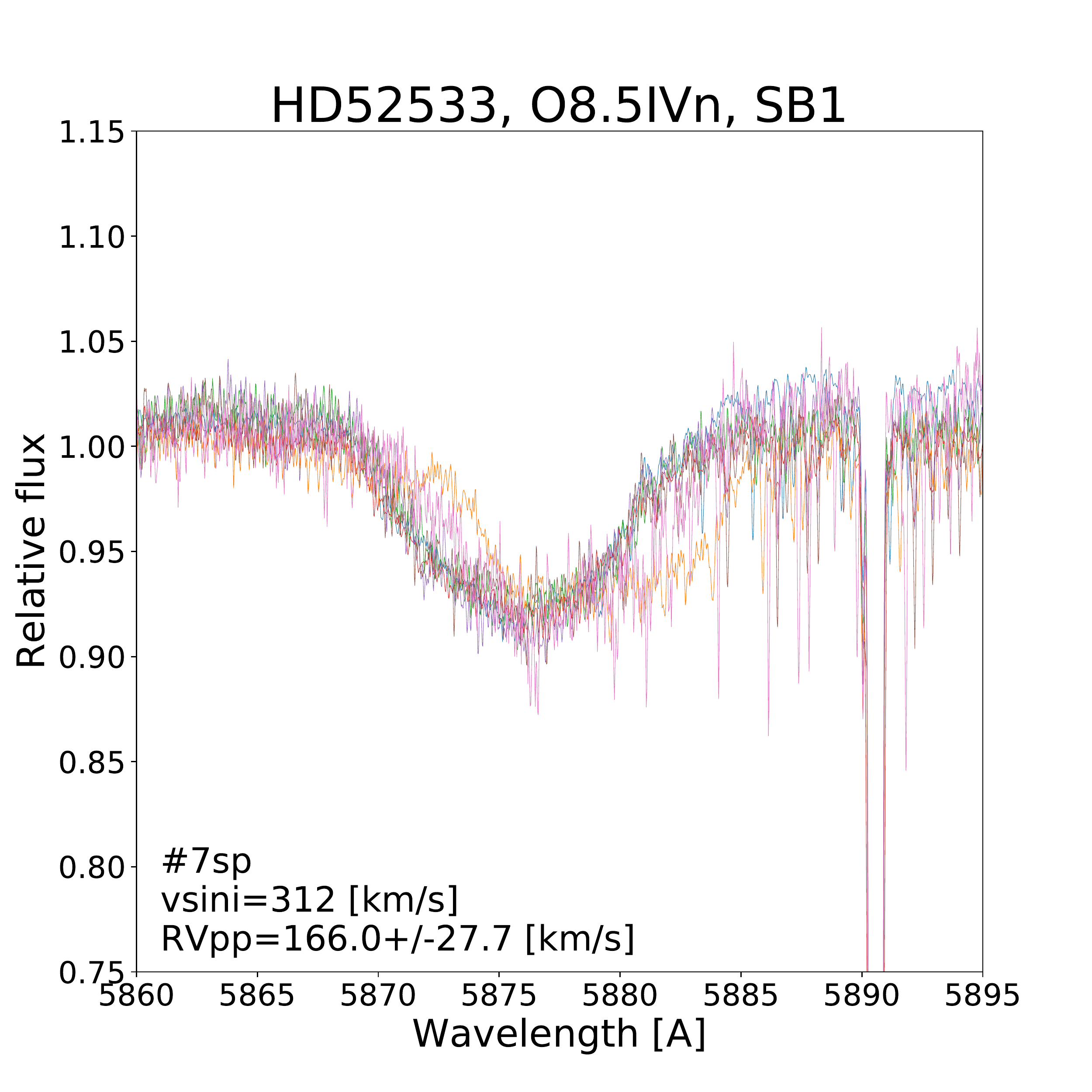}
\includegraphics[width=0.24\textwidth]{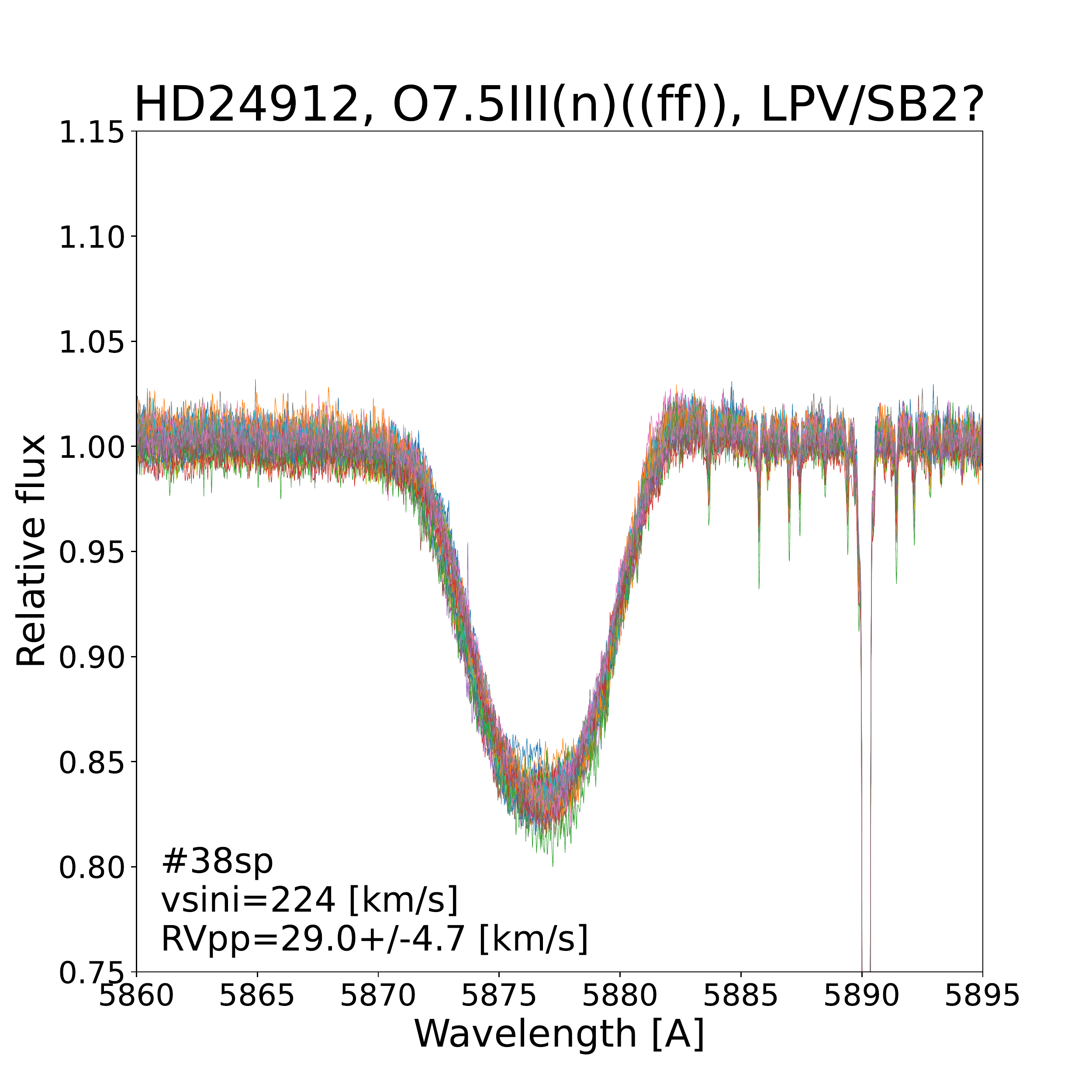}
\includegraphics[width=0.24\textwidth]{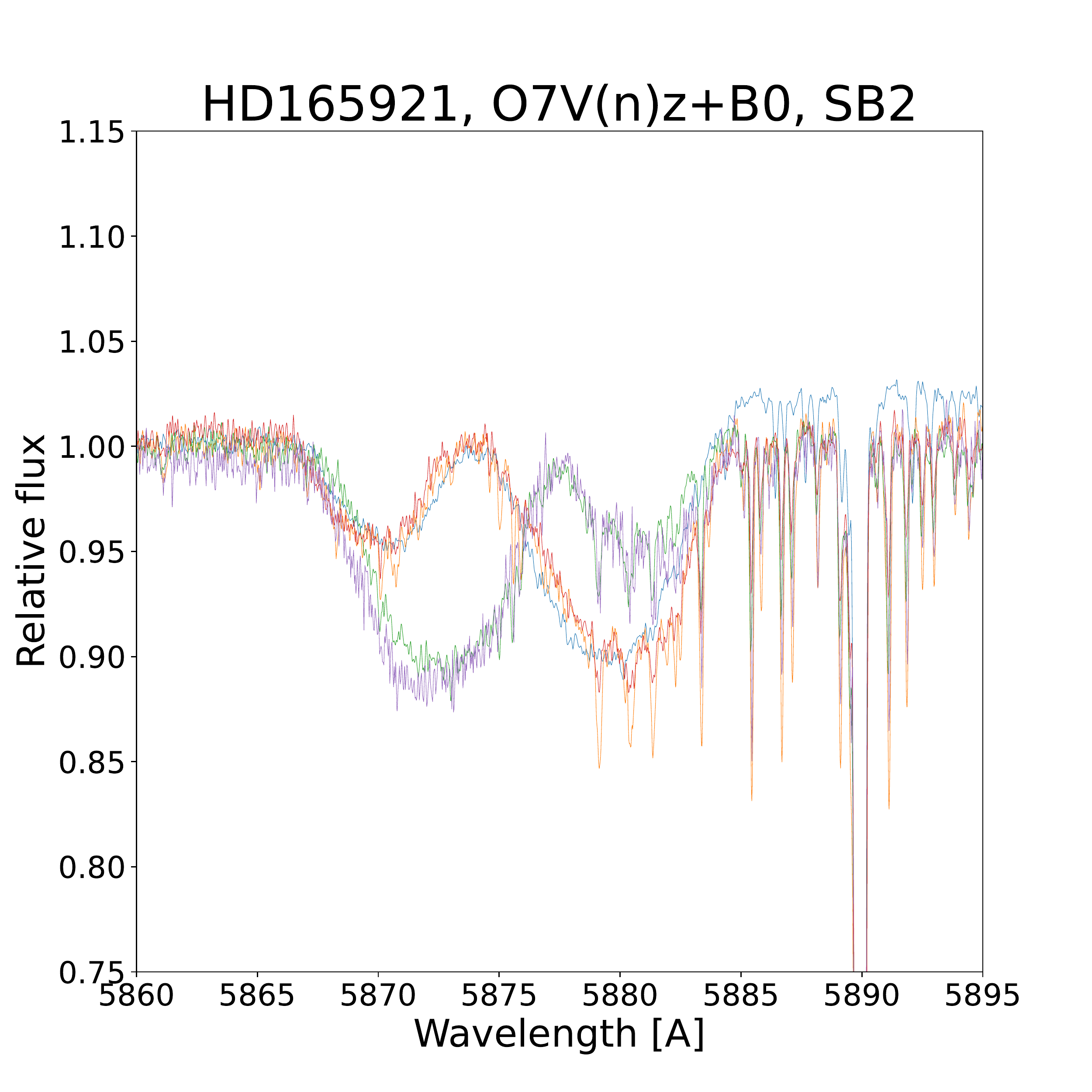}
\caption{Eight examples of the type of line-profile variability detected in our sample of 54 fast-rotating O-type stars. In each panel, we indicate the name of the star, its spectral classification, the number of available spectra, the measured \vsini\ and \RVpp, and the visually assigned classification within the following categories: LPV, SB1, LPV/SB1?, LPV/SB2?, and clear SB2 systems. We also include in the various panels the (much narrower) \ion{Na}{i}~$\lambda$5890 interstellar line that we use as a sanity check to make sure that the heliocentric correction has been adequately applied to each spectrum in the time-series. Stars are sorted by spectral classification.}
\label{spectra_example}
\end{figure*}

As a result, we obtained $RV$ measurements relative to the first spectrum and their associated uncertainties for all available spectra. Based on these values, we calculated the peak-to-peak amplitude of radial velocity variability (\RVpp) following the various steps described below\footnote{Tables xx and xx with individual $RV$ measurements for each star and \RVpp~ estimates are only available in electronic form
at the CDS via anonymous ftp to cdsarc.cds.unistra.fr (130.79.128.5)
or via https://cdsarc.cds.unistra.fr/cgi-bin/qcat?J/A+A/}.

First, to make sure that only reliable $RV$ measurements are considered, we calculate the S/N of our diagnostic line for each spectrum in the time-series. This quantity is defined as:
\begin{equation}
{\rm S/N}_{\rm L} = \frac{1-{\rm min}(F_{5875})}{\sigma(F_{5855 - 5865})}
,\end{equation}
where min$(F_{5875})$ is the flux at the core of the line-profile and $\sigma(F_{5855 - 5865})$ is the standard deviation of the normalised flux between 5855 and 5865~\AA, a spectral window which only includes continuum points. Basically, this quantity indicates how prominent the spectral line is with respect to the continuum noise.

After evaluating several options, we decided to exclude all RV estimates for spectra with a  ${\rm S/N}_{\rm L}$\,$<$\,5. Even with this conservative threshold, most of the initially available spectra  endure to the end of this cleaning process, indicating the good quality of our spectroscopic data~set. The estimates of ${\rm S/N}_{\rm L}$ and overall ${\rm S/N}$ for all measured spectra as a function of uncertainty in \RVpp~ are presented in Fig.~\ref{figure_p2p_err}.

Then, in order to compute a final estimate of \RVpp, we use the pair of $RV$ measurements that maximises the quantity:
\begin{equation}
\frac{|RV_{i}-RV_{j}|}{\sqrt{(\sigma_{i}^{2}+\sigma_{j}^{2})}}
,\end{equation}
defined in \citet{Sana_2013}, where $i$ and $j$ refer to two different spectra from the time-series of a given star, and $\sigma_{i}$ and $\sigma_{j}$ are the individual uncertainties of the radial velocity measurements $RV_{i}$ and $RV_{j}$, respectively. This approach helps to identify \RVpp\ from the most accurate measurements by taking into account the individual uncertainties of radial velocity measurements. Also, the uncertainty associated with the final estimate of \RVpp\ is computed as $\sigma$(\RVpp)\,=\,$\sqrt{(\sigma_{i}^{2}+\sigma_{j}^{2})}$.

Both quantities are presented in column 4 of Table~\ref{table:b_fast}, while the number of spectra that are finally considered in computing them are quoted in column 7 of Table~\ref{table:sample}. Generally speaking, the accuracy of our \RVpp\ estimates is better than 10~\kms\ in most of the stars in our programme sample.


\subsubsection{Visual inspection of line-profile variability}\label{visual_inspection}

Orbital motion is not the only effect that can led to the detection of large radial velocity variations in O-type stars, especially in the case of fast rotators. As shown elsewhere \citep[e.g.][]{Fullerton_1996, Aerts_2014, simondiaz2017}, the line profiles of most O- and B-type stars are often subject to various types of variability due to, for instance, stellar pulsations or spots. In addition, in the case of the \ion{He}{i}\,$\lambda$5875 line a non-spherically symmetric wind could also produce some variability which could be erroneously interpreted as empirical evidence of the star being a spectroscopic binary (SB) if only the $RV$ measurements are taken into account.

We illustrate this argument in Fig.~\ref{spectra_example}, where we show the detected line-profile variability in six illustrative examples including two unambiguously identified SB1 systems among our working sample, plus another four cases in which, despite a relatively high \RVpp\ having been measured (reaching up to 48.5~\kms\ in one of the cases\footnote{HD\,93521, an O9.5~\,IInn star which has been extensively studied in the literature \citep[e.g.][and references therein]{Howarth_1993, Rauw_2012, Gies_2022}, but never identified as spectroscopic binary.}), the detected variation in $RV$ is likely produced by intrinsic line-profile variability. 



Taking into account these ideas, we decided to complement the information about the measured \RVpp\ with the outcome from a visual inspection of the detected variability of the \ion{He}{i}\,$\lambda$5875 line profile in all stars in our programme sample, classifying each star as being part of one of the following subgroups: 'LPV', that is,   line profile variables 'SB1', 'LPV/SB1?', and 'LPV/SB2?'. This information is provided in 'SB tag'
 column of Table~\ref{table:b_fast}. 


We used the LPV tag to identify stars which are likely single (LS) as we only detected small variations in the shape of the \ion{He}{i}\,$\lambda$5875 line-profiles. In an opposite situation, if a significant shift of the entire spectral line is detected visually, we classified the star as SB1. If the detected line-profile variability is not prominent, but there seems to be an overall shift of the line, we consider the star as an unclear SB1 and marked it as 'LPV/SB1?'.

In addition to this, we identified five SB2 candidates for which we were not completely sure about their spectroscopic binary nature. Hence, these stars were provisionally labeled as 'LPV/SB2?'. One interesting example of this type is HD\,24912 ($\xi$~Persei). In this case, the line-profile variability is most probably a consequence of the existence of co-rotating bright spots on the surface of the star, as detected by \citet{Ramiaramanantsoa_2014}, using photometric observations provided by the MOST satellite, and, hence, it is not an SB2 system. We excluded this star from the final statistics of 'LPV/SB2?'. Further notes on the other four targets can be found in Appendix~\ref{sb2_appendix}, along with a final decision of their revised spectroscopic binary status.

When presenting the final statistics of detected spectroscopic binaries (Sect.~\ref{final_statistics} and Table~\ref{table:fast_stat}) among our sample, the 'surviving' SB2 candidates will be added to the other four clearly detected SB2 systems\footnote{Indeed, one of them is detected as a SB3 system, see Appendix~\ref{sb2_appendix}.} included in the initial sample of 415 O-type stars for which we have found that at least one of the two components has a \vsini\ larger than 200~\kms. For reference purposes, we also show two illustrative examples of the detected line-profile variability in the case of these double-line spectroscopic binary systems (see Fig.~\ref{spectra_example}).

In this regard, we also provide here some further comments about the strategy we followed to identify SB2 systems. Again, our main diagnostic line for visual detection of double line spectroscopic binaries has been the \ion{He}{i}\,$\lambda$5875 line. This is not only one of the stronger lines in the spectra of O-type stars, but is also a line which remains strong in the full B star domain. This characteristic makes the line perfect to detect any secondary component hidden in the spectrum, even in cases where the \vsini\ of this second component is large and, hence, the line is greatly diluted. As a consequence, we can state with a high degree of confidence that given the quality of our spectroscopic data\,set (in terms of resolving power and S/N), this first visual inspection allows us to detect all possible companions contributing down to $\sim$10\,--\,20\,\% of the total flux of the system in the optical range, especially when we have enough epochs and the amplitude of $RV$ of this (fainter) secondary component is larger than $\sim$\,70\,--\,80\% of the \vsini\ of the primary.

Obviously, there will be certain cases in which this visual inspection will fail, specifically when the amplitude of $RV$ variability of the fainter companion is less than $\sim$\,50\% of the \vsini\ of the more luminous star. This case is expected to affect more importantly to stars with larger \vsini\ (i.e. the fast-rotator domain), thus leading to situations as those described above, where we are not sure if the detected line profile variability is due to any type of intrinsic variability in a single star or the presence of a companion (the LPV/SB2? case). To minimise the number of such cases, we specifically increase the number of available epochs in the sample of fast rotators and also explored more carefully other diagnostics lines which could help us to decide if we have a SB2 system or a LPV case. Also, we explored the possible detection of eclipses in the available {\it TESS} light curves (see Sect.~\ref{tess}), as well as the potential identification of close-by companions using high angular-resolution images (see Sect.~\ref{visual_components}) in order to complement the spectroscopic information with the aim of minimizing as much as possible the effect of observational biases in our detection of both single- and double-line spectroscopic binary systems.

Some additional notes about how the above-mentioned possible observational biases could be affecting the resulting statistics of detected spectroscopic binaries in stars in both the faster and slower rotating samples can be found in Sect.~\ref{final_statistics}.

\subsection{Empirical information extracted from {\it Gaia} and {\it TESS} data}\label{Extra_info}

The $Gaia$ and {\it TESS} missions have provided a unique opportunity to have access to very valuable information of interest for our study in a homogeneous (and almost unbiased) way.

On the one hand, the proper motions delivered by $Gaia$-EDR3 \citep{gaia_edr3} can be used to identify runaway stars. Some of these runaways are expected to be produced by the dynamical ejection of the surviving companion in a high-mass binary system after a supernova explosion event. Following \cite{deMink_2013}, an important fraction of the O-type stars with \vsini\,$>$\,200~\kms\ are the mass gainers in binary systems after Roche-lobe overflow of the initially more massive star. In this sense, it is interesting to know, not only which of the stars in our programme sample are identified as a runaway star, but also to compare the percentage of runaways detected in the slow- and fast-rotator samples of O-type stars investigated in \citet{Holgado_2022}.

On the other hand, the high quality light curves provided by the {\it TESS} mission for almost all our programme stars (with the caveats described in Sect.~\ref{tess}), allow us to search for signatures of hidden companions not detected through our multi-epoch spectroscopy. Also, although such study is out of the scope of this paper, a thorough investigation of the detected photometric variability (by means of standard asteroseismic data analysis techniques) can provide new insights about the evolutionary nature of fast rotators.

\subsubsection{Detection of runaway star candidates among Galactic O-type stars using {\it Gaia} ED3 proper motions}\label{gaia}

It has been known since the 1950s that some OB stars move at high speeds through the Galaxy as a consequence of dynamical interactions between three or more bodies in stellar clusters or of supernova explosions in binary systems \citep{Zwic57,Blaa61,Poveetal67}. Some of the stars are ejected with velocities higher than 30 \kms. Those are called runaway stars \citep{Hoogetal01} and can be easily found with {\it Gaia} astrometry \citep{Maizetal18b}. The slightly slower ones defined as walkaway stars are more common but more difficult to identify \citep{Renzo_2019runaways}.

In order to identify runaway stars one needs to calculate their 3D velocity with respect to their local standard of rest (LSR). Strictly speaking, one should differentiate between the ejection velocity and the current velocity due to the possible differences between them caused by the different locations in the Galaxy and the effect of the Galactic potential on the trajectory (see \citealt{Maizetal21f} for examples). However, such differences are usually small (especially for recent ejections) and their calculation requires knowledge about the location of the ejection event, something that we do not currently have for most runaway candidates. For that reason, we consider only the current velocities here. 

The 3D velocity is calculated from the 2D components of the tangential velocity and the radial velocity. To obtain the tangential velocity ($v_{t,lsr}$) of a star we need to know its distance and proper motion. In this domain, {\it Gaia} has opened up the door to a revolution in our knowledge. Radial velocities are a different story. {\it Gaia} will provide radial velocities for many stars using its radial velocity spectrometer, which operates in the calcium triplet window. Unfortunately, O-stars have few lines in that wavelength range and the most prominent ones there belong to the Paschen series, which are too broad to obtain precise radial velocities. Furthermore, O-stars suffer from their multiplicity that makes many of them spectroscopic binaries, hence requiring multiple epochs to determine their average radial velocities accurately. To complicate matters further, the spectral lines of O-stars are broad and affected by winds, pulsations, and other effects that lead to disagreements between measurements by different authors \citep{Trigetal21}. 

\begin{figure}[!t]
\centering
\includegraphics[width=\linewidth]{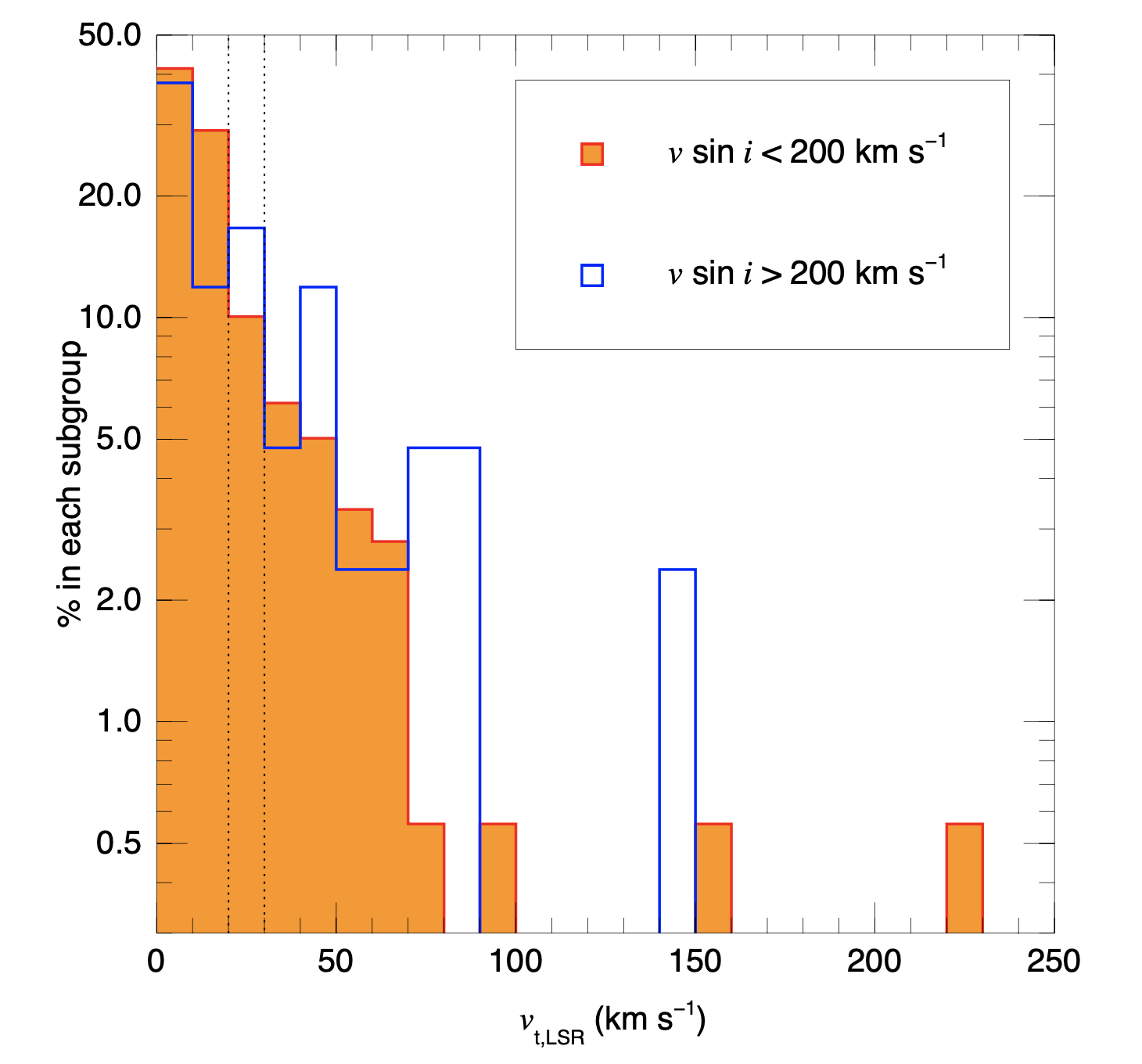}
\caption{Histogram (in percentage of the sample) of tangential velocities with respect to their LSR for the slow- and fast-rotator samples. The two dotted vertical lines indicate the critical velocities $v_{t,lsr}$ of 20~\kms~ and 30~\kms, respectively.}
\label{vstats}
\end{figure}

Given the limitations described above, we estimated the number of runaways in the initial sample of 285 O-type stars not identified as SB2 stars by \citeauthor{Holgado_2022} (see Sect.~\ref{sample}) -- splitting the information between the two samples of fast and slow rotators, respectively (see Fig.~\ref{figure_hr_brott}) --  using only the information available from {\it Gaia}-EDR3 data, that is, the parallaxes and proper motions. We applied the astrometric calibration of \citet{Maiz21}, which includes a zero point for the parallaxes that depends on magnitude, color, and position in the sky and a correction on the parallax and proper motion uncertainties that depends on magnitude and the correction to proper motions of \citet{CanGBran21}. Distances are calculated with the prior of \citet{Maiz01a,Maiz05c} and the parameters of 
\citet{Maizetal08a}, as those are the most appropriate for Galactic O-type stars. We use the Galactic rotation curve described in \citet{Maizetal21f} and the velocity of the Sun with respect to its LSR from \citet{Schoetal10b}. Using those parameters, we calculated the velocity of the stars in the plane of the sky with respect to their LSR.

The uncertainties on the calculated velocities depend first on the parallax uncertainties: a large value implies a large uncertainty on the distance and from there on the tangential velocities. They also depend on the distances themselves, as for stars that are far away the subtraction of the LSR velocity may be biased by our assumed Galactic rotation curve. Hence, in order to avoid objects with large velocity uncertainties, we restricted our samples to those objects with relative distance uncertainties ($\sigma_{\varpi, {\rm ext}}/\varpi_{\rm c}$, see \citealt{Maizetal21c} for the notation) lower than 1/3 and (average) distances smaller than 4~kpc. With those restrictions, the sample now contains 179 slow rotators (out of 235) and 41 fast rotators (out of 50).

Figure~\ref{vstats} compares the histograms of tangential velocities for the two samples. Overall, 14/41 (34.1\%) of the fast rotators have tangential velocities above the runaway threshold of 30 \kms, while for slow rotators, only 35/179 (19.6\%) are above the threshold. Those objects are secure runaways but it is possible that some objects have a radial velocity with respect to their LSR large enough to bring them across the threshold when combined with their tangential velocity. To estimate how many runaways we are missing, we can count how many have tangential velocities in the 20-30~\kms\ range, as those are the most likely candidates to shift status. We label those as `possible runaways' and their numbers are 7/41 (17.1\%) for the fast rotators and 18/179 (10.1\%) for the slow rotators.

Jumping to a more detailed investigation of our working sample of fast rotators, column 10 of Table~\ref{table:b_fast} (under the heading 'runaway?') quotes the 26 stars identified as runaways. We note that, in this case, we also take into account previous findings in the literature, not necessarily based on {\it Gaia}-EDR3 data \citep[see e.g.][and references therein]{Maizetal18b}, but also reported by other surveys: HD\,191423 \citep{LAMOST_2020}, HD\,117490 \citep{Li_2020ApJ}, and HD\,15642 \citep{deBurgos_2020}. Interestingly, six out of the nine stars not fulfilling the distance criteria mentioned above are recovered as confirmed runaways, with a couple of them (BD+60$^{\circ}$2522 and HD\,41161) being tagged as runaways due to the existence of an associated bow shock, despite having tangential velocities in the 20-30~\kms\ range \citep{Green_2019}. These additions increases the number of 'bona fide' runaways among the sample of fast rotators to 52\%, namely, a value closer to that obtained from {\it Gaia}-EDR3 data, assuming a threshold in tangential velocity of 20~\kms\ (instead of 30~\,\kms).

Some implications of these findings regarding the global sample, as well as further notes about the connection between the spectroscopic binary and runaway status of the sample of fast rotators can be found in Sections.~\ref{rw_discussion} and \ref{final_statistics}.

\subsubsection{Using the {\it TESS} light curves to identify eclipsing binaries and B-type hidden companions}\label{tess}

\begin{figure*}
    \centering
    \includegraphics[scale = 0.6]{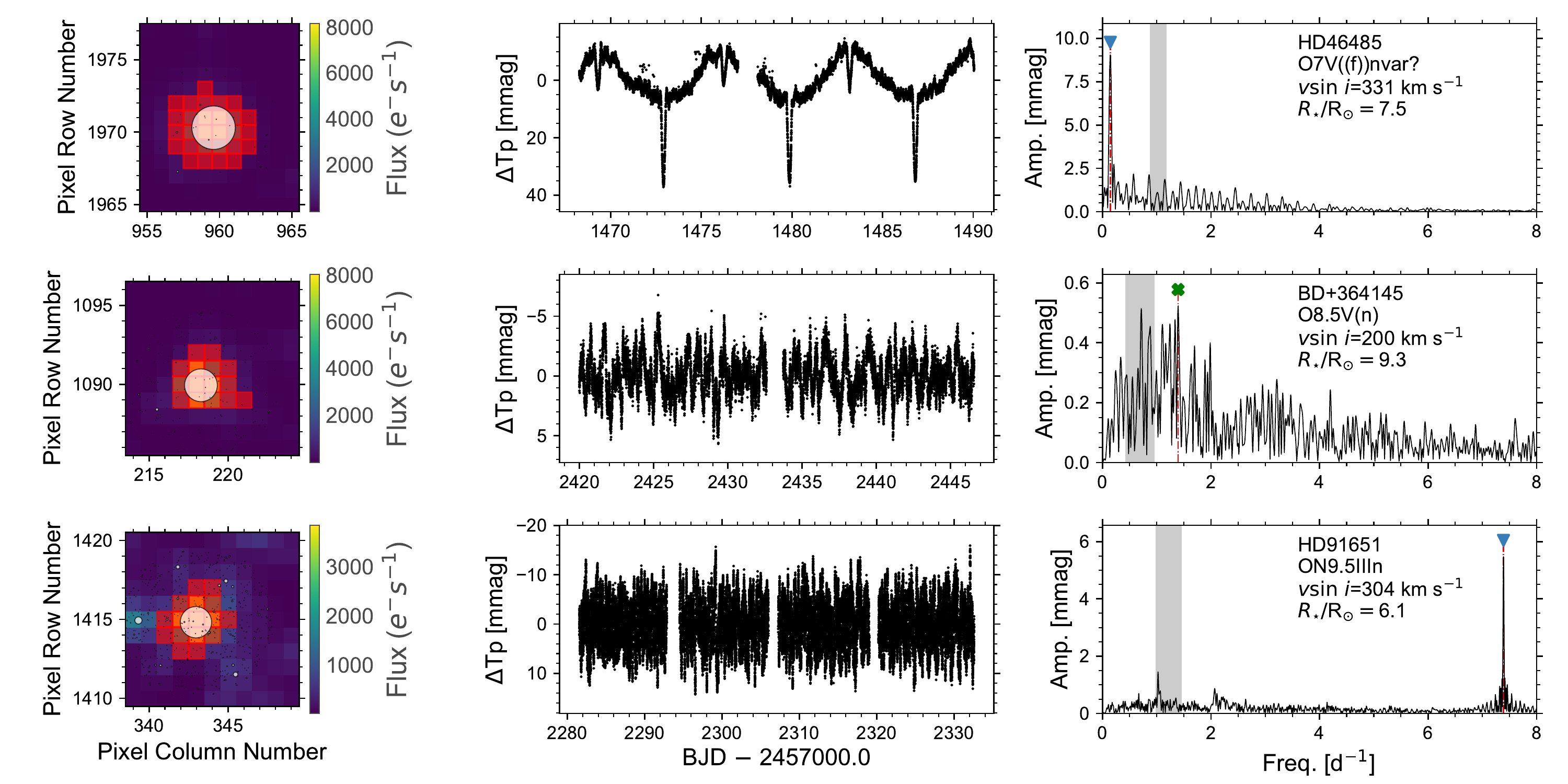}
    \caption{Three examples of the typical {\it TESS} pixel maps, light curves, and periodograms for fast-rotating O-stars. In the left panels, we show the {\it TESS} pixel masks, with the pipeline mask in red and {\it Gaia} sources marked by white circles. The markers are scaled logarithmically with the {\it Gaia} magnitude of the source. In the middle panels, we show the one (or two) sector(s) {\it TESS} light curves extracted from the {\it TESS} pixel maps. In the right panels, we show the periodograms of the light curves. Significant frequencies, satisfying a signal to noise of $5$ in a $1$~d$^{-1}$ window, are given by blue triangles. In the case of BD+36$^{\circ}$4145 the highest amplitude frequency was not significant, and we therefore mark it by a green cross. The rotational modulation frequency range is given in gray. This range is estimated using the measured $v\sin\,i$, an upper limit of $v_{\rm eq}=450$~km~s$^{-1}$ to account for the inclination and the stellar radius inferred from {\it Gaia}-EDR3 (cf. Table~\ref{table:phys_parameters}).}
    \label{fig:example_tess_plot}
\end{figure*}

The {\it TESS} mission \citep{tess} is delivering an enormously rich amount of high quality data for the asteroseismic analysis of large samples of stars covering a broad range of masses and evolutionary stages. In addition, the light curves delivered by {\it TESS} serve, among other things, for the detection of eclipsing binary systems (not necessarily transiting exoplanets) and the identification of specific variability patterns associated with the presence of spots at the stellar surface, variable winds, disks, or magnetospheres.

Our interest in getting access to the {\it TESS} lightcurves for a good fraction of our sample of fast-rotating O-type stars was mainly twofold. On the one hand, we wanted to identify signatures of eclipses in the photometric data, which could be indicating the presence of a companion not necessarily detected via time-series spectroscopy. On the other hand, by analyzing the resulting periodograms, we would be able to detect frequency patterns associated with fainter, less massive companions (i.e. $\beta$~Cep and SPB type pulsators, see \citealt{Aerts2010} for an overview) whose identification would not be straightforward in the spectra.

An example of this latter situation was presented in \citet[][their Fig.~10]{Burssens2020}, where it was shown that the periodogram obtained from the {\it TESS} lightcurve of the O7~V star HD\,47839 -- known to be $\sim$25 years period SB1 system \citep{Gies_1993,Gies_1997}. In addition, one of the longstanding standards for spectral classification clearly shows a high frequency peak (at $\sim$12.5~d$^{-1}$), which most likely corresponds to a faint B-type companion and not to the bright O7~V star.

With these two ideas in mind, we first searched and extracted the {\it TESS} 2-min short-cadence data from the Mikulski Archive for Space Telescopes (MAST\footnote{\url{https://archive.stsci.edu/}}) for all stars in Table~\ref{table:b_fast}, if available (20 stars). We retrieve two light curves, the light curve extracted using simple aperture photometry (referred to as SAP) and the pre-conditioned light curve (Pre-search Data Conditioning Simple Aperture Photometry, PDCSAP). The latter has systematics removed that are common to all stars on the same CCD \citep{Jenkins2016}. Nonetheless, the SAP light curve may be preferable for certain stars as the pipeline is not optimised for OB stars. 

By means of a visual comparison and based on predicted variability in the O-star regime, we selected the preferable light curve for each star. We additionally inspected the light curve aperture masks using the \texttt{lightkurve} software package \citep{lightkurve2018} to rule out any contamination by nearby sources, large sector-to-sector mask variations, or the presence of oversaturated pixels. Light curves for which this was the case were removed from further consideration (1 star).

For stars with no available 2-min short cadence data, we extracted 30-min long cadence data using the \texttt{lightkurve} software package. If available (21 additional stars), we performed simple aperture photometry using a watershed method. That is, we included pixels with a light contribution within $8\sigma$ in flux of the light contribution of the central pixel of the source of interest. The extracted light curve was then detrended using principal component analysis, following \citet{Garcia2022}. Again, problematic light curves were removed from further consideration (HD\,216532).

The extraction procedure yielded 39 {\it TESS} light curves, all of which show some form of variability. This includes stochastic low-frequency variability, coherent pulsation modes, and rotational modulation and eclipses, which are in line with general findings for the O-star regime \citep{Pedersen2019, Burssens2020}. We show three examples of typical pixel maps, light curves and periodograms in Fig.~\ref{fig:example_tess_plot}.  A detailed discussion about the photometric variability for each interesting target is presented in Appendix \ref{Tess_appendix}.  All information regarding  photometric variability is presented in Table \ref{table:b_fast} (column 'Phot. var.').

\subsection{Compiling extra information from the literature}\label{literature_info}

In addition to our own spectroscopic, photometric and astrometric analysis, we also performed a careful search in the literature for extra relevant information about our programme sample of stars. In particular, we wanted to know if any of the stars in our sample had been previously identified as a spectroscopic binary, as well as to compile any type of orbital and dynamical information resulting from any existing (more detailed) study of specific targets. Also, we gathered published information about close-by companions as resulting from high angular resolution surveys. Lastly, we looked for papers investigating whether any X-ray emission and/or magnetic feature had been identified among our sample of fast rotators.

\subsubsection{Spectroscopic binaries}\label{spec_binaries}

An important fraction of the stars considered in this work have been previously studied elsewhere. For example, 25 of them were included in the investigation of chemical abundances in fast-rotating massive stars by \citet{Cazorla_2017a}. The authors also searched for spectroscopic signatures of binarity in their sample. 
In addition, \citet{Trigetal21,Mahy_2022} have studied several targets in common with our sample. A detailed discussion about the stars in common and some further notes about how this cross-match between our results and those obtained by previous authors have helpped us to fine-tune and/or reinforce our classification of stars in the sample of fast rotators between SB1, LPV, LPV/SB1? and LPV/SB2? (see column 'SB tag' in Table~\ref{table:b_fast}) is presented in Appendix \ref{sb2_appendix}.

\subsubsection{Hunting for visual companions using high angular-resolution images}\label{visual_components}

In order to check for any possible contamination of {\it TESS} data from other visually close stars to our programme targets, we searched for visual companions in different photometric surveys.

Taking into account that during the extraction of flux from the {\it TESS} full frame images, we applied a threshold mask, at the end we collected the flux from a different number of CCD pixels for each star (see Sect.~\ref{tess}). During the flux extraction, typically we chose the pixel with the greatest light contribution and select all pixels with light contributions within 8$\sigma$ of flux. Thus we collected the flux, typically  between four and five pixels; however, in some crowded areas, the final mask of pixels consists of a larger amount of pixels. The size of {\it TESS} pixel is 21 arcsec, then we should take into account any source contamination within 2 arcmin (a bit more than five pixels). 

This search for any possible contaminants was initiated with {\it Gaia}-EDR3 data. We identified all stars within 2 arcmin that demonstrate a difference in G magnitude smaller than 3 mag. Then, we complemented the available information in {\it Gaia} by performing an additional search in the Washington Double Star Catalog \citep[WDSC,][]{Mason_2001}, which provides updated information about any identified companion in the literature. We also included results form specific high angular resolution surveys such as SMASH and Astralux, among others. 

Generally speaking, we consider that if there is a nearby star with a difference in magnitude of  less than 3 mag, the {\it TESS} lightcurves for a given target can be contaminated by the flux of a companion, which is not necessary gravitationally bound. This identification of the visual neighbours serves as a warning when extracting and interpreting the {\it TESS} photometric data. We note, however, that we also performed a thorough identification of potential contaminants which could be avoided when deciding on the final pixel mask used to extract the {\it TESS} light curve for each individual investigated target (see Sect.~\ref{tess}).

In the 'Potential contaminants' column of Table~\ref{table:b_fast}, we indicated the outcome of this search. Namely, we indicated number of companions found within 1 arcmin, and between 1 and 2 arcmin which are separated by the symbol '+'. More detailed information about the cases in which at least one visual component has been identified is summarised Table~\ref{table:info_imaging}, including their angular distances, difference in magnitude, and the corresponding reference from which the information has been extracted.

\subsubsection{Incidence of X-ray emission}

Close compact companions to massive stars may lead to the emission of hard and bright X-rays, as demonstrated in X-ray binaries \citep[e.g.][]{Reig_2011}. Amongst our targets, 17 objects (including four SB1 systems) have been detected at X-ray wavelengths. For 14 targets, the $\log(L_{\rm X}/L_{\rm BOL})$ estimates are available and have been published in various works. The corresponding flux ratio is listed in our Table \ref{table:b_fast} (see X-ray column). 
For the following targets, the measurements of  X-ray fluxes are available in the literature, however, the $\log(L_{\rm X}/L_{\rm BOL})$ estimates are not: HD\,15137 \citep{McSwain_2010}, HD\,149452 \citep{Fornasini_2014} HD\,46485 \citep{Wang_2007}.
Two additional ones, HD\,76556 and HD\,228841, are listed in the 4XMM source catalog. Their spectra were downloaded\footnote{https://xcatdb.unistra.fr/4xmmdr10/index.html} and analyzed within Xspec. All X-ray sources have a soft emission with $\log(L_{\rm X}/L_{\rm BOL})\sim-$7, that is, their X-ray emission can fully be explained by the usual embedded wind-shocks of massive stars. There is therefore no indication for the presence of an accreting compact companion in any of our targets, nor of X-ray bright colliding winds arising in massive binary systems.

\subsubsection{Incidence of magnetism}

Another parameter that can be useful for a characterization of our sample is the presence of a magnetic field. The vast majority of O-type stars are known to be non-magnetic, but $\sim$7\% of O-stars display strong, dipolar magnetic fields \citep{Grunhut_2017,Petit_2019}. Several theories have been proposed to explain this magnetism. In particular, \citet{Ferrario_2009,Schneider_2016} suggested that large-scale magnetism among massive stars originates in mergers.
Because of the presence of a magnetic field in one component of Plaskett's star, close binary interactions were also thought to be possible generators, but this assumption was discarded based on actual observations \citep{Naze_2017}.

Nevertheless, such channels (merging, binary interactions) may produce fast-rotating stars as an end product; thus, we decided to check available literature for the existence of magnetic field in our targets. 
Within the MiMeS survey \citep{Grunhut_2017, Petit_2019}, the Stokes $V/I$ profiles have been modelled for seven stars from our sample: HD\,210839, HD\,36879, HD\,24912, HD\,192281, HD\,203064, HD\,46056A, and HD\,149757 -- and none of these stars were found to be magnetic.  Given the small number of studied stars, we could not draw any conclusion regarding the correlation between the existence of a magnetic field with binary or runaway statuses of our fast rotators. 

\section{Results and discussion}\label{discuss}

\subsection{Preliminary considerations}\label{preliminary}

In this section, we use the empirical information compiled in Tables~\ref{table:b_fast} and \ref{table:phys_parameters} to evaluate the validity of the binary interaction scenario to explain the existence of a tail of fast rotators among Galactic O-type stars. To this aim, we compare some of the global properties of this working sample with those extracted from a complementary sample of stars with \vsini\,<\,200~\kms.

Starting with the original sample of Galactic O-type stars investigated by \citet[][see also Sect.~\ref{sample} and Fig.~\ref{dist_fast}]{Holgado_2020, Holgado_2022} and guided by the objective of trying to minimise as much as possible the potential effect of observational biases when performing this comparison, we decided to exclude all those stars fainter than V\,=\,10~mag and/or located at a distance from the Sun larger than 3~kpc. After considering this filter -- which was not only applied to the initial sample of 285 LS and SB1 stars, but also to the 113 SB2 systems detected by \citeauthor{Holgado_2020} --, we ended up with 179 and 47 LS or SB1 stars\footnote{In the case of the fast-rotating sample, the number also includes 2 stars classified as LPV/SB2?.} having a \vsini\ below or above 200~\kms, respectively, plus 93 SB2 systems (four of them having at least one of the components with \vsini\,$>$\,200~\kms). Hereinafter, we refer to them as the slow- and fast-rotating samples, respectively.

\begin{figure*}[!t]
\centering
\includegraphics[width=0.80\textwidth]{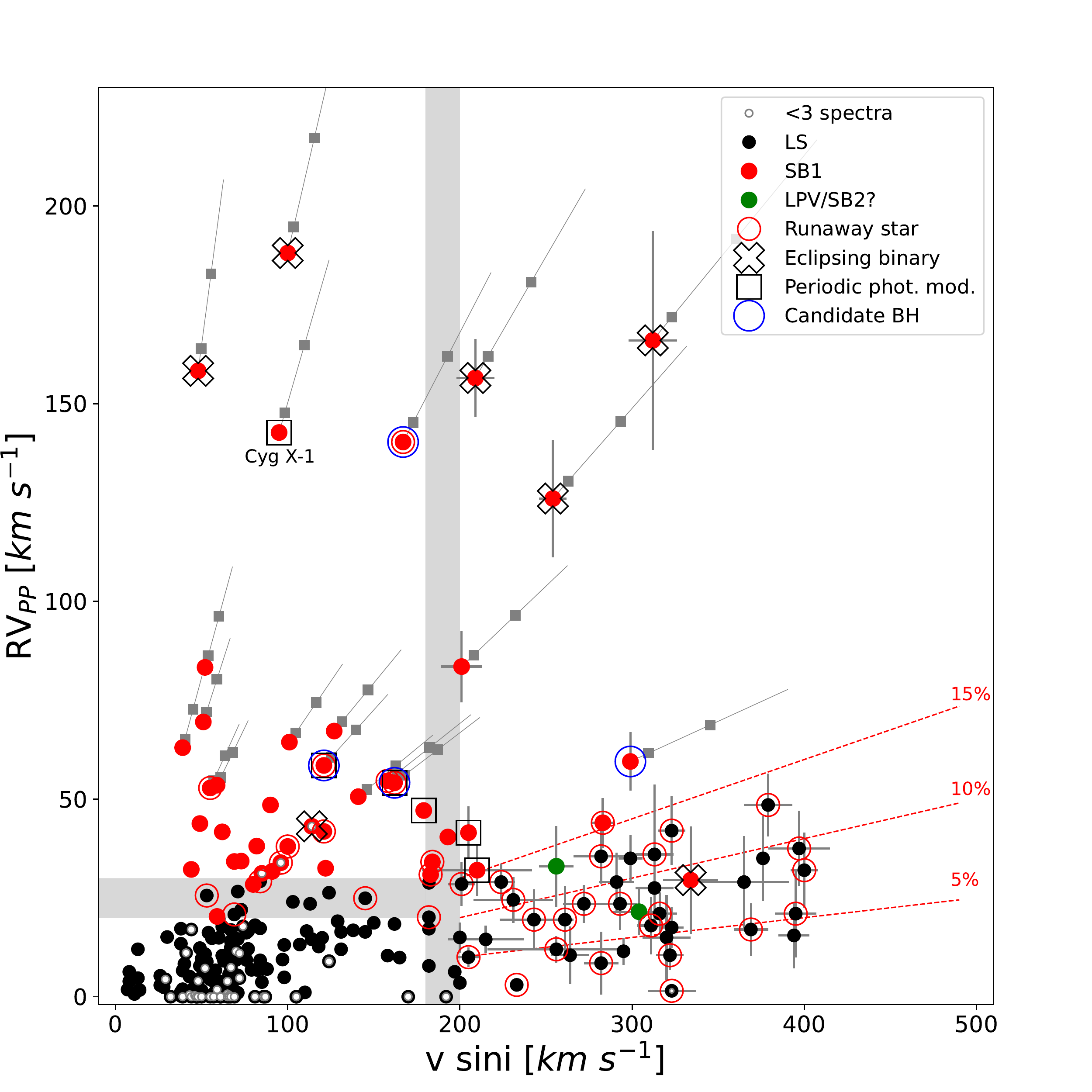}
\caption{Distribution in the \RVpp\ vs. \vsini\ diagram of the sample of 226 LS and SB1 O-type stars described in Sect.~\ref{preliminary}. The vertical gray area represents the artificial threshold between slow- and fast-rotating stars (180\,--\,200~\kms). The horizontal gray area represents the \RVpp\ threshold (20\,--\,30~\kms) in the slow-rotating domain below which the identification of SB1 systems is more difficult due to the effect of intrinsic variability. We note that the small sample of 22 stars for which we only have one spectrum available (i.e. \RVpp\,=\,0) is also included for completeness. See Sect.~\ref{RVppvsini} for the meaning of the inclined red and gray lines and other details. We also note  that while we do not include results in this figure for the additional sample of 93 SB2 systems fulfilling the criteria indicated in Sect.~\ref{preliminary}, the corresponding number and percentage of stars of this type in both the fast- and slow-rotating domain can be found in Table~\ref{table:fast_stat}.}
\label{figure_pp_fin_zoom}
\end{figure*}

\begin{figure}[!t]
\centering
\includegraphics[width=0.54\textwidth]{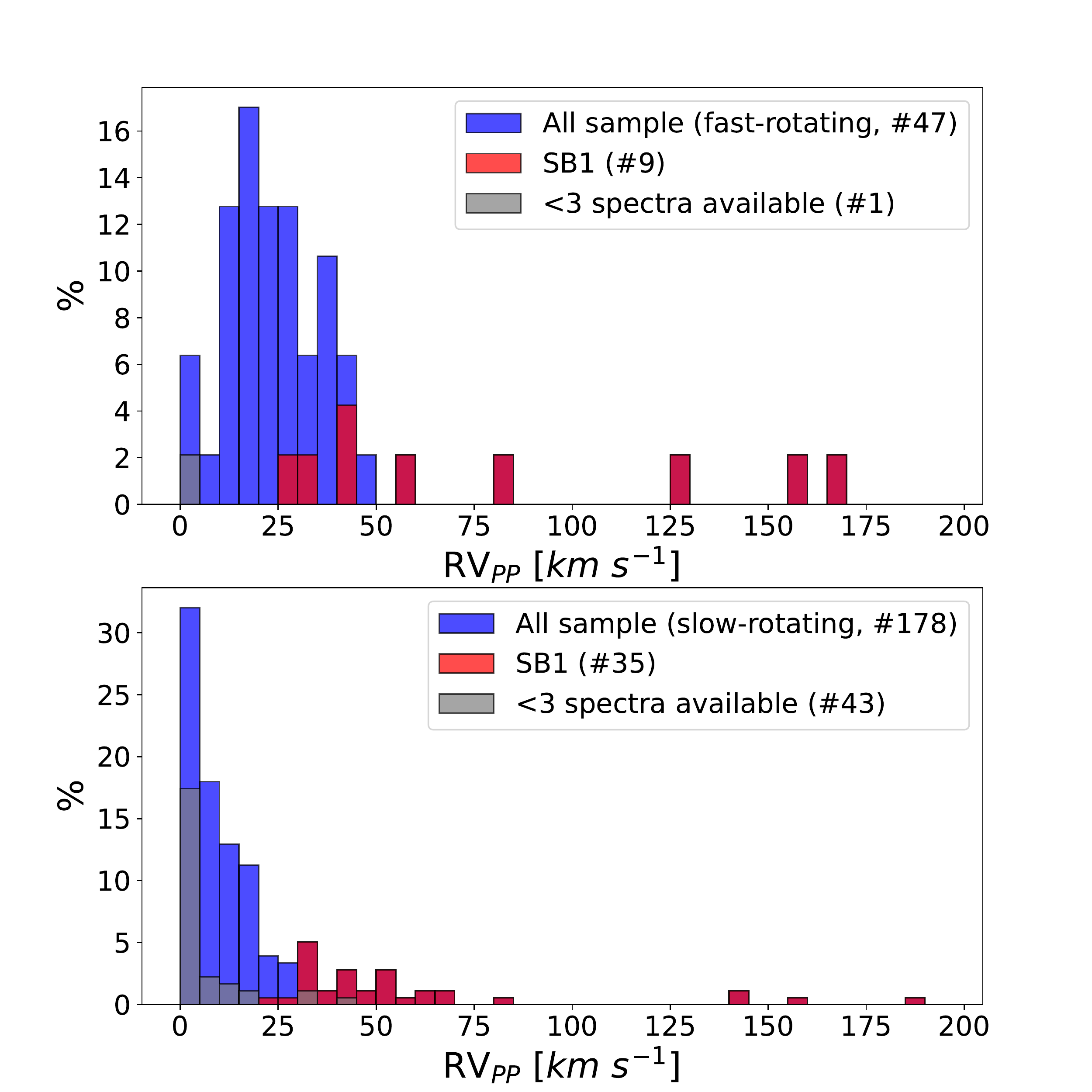}
\caption{Histograms (in percentage, with respect to the total number of stars in each subsample) of \RVpp\ estimates for the fast- (top) and slow-rotating (bottom) O-type star samples. Number of SB1 systems and stars for which we have only one or two spectra available are also indicated in each panel in red and grey, respectively.}
\label{figure_pp_hist_sb1}
\end{figure}

\subsection{The \RVpp\,--\,\vsini\ diagram of Galactic O-type stars}\label{RVppvsini}

Figure~\ref{figure_pp_fin_zoom} depicts the distribution of the above-mentioned sample of 226 LS and SB1 stars in a \RVpp\ vs. \vsini\ diagram.
Both the \RVpp\ and \vsini\ measurements for those targets in the slow-rotating sample are directly taken from \citet{Holgado_phd} and \citet{Holgado_2022}. In the case of the fast rotators (located to the right of the vertical grey shadowed band), we updated the estimates of these two quantities following the guidelines presented in Sections~\ref{rotation} and ~\ref{radial}. As a result, we are also able to indicate the associated uncertainties in both \vsini\ and \RVpp\ for the fast rotators.

Except for those stars with \vsini\,<\,200\,\kms\, and \RVpp\,<\,20\,\kms\, (see reasoning below), wed use different colored symbols to identify the spectroscopic and eclipsing binaries, the periodic photometric modulation variables (including ellipsoidal and reflection modulations, labelled as 'periodic phot. mod.'), and those stars labelled as runaways. In the case of the fast-rotating sample, we take this information directly from Table~\ref{table:b_fast}. For the sake of the interest of the discussion, we have also performed a quick evaluation of the above-mentioned characteristics -- following a similar strategy as in our main working sample -- in the group of stars located in the upper left region of the diagram. 

Red inclined lines depict the regions where \RVpp\ is 5$\%$, 10$\%$, 15$\%$ of the \vsini\ value in the fast-rotating domain. These lines have been drawn to illustrate how intrinsic variability can lead to values of \RVpp\ up to $\sim$15\% of the corresponding \vsini. This spectroscopic variability does not only hamper the identification of low amplitude single-lined spectroscopic binaries among fast-rotating O-type stars\footnote{See also the case of O-stars and B supergiants with lower values of \vsini\ in \citet{simondiaz_2023}.} -- especially when only a few number of epochs is available -- but can also lead to a spurious identification of SB1 systems among these stars (see e.g. the case of HD\,203064 in Sect.~\ref{literature_info}). However, based on our analysis we can suggest to use the threshold of \RVpp\,>\,0.15\,\vsini~ to separate the clear binaries from apparently single stars in the fast-rotating domain.

In the same vein, and following the guidelines presented in \citet{simondiaz_2023}, we depict a horizontal grey shadowed band at \RVpp\,$\sim$\,20\,--\,30~\kms. This approximate threshold separates those targets below \vsini\,=\,200~\kms\ which can be clearly marked as SB1 from those whose detected line-profile variability could originate from intrinsic variability. In particular, we mark all those stars with \RVpp\,$>$\,30~\kms\ as SB1, but only those clearly detected as SB1 among the \vsini\,$<$\,200~\kms\ and 20\,$<$\,\RVpp\,$<$\,30~\kms\ sample after performing a more thorough evaluation of their spectroscopic binary status. The remaining targets (with \RVpp\,$<$\,20~\kms) are excluded from a similar study since most of them do not have enough number of epochs available for a reliable investigation of the spectroscopic variability.

Along this line of argument and taking into account the fact that the measured \RVpp~ is expected to importantly depend on the number of available spectra when this number is low, we highlight stars for which we count on less than three spectra using empty circles. As depicted in Fig.~\ref{figure_pp_fin_zoom}, thanks to the specific observational efforts motivated by this study (see Sect.~\ref{spectra-obs}), the new compiled spectra have made us possible to avoid the low number of epochs caveat for a large percentage of stars in the fast-rotating sample (even reaching more than 5 epochs in most of them). However, as mentioned above, this is not the case for the \vsini\,$<$\,200~\kms\ sample. Therefore, in our working sample of fast rotators, we are quite confident that the measured \RVpp\ is not going to importantly change by increasing the number of epochs and the percentage of identified SB1 systems will remain basically unaffected. On the contrary, in the low \vsini\ sample, those targets with less than five epochs (or even three) and \RVpp\ estimates below 20~\kms\ could actually be unidentified SB1 systems. While this limitation will be taken into account for the discussion presented in Sect.~\ref{final_statistics} (where we compare the multiplicity statistics in the slow- and fast-rotating domains), it is interesting to also remark that there is a non-negligible number of stars (7) in the low \vsini\ sample with two to four epochs and a measured \RVpp\ larger than 30~\kms. Therefore, this fact seems to indicate that large amplitude SB1 systems are easily detected even with such a low number of epochs.

Grey inclined lines initiated in those data~points corresponding to stars with \RVpp\,>\,50~\kms\ represent deprojected values of both velocities assuming an inclination angle ranging from 90$^{\circ}$ to 50$^{\circ}$ (and perfect alignment between the spin and orbit axis). Filled gray squares along those lines represent the values at $i$\,=\,75$^{\circ}$ and 60$^{\circ}$, respectively. This analysis actually shows us that the inclination effect cannot significantly affect the statistics of spectroscopic binaries in the fast-rotating domain.

Globally speaking, in addition to the nine previously mentioned fast-rotating SB1 stars, we identify up to 38 clear single-line spectroscopic binaries with \RVpp\,$>$\,20~\kms\ among the slow-rotating sample.
Among them, and thanks to the availability of {\em TESS} data, we identify 3 eclipsing binaries (HD\,36486, HD\,152590, and BD+60$^{\circ}$498), as well as another 4 stars showing clear signatures of ellipsoidal (or reflection) modulation in their light curves (HD226868 -- aka Cyg\,X1 --, HD\,12323, HD\,53975, and HD\,94024)\footnote{We refer the reader to Sect.~\ref{discussion_about_sb1} for a more detailed discussion about some of these targets (see also some information of interest in Table~\ref{tab-app-eb-ev}), along with those SB1 systems identified in the fast-rotating sample.}. In addition, within the full sample of stars with \vsini\,$<$\,200~\kms, there are 50 secure runaways (9 of them also tagged as SB1).

Inspection of Fig.~\ref{figure_pp_fin_zoom} allows us to highlight several results of interest about the distribution of the global sample in the \RVpp\,--\,\vsini\ diagram. Firstly, all SB1 systems in the fast-rotating domain have \RVpp\,$\gtrsim$\,30~\kms, and are mainly distributed among two main groups -- with high ($\sim$120\,--\,170~\kms) and low ($\sim$30\,--\,60~\kms) \RVpp\ amplitudes, respectively (see also Fig.\ref{figure_pp_hist_sb1}). Indeed, a similar distribution is found in the slow-rotating sample, where a clear gap in \RVpp\ is also detected. In addition, we confirm the result previously found by \citet{Holgado_2022} that the percentage of SB1 stars with \vsini\,$>$\,300~\kms\ is very small, and basically zero for stars in the extreme tail of fast rotators. Further notes about the SB1 systems, along with an evaluation of the possible nature of the hidden secondary components (also using information from the literature and the {\em TESS} light curves) can be found in Sect.~\ref{discussion_about_sb1}.

For completeness, we remind that, in addition to the mentioned difficulties to separate $RV$ variations due to orbital motions in a binary system from intrinsic stellar variability in those stars with \RVpp\,$<$20~\kms, there are 40 stars in the low \vsini\ sample for which we have less than three spectra (i.e. $\sim$15\% of the sample). This clearly explain why we do not see a normal distribution in the low \RVpp\ domain of the corresponding histogram in Fig.~\ref{figure_pp_hist_sb1} (contrarily to the case of the fast-rotating sample). Again, this will be taken into account when discussing the comparison of percentages of detected spectroscopic binaries in both samples.


Interestingly, all but two SB1 systems with \RVpp\,$>$\,100~\kms\ are identified as eclipsing binaries. One of the stars in this subsample not identified as EB is Cyg~X-1 (HD\,226868, O9.7\,Iab\,p\,var, \vsini\,=\,95~\kms), a well-known binary system (P$\sim$5.59~d) hosting an accreting stellar-mass black hole \citep[see][an references therein]{CaballeroNieves_2009}; the other one is HD\,130298 (O6.5\,III(n)(f), \vsini\,=\,167~\kms, \RVpp\,=\,143~\kms, P$\sim$14.63~d) recently proposed by \citet{Mahy_2022} to host a quiescent stellar-mass black hole. 

If we now concentrate on the sample of SB1 stars with \vsini\,$>$\,150~\kms\ and \RVpp\,$<$\,100~\kms, there is only one star detected as eclipsing binary (HD\,46485). Taking into account that the higher the measured \vsini, the most likely the binary system is observed at a high inclination (i.e. a configuration which favors the presence of eclipses if the orbital and rotational axes are aligned, as often assumed), those fast-rotating SB1 stars for which {\em TESS} do not show any signature of eclipses are potential candidates to host a compact object, as will be further discussed in Sect.~\ref{discussion_about_sb1}. Other SB1 systems of interest (in both the slow and fast-rotating samples) for which we should be able to provide further information about the evolutionary nature of the hidden companion (see Sect.~\ref{discussion_about_sb1}) are those stars identified to show ellipsoidal variability in the {\em TESS} light curves.  

Lastly, most of the LS stars in the fast-rotating domain are runaways (see also Table~\ref{table:fast_stat}), and only one fast-rotating SB1 star is also identified as runaway. The higher fraction of runaway stars among fast rotators was already highlighted in Sect.~\ref{gaia}, while further insights about this result are presented in the next sections.

\subsection{Comments on the global statistics of detected runaways}\label{rw_discussion}

In Sect.~\ref{final_statistics}, we discuss in more detail our findings about the multiplicity and runaway incidence among fast-rotating O-type stars. We also evaluate to what extent the obtained empirical results can be used to confirm or reject the binary evolution scenario proposed to explain the existence of a tail of fast rotators. Prior to this, we consider it of interest to briefly discuss the global statistics of detected runaway candidates among Galactic O-type stars\footnote{A more extensive study using the full list of OB-type stars in the ALS3 catalog \citep{Pantetal21}, comprising several thousand of targets and not excluding the SB2 systems, will be presented in Ma\'iz Apell\'aniz et al. (in. prep.).} presented in Sect.~\ref{gaia} (see also Fig.~\ref{vstats}).

Two important results can be derived from these global statistics. The first one is that the fraction of runaways we find is very high. Combining the fast and slow samples, there are $\sim$22\%
certain runaways and an additional $\sim$11\%
possible runaways. Those numbers indicate that up to 1/3 of the population of O-stars in the solar neighborhood may be runaways. These numbers are in rough agreement with the runaway fraction of 27\% derived by \citet{Tetzetal11}, but we note that their sample and methods are quite different. Their sample is dominated by B stars of lower mass than our stars and they used Hipparcos astrometry of much lower precision than that of {\it Gaia}-EDR3, so they were only able to assign probabilities to each star. On the other hand, the fraction is significantly higher than the value obtained by \citet{Maizetal18b}, which was likely a consequence of the conservative nature of their methods.


The second result, of greater relevance for the present study, is that there is a significant difference between fast and slow rotators: for the first type we find that $\sim$35\,--\,50\% are runaway stars, while for the second, we find a smaller number of 20-30\%. This result confirms the finding of \citet{Maizetal18b} that Galactic runaway O-stars rotate significantly faster on average than their non-runaway counterparts and is also in agreement with the recent study by \cite{Sana_2022runaways}, who found that the runaway population of (presumed) single O-type stars in the 30 Doradus region of the Magellanic Cloud presents a statistically significant overabundance of rapidly-rotating stars, compared to its non-runaway population. 

This non-negligible difference in the percentage of runaways between the slow- and fast-rotating samples of Galactic O-type stars is somewhat expected if we assume as valid the proposal that binary interaction plays a dominant role in populating the tail of fast rotators. In this case, an important fraction of fast-rotating runaways (if not all) would be originated by the disruption of a post-interaction binary after the first core-collapse in the system. Indeed, \citet{Renzo_2019runaways} predicted that $\sim$50\% of a population of high-mass interacting binaries will become a disrupted binary in which the initially less massive star is still on the main sequence. This percentage has been calculated following the information presented in Fig. 4 of \citet{Renzo_2019runaways} study; namely, starting from 78\% of binary systems which do not merge and considering that 86\% of those are predicted to be disrupted, 75\% of them including a high-mass main sequence object after core-collapse of the companion. In constrast, since this ejection mechanism is not expected to be operating so efficiently among the slow-rotating sample, the associated runaways would be more likely produced by dynamical ejections resulting from a multi-body interaction in a dense cluster.

There is, however, one important caveat that must be taken into account in the argumentation above. One of the main outcomes of the extensive numerical study of the evolution of massive binary systems performed by \citet{Renzo_2019runaways} is that, despite the large percentage of disrupted binaries resulting from the simulations, only a small fraction of them is predicted to acquire peculiar velocities above 20\,--\,30~\kms\ (i.e. becoming a runaway from an empirical point of view). Therefore, if the estimations by \citet{Renzo_2019runaways} are correct, only a minor fraction of the detected runaways among the fast-rotating sample would come from the disruption of a binary.

\citet{Renzo_2019runaways} claim that this is a robust outcome of their simulations, also indicating that similar findings have been previously found by other authors \citep[e.g.][]{De_Donder_1997,Eldridge_2011_run}. However, along the next sections, we will provide some arguments supporting the idea that this theoretical result seems to be in tension with our empirical findings. In particular, if all detected runaways in our sample of Galactic O-type stars would have been produced by dynamical ejections, there would no reason for the significantly larger fraction of runaways found between the slow and fast-rotating samples. Indeed, we note that even if we consider the most extreme runaways (with $v_{t,lsr}$\,>\,50~\kms), the fractions of stars with such tangential velocity are 19$\%$ in the fast-rotating domain and 8$\%$ in the slow-rotating domain (see Fig. \ref{vstats}). However, this is not the only argument and we present more details in the next section, where information about the detected binary status is also taken into account.

\subsection{Multiplicity and runaway incidence amongst fast-rotating O-type stars}\label{final_statistics}

\begin{table}[!t]
\centering                 
\caption{Some statistics of interest for the cleaned slow- and fast-rotating samples (i.e. excluding three stars with V\,$>$\,10 and/or d\,$>$\,3~kpc from the initial sample, see text). Percentages in the 'all' columns refer to total in each group (268 and 51, respectively), while these are computed with respect to the total number of fast-rotating stars per subgroup in the case of runaways (51, 9, and 38, respectively). See also the note about the relative percentage of LS and SB1 runaways for stars with \vsini\ below 200~\kms\ in Sect.~\ref{final_statistics}. 
}              
\label{table:fast_stat}     
\begin{tabular}{l | r r c r r | r r} 
\hline
\hline
\vsini               & \multicolumn{2}{c}{$<$\,200~\kms} & || & \multicolumn{4}{c}{$>$200\,\kms} \\
\hline
               &  \multicolumn{2}{c}{All}  & || & \multicolumn{2}{c}{All}  &  \multicolumn{2}{c}{Runaways} \\
\cline{2-8}
               & \#     &    \%        &  || & \# & \% &  \# & \%    \\
\hline
Total          & 268 &                 & || & 51     &        &  24   &  47$^{+4}_{-4}$ \\
\hline 
SB1            & 35  & 13$^{+1}_{-1}$   & || & 9     &  18$^{+4}_{-3}$ &   1  &   11$^{+1}_{-1}$  \\
\hline
SB2            & 89  &  33$^{+2}_{-2}$ & || & 4      &   8$^{+3}_{-2}$ &   0  &   ...    \\
LPV/SB2?       & 0 & ...              & || & (2)    &   (4$^{+2}_{-2}$) & 0  &  ...   \\ 
\hline
LS             & 143 & 54$^{+2}_{-2}$   & || & 36     &  70$^{+4}_{-4}$ & 23  &  64$^{+6}_{-4}$    \\
\hline
\end{tabular}

\end{table}

Table~\ref{table:fast_stat} summarises the number and relative percentage (with respect to the total number of stars in each group) of the various types of identified spectroscopic binaries within the slow- and fast-rotating samples. In addition, we indicate the number and percentage of runaways detected among the fast rotators\footnote{We remind that, in the quoted statistics, we have excluded those stars from the original sample studied by \citet{Holgado_2020, Holgado_2022} located at a farther distance than 3~kpc and/or which are dimmer than $V$\,=\,10~mag (namely: BD+60$^{\circ}$134, HD\,124979, ALS12370).}. Lastly, for reference purposes, we indicate that a preliminary estimation of the runaway fraction of stars in the slow-rotating SB1 and LS subsamples (based on the sample of stars considered in Section \ref{gaia} and including 'possible runaways') results in a similar percentage of $\sim$30\%, respectively.

Concentrating first on the fast-rotating sample, it can be noticed that the sample is mainly dominated by likely single stars ($\sim$70\%), with a considerable smaller contribution of the single-lined spectroscopic binaries ($\sim$18\%) and only $\sim$8\% comprising the confirmed SB2 systems. These numbers assume that the two stars identified as "LPV/SB2?" are LS stars; however, if these were actually SB2 stars, the percentages of SB2 and LS stars would be slightly modified, but the main results remain valid.

In order to put these results in a wider context, we compare these numbers with the corresponding statistics for the stars with \vsini\,$<$\,200~\kms. Before presenting our results and making comparisons with the fast-rotating sample, we must indicate that (as mentioned in Sect.~\ref{RVppvsini}) there is a sample of 22 slow-rotating stars (8.2\% with respect to the total number of stars in the slow-rotating sample) for which we only have 1 spectrum available, plus another 18 (6.7\%) with \RVpp\ measurements based on two spectra. Although these stars will be labelled as LS hereinafter, there is some probability that some of them will be identified as SB1 or even SB2 systems when considering more epochs. Therefore, the percentage of SB1 and LS stars presented in the left column of Table~\ref{table:fast_stat} must be considered as lower (SB1 and SB2)  and upper limits (LS), respectively. However, these percentages are not expected to vary by more than $\sim$7\% (i.e. half of the sample with 1\,--\,2 epochs). In addition, since we have not explored in detail the SB1 status of those stars with \vsini\,$<$~200~\kms\ and \RVpp\,<~20~\kms, the percentages of SB1 and LS stars among the slow-rotating sample could again be somewhat larger (SB1) and smaller (LS), respectively, than those indicated in Table~\ref{table:fast_stat}.

Taking all this into account, we can nevertheless extract interesting conclusions from inspection of the statistics presented in Table~\ref{table:fast_stat}. The percentage of SB2 systems is smaller by a factor $\sim$4  in the fast rotating sample (with a difference between the relative percentages of the two samples of $\sim$25\%). In contrast, the percentage of LS stars behaves in the opposite direction (with a minimum difference in percentage of $\sim$20\% if we assume that no new SB1 systems in the low \vsini\ sample will be discovered whenever adding new epochs). Regarding SB1 systems, the percentage appears slightly larger among fast rotators. However, both percentages could become more similar if some SB1 stars are still hidden in our sample of stars with few epochs.

We can speculate that the observed fraction of SB2 systems in the slow-rotating domain is higher because these systems are pre-interaction binaries that have not yet evolved to a post-interaction binary system (including a fast-rotating O-type star orbiting a faint stripped star or a compact object), a merger, or a disrupted single runaway star after a supernova explosion event \citep[e.g.][]{deMink_2014}.

Alternatively, the explanation could be related to the difficulty to detect SB2 systems among stars with broad line profiles, or the fact that the majority of SB2 fast-rotating systems are short-period contact binaries \citep[see for example HD\,100213,][]{Penny_2008} and this evolutionary phase is relatively short with respect to the other stages. However, we consider these two latter explanations as less probable than the abovementioned one. On the one hand, the quality and number of epochs of our compiled observations are good enough to have been able to detect the vast majority of SB2 systems (we note that only 2 out of 54 stars in our sample of fast rotators have survived as 'LPV/SB2?'). On the other hand, we have found four (out of 51 stars, i.e. 8\%) clear SB2 systems, all of them having \vsini\ in the range of what is expected in terms of spin-orbit synchronization due to tides. 

Another result that seems to further support the binary evolution scenario is the following. Even if future analyses of high quality data with a more extended multi\,epoch data\,set could aid to detect faint companions among the SB1 samples, if we add together the percentages of SB1 and SB2 systems, we end up with 26\% and 46\% of secured spectroscopic binaries among the fast- and slow-rotating samples, respectively. Even if we add the two fast rotators labeled as LPV/SB2? as spectroscopic binaries, the percentage of this type of systems among fast rotators is still considerably smaller than in the case of stars with \vsini\,$<$\,200~\kms. In addition, as commented, some of the stars in our sample with 1 or 2 spectra could be discovered as SB1 systems in future investigations, thus increasing  the difference in percentages even further. 

Another point of interest refers to the statistic of runaway stars.
As commented prevously( in Sections~\ref{gaia} and \ref{rw_discussion}), 52\% of the total sample of fast rotators are identified as runaways. Interestingly, all of them but HD\,15137 (O9.5\,II-IIIn, \vsini\,=\,283~\kms, \RVpp\,=\,44, SB1) are likely single stars, a result which, assuming the theoretical scenario proposed by \citeauthor{deMink_2011}, would support the dominance of the binary evolution scenario (after first core-collapse and supernova explosion) over the dynamical ejection (from a stellar cluster) channel \citep[e.g.][]{Perets_2012}. As we discuss earlier in this paper, a significant difference in the observed fraction of runaways in the slow- and fast-rotating domain supports the scenario of disrupted binaries. However, we should note that the mentioned theoretical predictions \cite[including][i.e. simulations]{Renzo_2019runaways} are sensitive to many initial conditions, including the star formation history rate, etc. Thus, if we do not know at which age we are observing the population of stars (the case of our sample), we cannot directly predict the exact fraction of runaway stars at a given stellar age.

From a different perspective, if we concentrate on the (36) fast-rotating and likely single stars, we can highlight two main groups: i) the runaways (64\% of this subgroup versus 30\% in the slow-rotating subgroup of these stars), as noted (see comments below), are expected to be mainly dominated by the products of a binary disruption event; and ii) the rest, which could either be  the end-products of a merger event or binary systems for which we are not able to detect any clear spectroscopic binarity signature of a companion.

 \begin{figure*}[!t]
\centering
\includegraphics[width=0.46\textwidth]{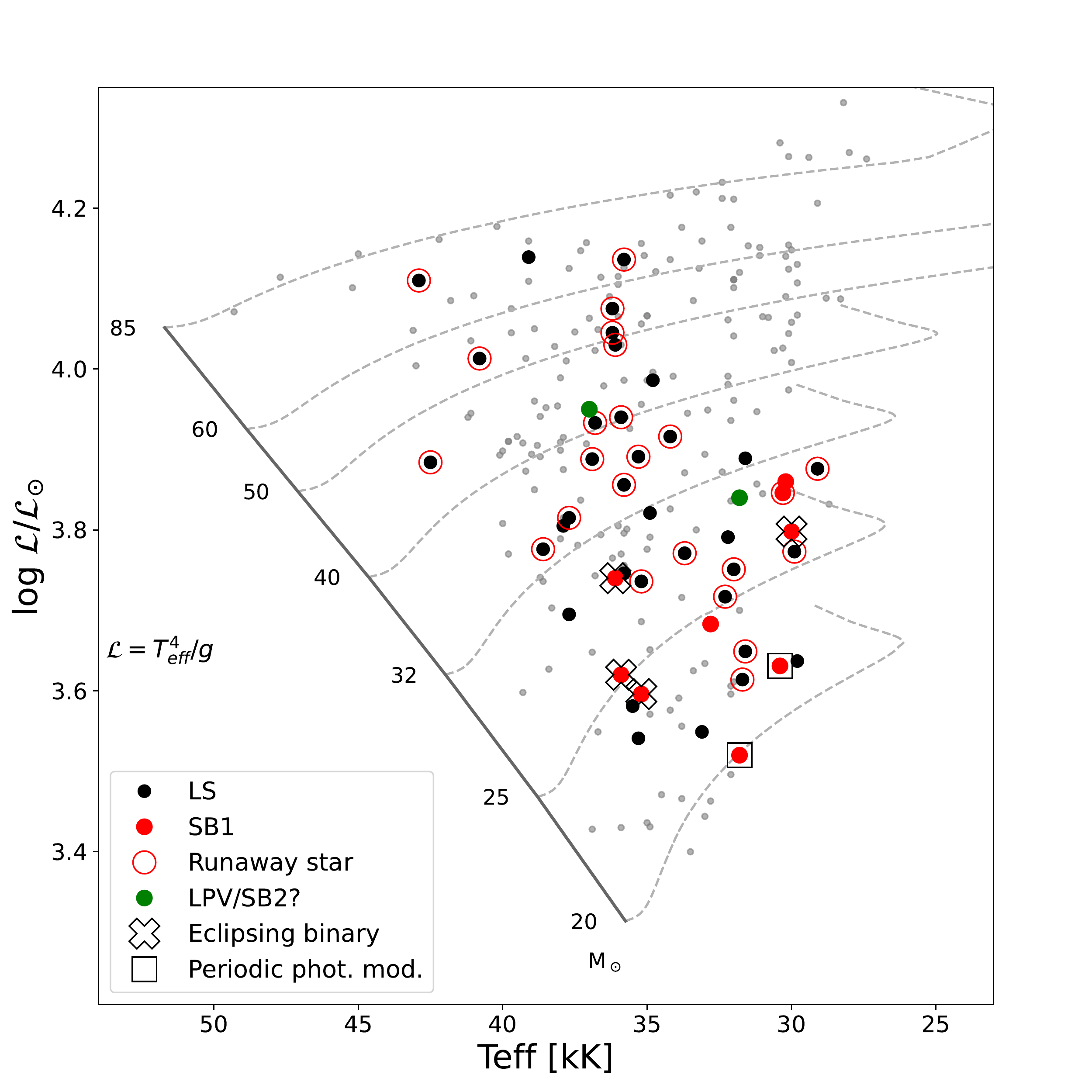}
\includegraphics[width=0.46\textwidth]{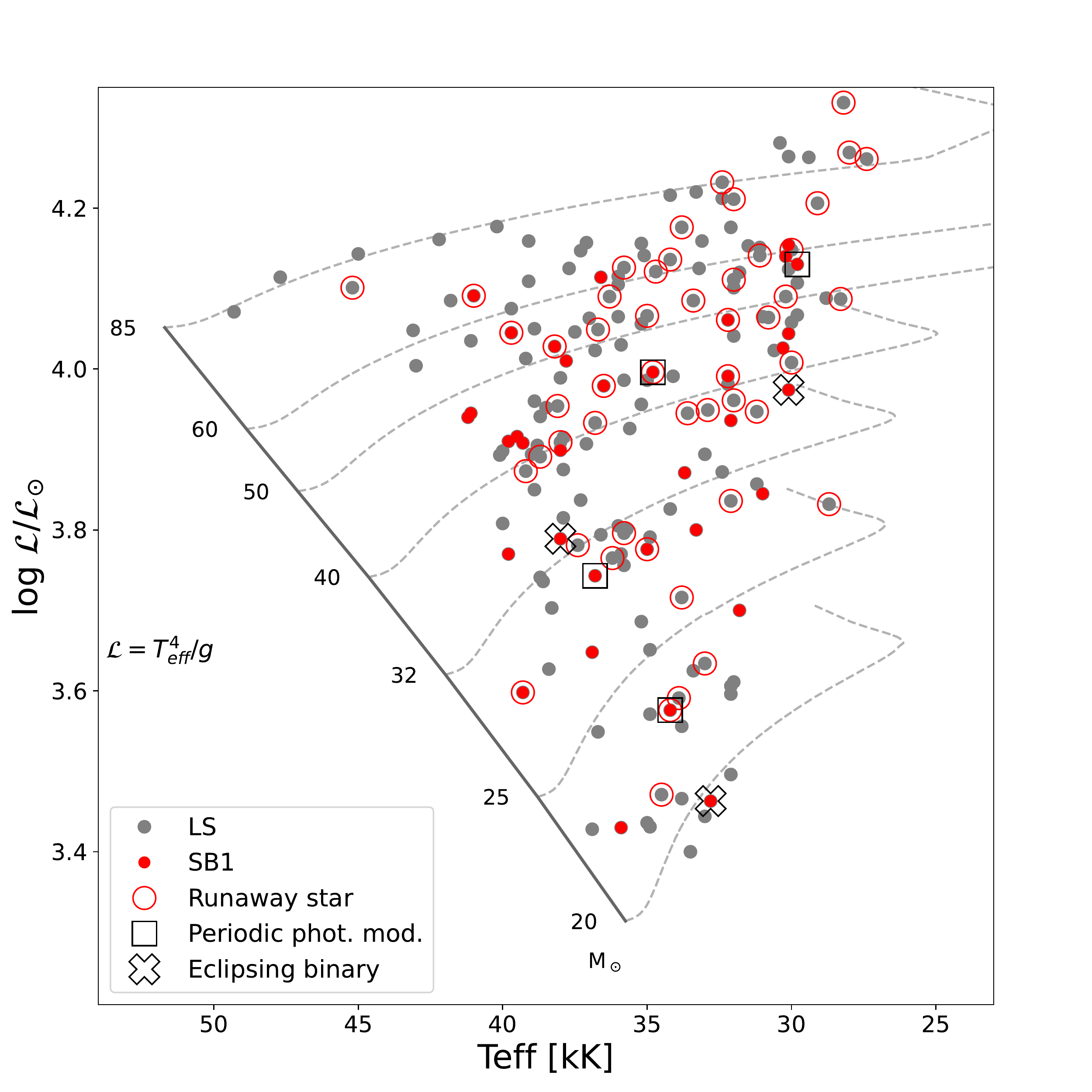}
\caption{Location of the investigated sample of fast-rotating stars on the spectroscopic HR diagram with the non-rotating solar metallicity evolutionary tracks and ZAMS from \citet{Ekstrom2012} (left). SB1 systems, runaway stars, and eclipsing binaries are labeled by different symbols in the same way as in Fig.~\ref{figure_pp_fin_zoom}. Gray dots represent the sample of slow rotating O-type stars \citep[from][]{Holgado_2022}. The same diagram but only showing the slow-rotating sample of O-type stars (right).}
\label{figure_hr_brott_sb1}
\end{figure*}

In reference to the runaway stars in the sample, there is still a possibility that some of them are fast-rotating single stars ejected from their parental stellar cluster by means of one or several dynamical interaction events during the star formation process \citep{Maizetal21f}. However, to distinguish the dynamical ejection from the binary supernova scenario for a sample of our stars is a tricky task that will require further work. An example of a star with peculiar kinematic properties is HD\,149757, the origin of which is still under debate. It has been suggested that this star is a post-interaction binary \citep{Villamariz_2005,Renzo_2021}. It was also suggested that this star has been ejected after the supernova explosion of its massive companion \citep{Tetzlaff_2010,Kirsten_2015}. Also, the runaway status of HD\,93521 possibly originates  from a close binary interaction, and in addition to this, it is a possible merger product \citep{Gies_2022}. Interestingly, we detected a single runaway SB1 system \citep[HD\,15137,][]{McSwain_2010} amongst them: the only possibility to explain its existence would then be a dynamical ejection from a triple system in the past. In this context, it is interesting to note that HD\,13268 and HD\,15642 are runaways and members of the Perseus OB1 association \citep{deBurgos_2020}. Thus, there is a possibility that they were dynamically ejected. 

Regarding the LS fast rotators, the majority of stars in this group have \RVpp<30~\kms, thus if any hidden secondary component would be present, it should have a relatively small mass compared to the O-star. Indeed, we recall that in this rapid rotation regime, we expect to see the stars with the rotational axis almost perpendicular to the line of view. Moreover, considering (zero order approximation\footnote{It could most likely change following a donor's supernovae kick.}) that the orbit of fast-rotating star is expected to be perpendicular to the rotation axis, as usually assumed in case of short period or over-contact binaries \citep[e.g.][]{Fabry_2022}, this implies that the velocity variations due to orbital motions should be close to their maximum values and would have been detected if the secondary masses were high enough. Therefore, it is only possible to have some hidden low-mass companion amongst them. One example is HD\,46485, for which we could not detect any clear spectroscopic signature of binary motion with the available data\,set (five spectra) but {\it TESS} data show eclipses.
In addition to this, \citet{deMink_2014} simulated the appearance of various binaries in terms of the orbital velocity of the components. According to these simulations, the systems with K\,$<$\,10~\kms~ (\RVpp\,$\lesssim$\,20~\kms) could be either products of binary interaction (i.e. mergers or post-interaction binary systems) or effectively single stars. However, these simulations are valid for the general case of main-sequence objects without taking into account the rotation regime (presented in Sect.~\ref{theory}).

\subsection{Distribution of the fast-rotating sample in the sHRD}\label{sHRD_dist}

The left panel of Fig.~\ref{figure_hr_brott_sb1} depicts the location in the sHRD of our working sample of fast rotators, also including (for reference) the rest of the O-star sample investigated by \citet{Holgado_2022}, as well as the zero-age main sequence (ZAMS) line and the non-rotating evolutionary tracks computed by \citet{Ekstrom2012}. Coloured symbols as in Fig.~\ref{figure_pp_fin_zoom} are also used here to identify the various types of binaries as well as the runaway stars among the fast-rotating sample. For comparison, a similar sHRD diagram is depicted in the right panel of Fig.~\ref{figure_hr_brott_sb1}, this time including only the slow-rotating sample.

The first interesting result \cite[already pointed out by][]{Holgado_2022} is that all (9) fast-rotating SB1 systems are located below the 32~\msun\ evolutionary track, where a mixture of presumably single stars identified both as runaways (8) as well as targets without peculiar kinematical properties with respect to their local environment (7) can be also found. Additionally, the higher mass sample of fast rotators is basically dominated by LS runaway stars, with only 4 of the 18 stars in this region being detected as non-runaway LS stars. Also remarkable is that no fast rotators are found in the more evolved region of the MS in the higher mass domain.

The situation is completely different for the case of SB1 systems in the slow-rotating sample, where the distribution is more homogeneous along the whole O star domain (although with a somewhat larger relative percentage above the 40~\msun\ track). In addition, there is a clear concentration of runaway stars (mostly LS, but also some SB1) in the top right region of the diagram, where no fast-rotating stars are found.  We should note that in a such comparison between fast- and slow-rotating domains we are dealing with target masses which are distributed uniformly; thus in our analysis and conclusions we do not have any bias toward specific mass domains.  
If we assume the scenario proposed by de Mink as valid, we could interpret the different distribution of runaways and SB1 systems in both panels in the following way. On the one hand, the fast-rotating higher mass runaways might be rejuvenated, spun-up gainers in a post-interaction binary system already disrupted after supernova explosion of the initially more massive companion. On the other hand, most of the more evolved, slow-rotating higher mass O-type runaways could be single stars which have been dynamically ejected from a cluster during the star formation process. Since there is no rejuvenation effect due to binary interaction, the probability of detecting more evolved O-stars among this latter sample is larger compared to the case of the fast-rotating sample. In the same vein, the SB1 systems also detected as runaways in the slow rotating sample could be interpreted as pre-interacting binaries which have been also ejected from their parental cloud during the star formation process -- or as binary systems which previously were part of a triple system and in which one of the components exploded at some point as a supernova, hence producing the anomalous velocity (with respect to the local interstellar medium) of the detected SB1 system.

In summary, roughly speaking, we have detected two dominant sub-groups of fast-rotating stars in the sHRD, which likely represent two different stages of binary evolution: i) the least luminous SB1 systems and ii) presumably single luminous fast-rotating runaway stars coming from post-interaction systems. Such dominance of luminous runaway stars actually contradicts the simulations that are expected to have BHs as a companion of these systems,  hence producing a smaller kick to the accretor \citep{Sukhbold_2016,Shenar_2022NatAs}. To explain this distribution of runaways and SB1 systems will require detailed evolutionary modelling of each of these systems.  In addition, these two main sub-groups are complemented with a third one, also mostly concentrated in the lower mass region of the O star domain, corresponding to the LS, non runaway stars, and representing $\sim$25\% of the whole fast-rotating sample. This latter group could correspond to merger products or stars with relatively low mass (dim) post-interaction accretors which do not produce detectable $RV$ variations nor eclipses in the {\em TESS} light curves. In this context, it is interesting to note that despite the superb quality of our spectroscopic dataset and the {\em TESS} photometry, the detection of small-$RV$ signatures of orbital motion in a potential binary system and/or faint eclipses below \RVpp$\sim$20\,--\,30~\kms\ and $\Delta T_{\rm p}\sim$5\,--\,20 mmag is hampered by the ubiquitous presence of intrinsic photometric and spectroscopic variability at the aforementioned level in the whole O-stars domain \citep[see also][]{simondiaz_2023}.


Interestingly, it is unlikely that some of the stars that are located on the upper part of the sHRD have some hidden SB1 components taking into account that almost all of them are classified as runaways. Thus, we do not expect to spot any other potential fast-rotating SB1 systems in that part of the sHRD, as a result, we do not have any observational bias on this sHRD distribution of fast rotators. As we have already pointed out, to explain the observed distribution of SB1 systems in the slow- and fast-rotating domain will require additional work in modelling the population of these systems. However, in any case it should serve as a constraint for further theoretical works that describe the evolution of binary fast-rotating systems.

\subsection{Further notes about the properties of SB1 systems and the characterization of hidden companions}\label{discussion_about_sb1}

\begin{table*}[!t]
{\tiny
\caption{Some information of interest for the nine 'bona fide' SB1 systems found among the fast-rotating sample, ordered by increasing \vsini\ estimates. The table includes two newly detected SB1 systems (HD\,152200 and HD\,46485), as well as two other ones for which \citet{Mahy_2022} could detect the weak lines of a fainter secondary companion (HD\,163892 and HD52533). Uncertainties on the stellar parameters can be found in Table~\ref{table:phys_parameters}, except for the case of the spectroscopic mass ($M_{\rm sp}$), which is on the order of 15\,--\,20\%. Uncertainties on the orbital parameters can be found in the quoted references.}              %
\label{table:sb1_fast}      %
\begin{tabular}{l c c c c c c c c c} 
\hline
\hline
Target    & HD\,163892  &  HD\,308813 & HD\,37737 & HD\,152200 & HD\,165246 & HD\,15137      & HD\,165174 & HD\,52533       & HD\,46485 \\
SpC    & O9.5\,IV(n)   &  O9.7\,IV(n) & O9.5\,II-III(n) &  O9.7\,IV(n) &  O8\,V(n)  & O9.5\,II-IIIn       & O9.7\,IIn  & O8.5\,IVn       & O7\,V((f))nvar?  \\
\hline 
\vsini~ [\kms]& 201 & 205 & 209 & 210 & 254 & 283 &  299 & 312 & 334\\
\RVpp\ [\kms]  &  83.5  & 41.5 & 156.5 & 32.0 & 126.0 & 44.0 &  59.5 & 166.0 &  29.5 \\
\hline 
\Teff [kK]  & 32.8 & 31.8 & 30.0 & 30.4 & 35.9 & 30.3 &  30.2 & 35.2 & 36.1 \\
log($\mathcal{L/L_{\odot}}$)  & 3.68 & 3.52 & 3.8 & 3.63 & 3.62 & 3.84 &  3.86 & 3.6 & 3.74\\
$R$ [$R_{\odot}$] & 9.3 & 6.9 & 10.3 & 6.8 & 7.8 & 10.7 & 10.7 & 7.9 & 7.5\,--\,11 \\
log($L$/$L_{\odot}$)  &  4.95 & 4.65 & 4.88 & 4.54 & 4.96 & 4.94 &  4.93 & 4.92 & 4.93 \\
$M_{\rm sp}$ [M$_{\odot}$] & 19 & 12.3 & 12.4 & $<$8 & 22 & 13 & $<$15 & 22 & 17  \\
\vsini/$v_{\rm crit}$ & 0.3 & 0.3 & 0.4 & 0.4 & 0.4 & 0.5 & 0.6 & 0.4 & 0.5 \\
\hline 
Runaway tag &  n & n & n & n & n & y & n & n  & n  \\
EB tag   &  -- & ? & EB & RM & EB & n & -- & EB & EB+RM \\
\hline
P [d] & 7.8 & 6.3 & 7.8 & 4.5$^c$ & 4.6$^b$ & 55.3 &  23.9 & 22.0 & 6.9$^c$ \\
$K_{1}$ [\kms]   & 41 & 32 & 70 & 23$^c$ & 53$^b$ & 16 &  30$^a$ & 88 & 15$^c$\\ 
$e$ &  0.04 & 0.38 & 0.38 & 0.17$^c$ & 0.03$^b$ & 0.66 & 0.16 & 0.27 & 0.01$^c$\\
$f(m)$ [M$_{\odot}$] &  0.0708$^e$ & 0.0198 & 0.2224 & 0.004$^c$ & 0.071$^b$ & 0.0092 & 0.0313 & 1.02$^d$ & 0.002\,--\,0.00091$^c$ \\
$i$ [$^{\circ}$] &  70 & 25 & 75 & (45)$^c$ & 84$^b$ & 45 & (45) & 90 & 80\,--\,75$^c$\\
$v_{\rm sync}$ [\kms] & 60 & 55 & 67 & 76 & 86 & 10 & 23 & 18 & 55 \\
$v_{\rm eq}$ [\kms] & 215 & 485 & 215 & 295 & 255 & 400 & 420 & 310 & 340 \\
\hline
\multicolumn{9}{l}{Rough mass estimation of the hidden secondary companion ($M_{\rm 2}$ [\msun])} \\
\hline
From $M_{\rm sp}$, $f(m)$, $i$ & 3.5 & 4.1 & 4.1 & $<$1.0 & 3.6 & 1.8 & $<$3.1 & 10.0 & 0.9 \\ 
From $M_{\rm ev}$, $f(m)$, $i$ & 3.4\,--\,4.2 & 5.8\,--\,7.5 & 5.6\,--\,7.2 & $<$1.7\,--\,2.3& 3.4\,--\,4.5 & 2.6\,--\,3.3 & $<$4.8\,--\,6.2 & 11.2\,--\,14.1 & 1.0\,--\,1.3 \\ 
\hline
\end{tabular}

{\bf Notes:} Information presented in the top half of the table has been extracted from Tables~\ref{table:b_fast} and \ref{table:phys_parameters} in this paper. The orbital parameters presented in the second half of the table have been mainly compiled from \citet{Mahy_2022} except from those quantities conveniently specified, extracted from: $^a$ This work; $^b$ \citet{Johnston2021}; $^c$ Sim\'on-D\'iaz et al. (in prep.) and \citet{Naze_2023_rot}; $^d$  \citet{Trigetal21}; $^e$ \citet{Cazorla_2017a}. Values of \vsini/$v_{\rm crit}$, $P$, $K_{\rm 1}$, $e$, $v_{\rm sync}$, and $v_{\rm eq}$ have been rounded to 0.1, 0.1~d, 1~\kms, 0.01, 1~\kms, and 5~\kms, respectively. 
}
\end{table*}

Table~\ref{table:sb1_fast} provides a summary of all information of interest for the nine definitely confirmed fast-rotating SB1 systems\footnote{See also Table~\ref{tab-app-eb-ev} for some info on those SB1 in the low \vsini\ sample identified as eclipsing binaries and/or ellipsoidal variables.}. In addition to the information obtained by ourselves for the purpose of this paper (top half of the table), we compile additional information about several orbital parameters (period, semi-amplitude of the $RV$ curve, inclination, eccentricity, and mass function; namely, $P$, $K_{\rm 1}$, $i$, $e$, and $f(m)$, respectively) in the middle part of the table. The information described above is also complemented with some estimates of the surface velocity the star would have if the spin and orbit were synchronised ($v_{\rm sync})$, as well as the quantity \vsini/$v_{\rm crit}$ (were $v_{\rm crit}$ is the critical rotational velocity), and $v_{\rm eq}$, namely, the estimated equatorial velocity of the star assuming the inclination angle quoted in the table. Lastly, at the bottom, we provide two estimations of the individual mass of the hidden companions by assuming that the mass of the primary is given by a) the spectroscopic mass resulting from the estimated surface gravity (corrected for centrifugal forces) and the radius derived taking into account the apparent magnitude of the star, its distance, and extinction, as well as b) an approximated evolutionary mass (25\,$\pm$\,5~\msun), as obtained from the location of the fast-rotating SB1 sample in the sHRD (see left panel in Fig.~\ref{figure_hr_brott_sb1}). 

Most of the information on the orbital parameters quoted in the table have been extracted from the recent study by \citet{Mahy_2022}, except for three stars; one of them (HD\,152200) is a newly discovered SB1 system, while HD\,46485 is confirmed 'bona fide' SB1 system and the third one (HD\,165246) has been studied in detail by \citet{Johnston2021}.

To first give a global overview of the properties of the sample, regarding the amplitude of $RV$ variability and the detection of eclipses and/or ellipsoidal modulation in the {\em TESS} light curves we have: (i) three large-amplitude (\RVpp\,$>$\,100~\kms) SB1 binaries, all of them being also eclipsing binaries (HD\,37737, HD\,165246, and HD\,52533); (ii) four small amplitude (\RVpp\,$<$\,50~\kms) SB1 binaries, including one eclipsing, non-runaway system (HD\,46485), two non-runaway targets whose {\em TESS} light curves show periodic photometric modulations (HD\,308813 and HD\,152200), and one runaway star with no signatures of eclipses or ellipsoidal variability (HD\,15137); (iii) two intermediate amplitude non-runaway SB1 binaries for which there is no {\em TESS} data available (preventing us from evaluating their EB/EV status, HD\,163892 and HD\,165174).

If we pay attention to the orbital periods of the systems, we find two main groups, one of them including six systems with orbital periods in the range $P\sim$\,4.5\,--\,8 d (most of them also detected as eclipsing binaries and/or ellipsoidal variables), plus another one comprising three systems with a somewhat longer period ($\sim$20\,--\,50~d). In this latter group, there is one eclipsing binary, one runaway star, and a third target tagged neither as runaway nor as eclipsing binary or ellipsoidal variable. Interestingly, there does not seem to be any correlation between \RVpp\ and $P$ within stars in any of these two main groups.

The first conclusion we can extract by inspecting the information compiled in Table~\ref{table:sb1_fast} (more specifically, by comparing the measured \vsini\ with the estimated equatorial velocities assuming spin-orbit synchronization) is that it is quite unlikely that the spectroscopically detected components of these nine fast-rotating SB1 systems have acquired their relatively large equatorial velocities through tidal effects, leaving again the binary interaction channel as the most probable one.


According to theory, the dim (hidden) companion of a post-interaction high-mass binary can be either a compact object \citep[e.g. black holes and neutron stars, BH/NS; see][]{Langer_2020,Mahy_2022} or a stripped helium star or subdwarf object \citep[e.g.][]{deMink_2013, Shao_2014,Gotberg_2018}. 

Among the nine investigated SB1 systems, there are five targets for which the presence of a black hole or neutron star can be discarded. This includes the four stars for which two clear eclipses are detected in the {\em TESS} light curve, plus\footnote{We remind that, by the time of submission of this paper, there was no {\em TESS} light curve available for HD\,163892; hence, we could not evaluated if this system is also an eclipsing binary.} HD\,163892, for which \citet{Mahy_2022} claim to have identified spectroscopically the secondary by means of a spectral disentangling technique\footnote{This is also the case for HD\,52533.}. Among the remaining four stars, there are two targets for which {\em TESS} data shows potential signatures of periodic photometric modulation (HD\,308813 and HD\,152200), a third one detected as a runaway SB1 system with an orbital period of $\sim$55~days and a quite large eccentric orbit (HD\,15137), and a fourth target (HD\,165174, $P$$\sim$24~d, $e$$\sim$0.16) not detected as runaway, but for which we do not have {\em TESS} data available to evaluate its photometric behaviour. All but HD\,152200 have been investigated in detail by \citet{Mahy_2022}, who do not propose any of them as candidates to host a quiescent stellar mass black hole. Additionally, they show that in case the secondary would have a mass exceeding $\sim$3~\msun\ (see last 2 rows of Table~\ref{table:sb1_fast}), it would have been detected in the spectrum using their disentangling technique.

HD\,308813 displays clear phase-locked modulations in flux and radial velocities, however, they are difficult to interpret especially in view of the low number of spectroscopic observations. Nevertheless, no evidence indicate that this binary system hosts a black hole or neutron star. Further characterisations of this system will require more spectroscopic observations.

Another situation occurs in the case of the two newly confirmed SB1 systems HD\,152200 and HD\,46485. The shape of their light curve shows the typical behavior of a binary system in which half of the stellar surface of the cooler component is illuminated by the hotter star (reflection modulation; i.e. maximum and minimum flux at conjunction moment). This leaves us with HD\,165174 as the only possible fast-rotating O-type star in our sample for which the presence of a quiescent compact object cannot be definitely ruled out (pending the acquisition of new {\em TESS} data or high-quality ground-based photometry and further ultraviolet observations).

Lastly and rather importantly, our only fast-rotating SB1 system also detected as a runaway (HD\,15137) deserves further attention. This star was investigated in detail by \citet{McSwain_2010}, finding that it may contain an elusive neutron star, in the ejector regime or a quiescent black hole with conditions unfavorable for accretion at the time the {\em XMM} observations used in that study were obtained. \cite{Mahy_2022} also included this star in their search for quiescent O+BH/NS systems, estimating the mass of the hidden companion to be in the 1\,--\,6~\msun\ range (with a more probable mass of $\sim$\,2.5~\msun) and indicating that they could have identified this dimmer companion in the spectra if it would have a mass above $\sim$3~\msun\ while the star is not actually detected. Our estimated mass of the secondary also points to this boundary limit which leaves the exact nature of the hidden companion still elusive. 

As commented in Sect.~\ref{RVppvsini}, there are another four stars among the low \vsini\ sample that deserve further attention in this  context. These refers to those SB1 stars with \vsini$<$\,200~\kms\ for which we have detected signatures of ellipsoidal (or reflection, depending on the considered orbital period of the system) variability ion the {\em TESS} light curves. One of them is HD\,226868 (aka Cyg\,X-1), a well studied binary system comprising a late-O supergiant orbiting an accreting stellar-mass black hole \citep[see][an references therein]{CaballeroNieves_2009}. Another two are: HD\,12323 (ON9.2\,V, \vsini\,=\,121~\kms, \RVpp\,=\,58~\kms) and HD\,94024 (O8\,IV, \vsini\,=\,162~\kms, \RVpp\,=\,54~\kms). By combining the available $RV$ and {\em TESS} data we can confirm that the period of these binary systems are 1.92 and 2.46~days, respectively, and that the folded light curves correspond to ellipsoidal modulation \citep[as also pointed out by][]{Mahy_2022}. Interestingly, these authors claimed as unlikely that any of these two systems host a stellar mass black hole by considering that the higher mass of the hidden companion is obtained when assuming spin-orbit synchronization. However, the situation changes if, instead of tidal synchronization and in view of the detected ellipsoidal modulation, we assume that these stars actually have a much faster equatorial velocity, reaching values on the order of $v_{\rm eq}$/$v_{\rm crit}\sim$0.4\,--\,0.6, as in the case of the other SB1 system in our fast-rotating sample (see Table~\ref{table:sb1_fast}). In this case, the estimated masses of the hidden companion would be in the range $\sim$3\,--\,6~\msun\ and 2.5\,--\,5~\msun, respectively; hence, leaving the door to be considered as binary systems hosting a stellar-mass black hole (or at least a neutron star) still open.

Regarding HD\,53975 (O7.5\,Vz, \vsini\,=\,179~\kms, \RVpp\,=\,47~\kms), the last star in this group, we do not have yet enough epochs to check if the orbital period is $\sim$12.5 or $\sim$6.25~days. Therefore, more $RV$ measurements are needed to decide if the photometric variability detected in the {\em TESS} light curve is associated with ellipsoidal or reflection modulation, an evaluate the evolutionary nature of the hidden companion.

\begin{figure*}[!t]
\centering
\includegraphics[width=1\textwidth]{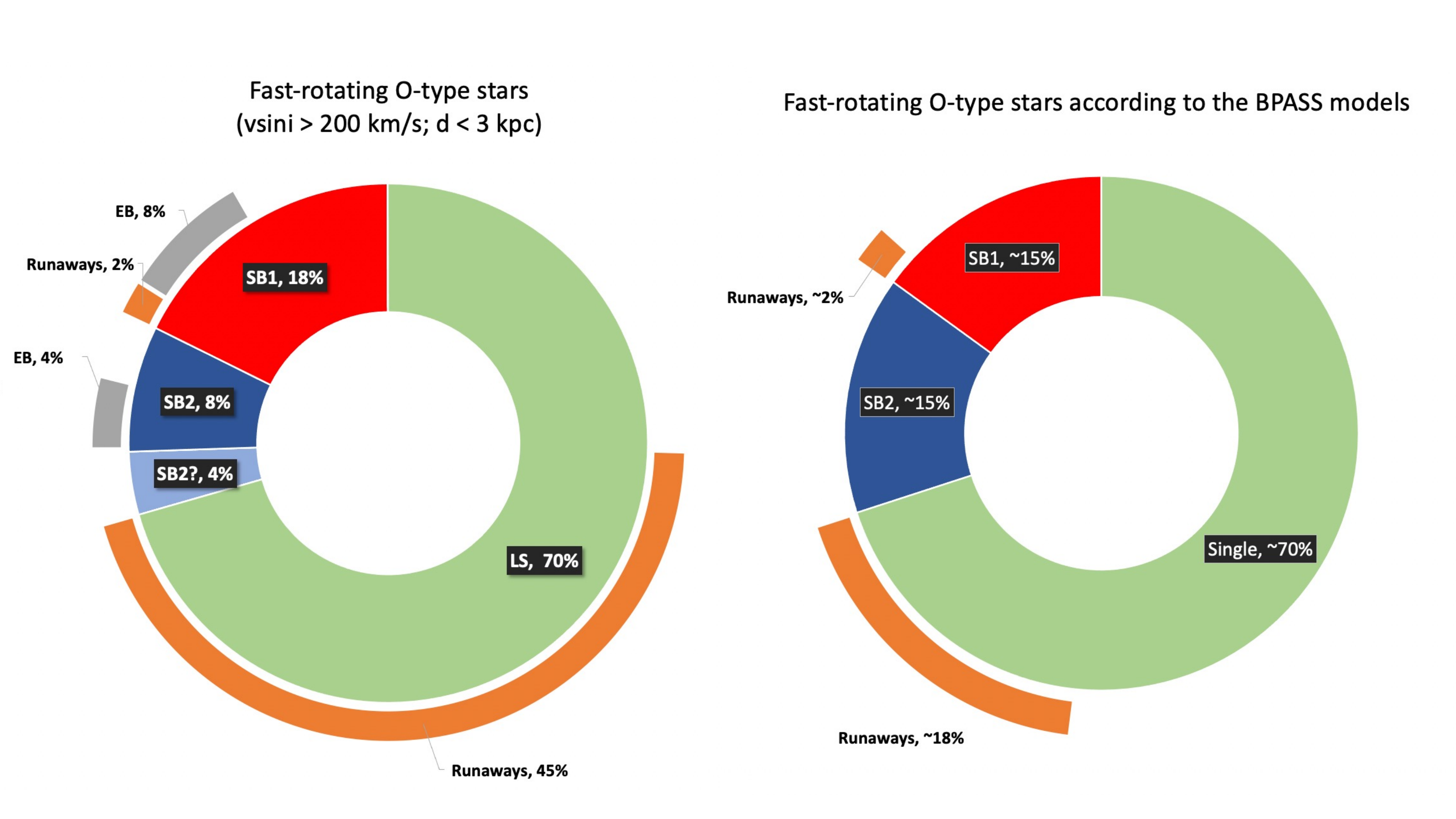}
\caption{Overall observed fractions of spectroscopic binaries, runaways (including walkaways) and eclipsing binaries for all working sample of fast rotators within 3 kpc distances (according to Table \ref{table:fast_stat}) shown on the left.  Simulated fractions of spectroscopic binaries and runaway (incl. walkaway) stars over all population of O-type fast rotators (based on the BPASS models) shown on the right.}
\label{figure_overall_dist}
\end{figure*}

\subsection{Comparison with theoretical predictions}\label{theory}

As we have pointed out, based on the observational properties of the investigated sample of fast rotators, it is likely that  the vast majority of them can be assumed to be post-interaction binary products. Using the information presented in Table \ref{table:fast_stat}, Fig. \ref{figure_overall_dist} provides a global overview of the main characteristics of our sample of 51 Galactic O-type stars with \vsini\,$>$\,200~\kms. In particular, it highlights the relative percentage of presumably single stars and spectroscopic binaries, as well as of detected eclipsing binaries and runaways in each subsample. As we can see from Fig. \ref{figure_overall_dist}, only 25\% of all fast rotators are likely single stars -- and not runaways. These targets could be mergers, following the theoretical predictions regarding
the origin of rapid rotators of  \citet{deMink_2013}. The detailed characterization of fast rotators presented in this work is a crucial testbed for further theoretical studies of stellar evolution. The next step is therefore to compare model predictions with our observational results. 

Recent simulations of OB binary system configurations by \citet{Langer_2020} suggest that fast-rotating systems with a range of orbital velocities from $K_{\rm 1}$ $\sim$ 50~\kms\ (i.e. \RVpp~$\sim$ 100~\kms, case A -- donor is on the main sequence) to $K_{\rm 1}$ $\sim$ 100~\kms\ (i.e. \RVpp~$\sim$ 200~\kms, case B -- donor is evolving to the Red Giant phase) most likely have black holes as companions (see Fig. 6 in the cited paper). These simulations can be placed in terms of rotational velocity as well. Thus, theoretically, the OB+BH systems have a higher probability to be detected with equatorial rotational velocity ($v_{eq}$)$\sim$100\,\kms, $K_{\rm 1}$ $\sim$100~\kms\ (slow-rotating domain) and $v_{eq}$$\sim$500~\kms, $K_{\rm 1}$ $\sim$50~\kms (fast-rotating domain, C. Schürmann, private comm.).
As we discuss in Sect.~\ref{discussion_about_sb1}, we did not detect any SB1 system with \vsini~>350~\kms. However, we did detect LS runaway stars with \vsini$\sim$400~\kms\ that could be proven to be disrupted systems following a supernova explosion. From another aspect, according to Fig.~\ref{figure_pp_fin_zoom}, the majority of SB1 systems with possible or confirmed BH companions are located in the slow-rotating domain at \RVpp~$\sim$ 100$\pm$50~\kms.  These observational results should be taken into account in future theoretical works regarding the prediction of the existence of OB+BH systems.

Moreover, within the present work, we also aimed to evaluate the observational appearance of fast-rotating O-type stars based on recent theoretical predictions.  
To do so, we have used the synthetic stellar populations from the Binary Population And Spectral Synthesis v2.2.1 results \citep{Eldridge_2017, Stanway_2018} to estimate the expected properties of fast rotators in various O-star subpopulations, i.e. SB1, SB2 and singles. We use the fiducial BPASS population, this includes a mix of single stars and binary stars as described by \citet{moe_2017}, although the majority of O-stars are in binary systems. We also use prescribed initial mass ratio and period distributions from the same source. The applied IMF is \citet{kroupa_1993} with an upper mass limit of 300~M$_{\odot}$.

BPASS stellar models calculate the detailed structure and evolution of each star in sequence using a custom version of the Cambridge \textsc{STARS} code. First, it evolves the more massive primary in detail and approximates the secondary evolution using the stellar evolution equations of \citet{hurley_2000}. It then calculates the evolution of the secondary in detail either as a single star or a binary with a compact companion. Thus, in making synthetic O-star sub-populations we use the following constraints to put stellar models into each group. We classify an O-star as having $T_{\rm eff}$ $\ge 30$~kK, a luminosity of $\log(L/L_{\odot})\ge 4.5$ and a surface hydrogen mass fraction above 0.2. 

In what follows, we use specific definitions of SB1, SB2, and single stars. It assumes that the SBs are all post-interaction products. This is an extreme assumption but it provides important constrain on what binary interaction can produce and, thus, help to interpret the observational results.

Indeed, to accurately predict the number of SB2 stars expected from an observation survey is difficult. It is a complex interplay between the relative luminosity of the two stars but also whether the orbital period is short enough that radial velocity variations can be observed. In the clearest cases of similar luminosities and short periods, a SB2 identification is simple. Actually, we performed the simulations of the line-profile appearance by varying the flux contributions of the secondary component with different \vsini~ regimes of the primary component (see our simulations in Fig.~\ref{spectra_simulations}). It becomes more difficult as the luminosities become more different and the periods become longer (i.e. smaller orbital velocities). In some ways determining if a star is a SB2 or SB1 depends more on the luminosity, while the LS sample will have some wide SB1 binaries. However, the error introduced by not taking all these factors into consideration is on the order of a few \% as the majority of our LS arise from unbound binaries and single stars. A true detailed reproduction of the observed population, taking into account all the complex selection effects, is beyond the scope of this paper. In a present work, for our BPASS modelling we used the following definitions. 
The SB2 O-stars are taken from the initial models when both stars in the binary are O-stars or when the luminosities of the two stars are both within an order of magnitude. The SB1 are all stars in binary systems that do not count as an SB2 system. This includes systems where the secondary star is an O-star and the primary is the post-binary interaction donor star or a compact remnant. It also includes systems with an O star as primary and a secondary being a lower mass main-sequence star with a difference in luminosity greater  than an order of magnitude. The single O-stars are either O-stars that were originally single (a small number), post-interaction binaries after a merger event or the initially less massive companions in a binary system that become unbounded after the supernova explosion of the star that was the primary when the system originated.

Current BPASS models do not include a detailed model of rotation. However, rapidly rotating O-stars are identified by tagging stellar models where the accretor increases its mass by more than 5\% of its initial mass or those O-stars that have merged with their companion. In both cases, such stars will be expected to be rapidly rotating and be far above the $v \sin i$ threshold assumed in our observations \citep{deMink_2013}. While this is an approximate treatment, it allows us to estimate a lower limit to the population of rapid rotating O-stars. For example, rotation will extend the life of the O-star and we did not include models in which there is a lesser degree of mass transfer, but where stars will still be boosted up the the extent that they end up being observed as rapidly rotating. However, these are second-order effects and our predictions provide a robust estimate of the importance of mass-transfer as the source of rapid rotation. In varying the parameters, we have chosen to define the synthetic O-star populations, such as the magnitude difference for SB2 or SB1 systems, leading to an uncertainty in our predicted sub-populations on the order of a few percent of the population for each predicted value.

We used two metallicity models appropriate for solar metallicity: $Z=0.014$ and 0.020. This allows for variations and uncertainties in the abundances of the observed stellar population. Amongst the modelled O-star populations, the fraction of fast rotators is $\sim$20-25\% among, which we can emphasise the next  sub-groups of targets depending on their binary status: (i) LS: $\sim$70\% (among which $\sim$14\% are runaways and $\sim$11\% are walkaways);
(ii) SB1: $\sim$15\% (among which $\sim$9\% are runaways, and $\sim$3\% are walkaways); (iii) SB2: $\sim$15\%.

The fraction of runaway stars was calculated by considering the impact of the first supernova. For single star runaways, these are assumed to be O-stars whose binaries have been unbound in the first supernova, while the SB1 runaways are all systems that have experienced a supernova, forming a neutron star and not a black hole. 

To calculate the O-star velocities, we modelled the effect of the primary star's supernova on the binary. We used the formalism of \citet{Tauris_1998} and \citet{Tauris_1999},   and assuming the neutron star receives a kick from the \citet{hobbs_2005} distribution. The O-star velocities are essentially the pre-supernova orbital velocity of the O-star in cases where stars are unbound. We estimate that the number of systems where a kick will be strong enough to accelerate the accretor up to 30~\kms~ is only 3\% (i.e. walkaways) and 9\% is the fraction of runaways with velocity more than 30~\kms~ of SB1 systems (or approximately 2\% of the entire rapid rotator population). For the single stars, we estimate the number of walkaways to be 11\% and 14\%  of runaways with a velocity of more than 30~\kms (in total this is $\sim$18\% of the entire population). As we may notice, according to our simulations and observations, the fast rotators will be more likely to be runaways than walkaways.


Schematically, these results are presented on the right panel of Fig. \ref{figure_overall_dist} with the same labels as on the left panel, to ease comparison with the observational results. These values are on the same order as the rates that we observationally inferred, especially if we will take into account that some of the 'LPV/SB2?' systems may  be shown to be LS objects after all. This good agreement implies that the majority of fast-rotating stars appear to be LS \citep{deMink_2011} and mass transfer is the primary physical process responsible for creating rapid rotation, as already suggested by \citet{deMink_2013}. This is also in line with expectations from \citet{Haemmerle_2017} and \citet{Bodensteiner_2020} regarding Be stars, that such a rapid rotation would be difficult to be already present at birth. We note that the prediction here of the number of mergers becoming rapid rotators (i.e. 52\%) should be treated as an upper limit. A recent work by \citet{Schneider_2019} suggests that merger products may be highly magnetic. This would then lead to a rapid spin down of such objects, suggesting some merger products may, in fact, be slow-rotating \citep{Schneider_2020}.

BPASS provides further detail on the nature of SB1 systems. In these models, a SB1 is any binary system for which there is only one O-star in the binary. If both stars are main-sequence stars, this occurs when the primary is $\ge$0.5 dex more luminous than the secondary since the secondary is not an O-star in this case. In addition to the case above, as SB1 systems  BPASS considers any O-type binaries with compact remnants (NS) as secondaries. Thus, BPASS yields: (i) the fraction of SB1 systems with a rapidly rotating star is about $\sim$6\% of all O-type SB1 systems; (ii) for SB1 systems with a slow-rotating O-star, $\sim$92\% are the pre-interacting stars of their systems, $\sim$5\% are systems in which the O-star was initially the secondary star but its status changed due to mass transfer, and $\sim$3\% are O-stars paired with a compact remnant; (iii) for SB1 systems with a fast-rotating O-star, mass transfer has occurred in all cases (from the definition of fast rotators in the models, see above). The fraction of O-star paired with a compact remnant is about $\sim$84\% of the systems, while the fraction of the O-star paired with a post-mass transfer object (e.g. stripped star) is ~$\sim$16\%. 

The illustration of these results is presented in Fig. \ref{figure_sb1_illustration}. A more detailed investigation of the SB1 system's companions requires the combination of photometric and radial velocity curve property studies in order to reconstruct the orbit of the system that could give a hint about which of the two predicted subgroups of companions we are dealing with. 

Given that these BPASS models explain many other varied observations of massive stars and related transients \citep[e.g.][]{Eldridge_2017,Stanway_2018,Massey_2021,Briel_2022}, it is positive to see that the same fiducial population can provide estimates that are of the same order as the observed population. However, the agreement is not perfect and, thus, refinements to the models of massive O-stars are required. 
Especially it concerns the predicted fraction of runaway stars -- 
the predicted runaway population includes runaways caused by supernova only -- if the dynamical runaways were included the total runaway population could be significantly increased \citep[see e.g.][]{Eldridge_2011_run}.


\section{Conclusions and future prospects}

In the present work, we studied a statistically meaningful sample of several tens of Galactic, fast-rotating O-type stars. By analyzing their multiplicity and runaway status, we can draw the following main conclusions:
\begin{itemize}
\item The fraction of runaway stars among fast-rotating O-stars is $\sim$33-50\%, which is significantly higher than for slow rotators ($\sim$20-30\%). Notably, we have detected several fast-rotating runaways with significant high tangential velocity, that is, $v_{t,lsr}$>50~\kms.
\item The fraction of SB1 systems in the fast-rotating domain is  $\sim$18$\%$.  If we assume that they are all post-interaction binaries, the comparison with models indicates that they are products of mass-transfer with a secondary component which can be a compact remnant (black hole or neutron star) or a post-mass transfer object (stripped helium star or subdwarf object).
\item The fraction of SB2 systems among O-type fast rotators is about $\sim$8-12$\%,$ which is significantly lower than in the slow-rotating domain $\sim$33$\%$. Most likely, this statistic indicates how binary systems are evolving from slow-rotating pre-interacting systems to the post-interaction system with rapid rotation. In addition, to support this hypothesis, we detected that the fraction of SB1 systems in the fast-rotating domain ($\sim$18$\%$) is slightly higher than the fraction of these systems in the slow-rotating domain ($\sim$13$\%$).
\item We found a couple of intriguing fast-rotating SB1 systems that will require further detailed investigation, namely: HD\,152200 and HD\,46485, which possibly host a post-interaction low-mass companion (compact remnant or stripped star or subdwarf object), as well as HD\,308813, which will require further characterization. 

\end{itemize}

Taking into account the fact that single fast-rotating runaway stars are post-mass transfer systems without a secondary component, whereas SB1 systems are actually post-interaction systems too, we can conclude that the majority of fast-rotating O-type acquired their fast rotation via mass-transfer ($\sim$ 65\%). In addition, some fast rotators are possibly passing by tidal synchronization scenario as the overcontact SB2 systems ($\sim$ 10\%). The rest of the stars could be considered as merger products (e.g. HD\,93521), although their exact fraction remains elusive. Indeed, the issue of how to distinguish the population of mergers from effectively single fast rotators (if any) or from fast rotators with hidden companions remains an open question.

The last point that we did not cover in this work is the study of the chemical evolution of fast-rotating stars which will require further investigation. We did not perform a detailed investigation for all fast rotators in our sample, however, helium and nitrogen abundances are available from the literature \citep{Cazorla2017b,Holgado_phd}. Notably, \citet{Cazorla2017b} showed that it is possible to reproduce the stellar atmospheric parameters and abundances for half of their sample of fast-rotating stars by using single-star evolutionary models but others require binary evolution models.  In order to perform a detailed investigation of the current evolutionary stages of fast-rotating stars from the chemical composition point-of-view, we need to derive the CNO abundances for the full sample of targets.

\begin{acknowledgements}
We thank the referee, Tomer Shenar, for an extremely constructive and respectful report which has helped us to improve the first version of the paper, and also provided interesting ideas for future development. N.B. acknowledge support from the postdoctoral programme (IPD-STEMA) of Liege University. N.B., S.S-D, G.H and A.H. acknowledge support from the Spanish Ministry of Science and Innovation (MICINN) through the Spanish State Research Agency through grants PGC-2018-0913741-B-C22, PID2021-122397NB-C21, and the Severo Ochoa Programe 2020-2023 (CEX2019-000920-S). This work has also received financial support from the Canarian Agency for Economy, Knowledge, and Employment and the European Regional Development Fund (ERDF/EU), under grant with reference ProID2020010016.
This paper is based on observations made with the Nordic Optical Telescope, operated by NOTSA, and the Mercator Telescope, operated by the Flemish Community, both at the Observatorio del Roque de los Muchachos (La Palma, Spain) of the Instituto de Astrof\'isica de Canarias. Based (partly) on data obtained with the STELLA robotic telescopes in Tenerife, an AIP facility jointly operated by AIP and IAC.
This work has made use of data from the European Space Agency (ESA) mission {\it Gaia} (\url{https://www.cosmos.esa.int/gaia}), processed by the {\it Gaia} Data Processing and Analysis Consortium (DPAC, \url{https://www.cosmos.esa.int/web/gaia/dpac/consortium}). Funding for the DPAC has been provided by national institutions, in particular the institutions participating in the {\it Gaia} Multilateral Agreement. 
J.~M.~A., G.~H. and M.~P.~G. acknowledge support from the Spanish Government Ministerio de Ciencia e Innovaci\'on through grant PGC2018-\num{095049}-B-C22.
Y.N. acknowledge support from the Fonds National de la Recherche Scientifique (Belgium), the European Space Agency (ESA), and the Belgian Federal Science Policy Office (BELSPO) in the framework of the PRODEX Programme (contract linked to XMM-Newton).

\end{acknowledgements}

\bibliographystyle{aa}
\bibliography{ref}

\begin{appendix} 

\section{Overview of {\it TESS} results}
\label{Tess_appendix}
In this section, we discuss in detail photometric peculiarities of some fast rotators, especially the ones we presented in Fig.~\ref{fig:example_tess_plot}. In the top row we show the {\it TESS} data for the HD\,46485 system, recently shown to be eclipsing by \citet{Burssens2020} using cycle 1 {\it TESS} data. The eclipsing signal is clearly visible in the light curve and we measure a frequency, $f =0.1453(4)$~d$^{-1}$, corresponding to the binary period $P=6.88(2)$~d. Because of the large {\it TESS} pixel size (21 arcsec), and therefore pixel masks, it is important to consider possible background contaminants. A detailed discussion about how we looked for background flux contaminants is presented in Sect.~\ref{visual_components}.
In the case of HD\,46485, possible contaminants in the pixel mask are much fainter (>3~mag) such that HD\,46485 is likely the source of the signal. We identified three eclipsing binaries (EB), HD\,37737, HD\,52533, and HD\,46485 (see Table \ref{table:b_fast}, 'var tag') for which we can confidently exclude a background origin. In addition, detailed investigation of {\it TESS} data of HD\,152200 reveals the presence of a small-amplitude periodicity of $\sim$4.5 days. It could be linked to the presence of a compact companion (e.g. ellipsoidal or reflection modulations) or be linked to rotational features. Our further spectroscopic observations were able to clarify the nature of this system. Thus, we classify HD\,152200 as a prominent photometric variable with reflection modulations. For some targets the {\it TESS} data is not available yet (marked as '$\_$ ' in Table~\ref{table:b_fast}); however, we indicate if these targets have been identified as EB based on other photometric surveys \citep[e.g. HD 165246;][]{Johnston2021}. 

In addition, we looked for signals in the light curves that might reveal a pulsating hidden companion. We demonstrate this by means of a comparison in the second and third row of Fig.~\ref{fig:example_tess_plot}. The second row shows the O8.5~V(n) star, BD+36$^\circ$4145, whose light curve and periodogram show low amplitude stochastic variability at low frequencies, typically found in O dwarfs. There are no significant frequencies in the typical SPB and $\beta$~Cep frequency regimes. By comparison, in the third row, we show the ON9.5~IIn star HD\,91651. In addition to low-frequency variability, we detect a signal with $f=7.3893(1)$~d$^{-1}$. Before we confirm the presence of a hidden companion, we first need to consider potential background contamination again. While there are about a dozen background sources in the {\it TESS} pipeline mask of HD\,91651 (see Fig.~\ref{fig:example_tess_plot}) they are much fainter than the central source. Moreover, a previous study by \citet{Pigulski2008} measured a similar frequency for HD\,91651 using 525 ASAS-3 observations over $\sim$2800 days. The signal is therefore likely intrinsic to the HD\,91651 system. However, since HD\,91651 is a late O9.5 star, we also need to account for the possibility that the signal may be from the O-star itself and not from a hidden lower mass companion. Indeed, some late O-stars do show $\beta$~Cep type pulsations, such as HD\,46202 \citep{Briquet2011}. The situation for HD\,91651 is therefore less clear than in the case of the O7~V star HD\,47839 mentioned earlier. We find that all other measurements of frequencies above 4\,--\,5~d$^{-1}$ in stars in this sample also occur in O9\,--\,9.5 stars: HD\,89137, HD\,28446A, HD\,117490, and HD\,102415. In all these cases there is therefore the possibility that the signal comes from a lower mass B-type companion but we cannot fully confirm this given the available data. On the other hand, we did not find clear evidence of SPB type pulsations (1\,--\,3~d$^{-1}$) in any of the stars in the sample. This is due to the ubiquitous presence of the stochastic low-frequency variability detected in non-eclipsing light curves, making it difficult to disentangle stochastic and coherent modes.

Finally, HD\,210839, HD\,14434, HD\,14442, HD\,192281, BD+60$^{\circ}$2522, HD\,24912, HD\,149757, and HD\,93521 have been already extensively studied in terms of photometric variability to reveal the nature of non-radial pulsations in On-type stars \citep[see][]{Rauw_2012,2021_Gregor}. Also, due to significant photometric variations, HD\,165174 has been proposed as a candidate $\beta$ Cep variable \citep{Chini_2012}. We confirm the variability in those stars for which a {\it TESS} light curve is available.


\section{Overview about individual targets regarding their spectroscopic binary status.}
\label{sb2_appendix}

In this section, we discuss in detail the spectroscopic binary classification of individual targets and the literature overview regarding it. 

Our first separation between SB1 and LPV stars (based on the measured \RVpp\ and the visual inspection of the line profiles) is in good agreement with the results presented in \citeauthor{Cazorla_2017a}, except for two cases: HD\,203064 and HD\,52266. While these authors claim that these two stars are SB1 systems with periods P $\sim$\,5.1 d and $\sim$\,75.8 d, respectively (see the appendix in the aforementioned article), we identified them as LPV/SB1? and LPV, respectively. 

For the case of HD\,203064, we count on 41 spectra covering a total time-span of more than 10 years \citep[compared to eight spectra over 6 years in][]{Cazorla_2017a}, and with some spectra distributed in blocks of several spectra per night during 4 nights. As illustrated in Fig.~\ref{hd_203064_rvs}, the detected variability is not compatible with a 5.1~d orbit, but more with the expected hourly variability resulting from stellar oscillations. Also, we do not find any clear peak at this period in the periodogram computed from the full dataset. Hence, we can exclude this star to be a SB1 system and modify its status to likely single (LS, see last column of Table~\ref{table:b_fast}). We note, in addition, that the same conclusion was reached by \citet{Trigetal21}, see below. 

Regarding HD\,52266, we did not find a very convincing phase-folded RV curve using the period quoted in \citet{Cazorla_2017a}. In addition, by performing various periodogram techniques using all the existing radial velocity data for this target, we could not detect any prominent peak in the associated periodogram. Thus, we consider this target as a likely single star with significant radial velocity variations. We will consider it as LS for the purposes of the discussion presented in Sect.~\ref{discuss}, but tag it as LPV/SB1? in the last column\footnote{Those targets for which we changed the final spectroscopic binary status after accounting from addition information are indicated with a "*" in the last column of Table~\ref{table:b_fast}.} of Table~\ref{table:b_fast}. 

The study of spectroscopic binarity among Northern Galactic O-type stars performed by \citet{Trigetal21} has five stars in common with our sample: HD\,229232, HD\,52533, HD\,192281, HD\,37737, and HD\,15137. Among these targets, HD\,15137 is quoted in that paper as a runaway SB1 system \citep[P\,$\sim$\,55.4~d, see also][]{McSwain_2010}. Our preliminary assessment of the detected line profile variability led us to propose it as LPV/SB1? despite measuring a \RVpp\ of 44~\kms. However, folding our $RV$ measurements with a period of 55.4~d, we found a nice $RV$ curve in agreement with the orbital solution proposed by \citet{Trigetal21}, \citet{McSwain_2010}, and the recent study by \citet[][see also below]{Mahy_2022}. Therefore, we decided to modify its binary status to SB1. 

Aiming at the identification of quiescent compact objects in massive Galactic single-lined spectroscopic binaries, \citet{Mahy_2022} have investigated 32 O-type stars reported as SB1 in the literature. We have six stars in common, all of them also identified as SB1 by us. Interestingly, for two of these systems,  HD\,52533 and HD\,163892, \citeauthor{Mahy_2022} could identify a secondary component using spectral disentangling. However, for consistency with our own classifications (since we do not detect any clear signature of the secondary component in our dataset) we keep these stars labelled as SB1, but mark them with a ($\dagger$) in the last column of Table~\ref{table:b_fast} to highlight the findings by these authors.

We have also two additional stars in common with \citet{Williams_2013}, namely HD\,308813 and HD\,229232, both identified as SB1 systems with periods of 6.34 and 6.2~d, respectively. The binary status of HD\,229232 has been more recently revised by \citet{Trigetal21} to a single star status, which we confirm. Regarding HD\,308813, we confirm its single-lined spectroscopic binary status \citep[see also][]{Mahy_2022}. In the same vein, HD\,165246, which we identified as SB1, was also found by \citet{Johnston2021} as an SB1 eclipsing binary system\footnote{This star is hence marked as EB in Table~\ref{table:b_fast}, despite we could not check the {\it TESS} light curve.} with a period of 4.5927~d. After revisiting the literature, we can highlight that  (to our best knowledge) we have identified one clear SB1 system (HD\,152200) and another potential SB1 system (HD~46485) previously studied in \citet{Burssens2020}.

HD\,152200 was already highlighted by \citet{Pozo_2019} as a possible high-mass eclipsing binary; however, we note that given the photometric variability detected in the {\it TESS} light curve, it is more likely that the star is a variable with a reflection modulation (RM). Indeed, the RM nature of variability for this target is confirmed by our on-going FEROS observations of this target, whose preliminary radial velocity measurements seems to indicate that the orbital period is $\sim$4.5\,d \citep[instead of the 8.89\,d tentatively suggested by][when interpreting the photometric variability as two eclipses]{Pozo_2019}.

HD\,46485 is one of those cases in which, despite we did not identified the target as clear SB1 from the compiled radial velocities (\RVpp\,=\,29.5~\kms), the {\it TESS} light curve shows two clear eclipses \citep[see top right panel of Fig.~\ref{fig:example_tess_plot} and ][]{Burssens2020}. Thus, we modified its spectroscopic binary classification to SB1. This star is presently being studied in more detail in a separated papers using an extended spectroscopic dataset Sim\'on-D\'iaz et al., in (prep.) and \citet{Naze_2023_rot} . 

Regarding the four double-line fast-rotating spectroscopic binaries we identified among the whole list of Galactic O-type stars investigated by \citet{Holgado_phd},  two of them, HD\,100213 -- O8\,V(n)z\,+\,B0\,V(n) -- and HDE\,228854 -- O6\,IVn\,+\,O5\,Vn, have been previously studied in detail by \cite{Penny_2008} and \cite{Abdul_2021}, respectively.
Both are (eclipsing) over contact binaries -- with an orbital period of 1.387 and 1.886~d, respectively -- in which the two components fill their Roche lobe. The third object, HD\,175514, for which we measured \vsini\,=\,288 and 108~\kms, respectively, is actually a triple system comprising an inner eclipsing binary \citep[O5.5 V((f))\,+\,B0.5: V with a period of $\sim$1.62~d;][]{Mayer_2005} and a third outer component -- O7.5 IV((f)) -- orbiting the other two with a period of at least 50~years \citep{Maizetal19b}. For the fourth one  (HD\,165921, O7\,V(n)z\,+\,B\,0:V:) we obtained \vsini\,=\,224 and 151~\kms, respectively, and did not find any detailed study, nor information about whether it is an eclipsing system, in the literature. A preliminary analysis of our spectroscopic data~set for this star\footnote{Also including new observations we are presently obtaining with HERMES} indicates that the period of this system is $\sim$1.7442\,d, hence another newly discovered over contact binary. Example of line profile variability for some of the SB2 systems is presented in Fig. \ref{spectra_example}.

Lastly, among five targets that we initially labelled as LPV/SB2?, as already mentioned in Sect.~\ref{visual_inspection}, one of them (HD\,24912) was detected by \citet{Ramiaramanantsoa_2014} to present co-rotating bright spots, something that might explain the peculiar spectroscopic variability. Thus, we label it as LS in last column of Table~\ref{table:b_fast}. 

Regarding the other four targets, BD+60$^{\circ}$2522 is the main ionizing source of the Bubble Nebula and its classified as O6.5(n)fp, indicating that it has peculiar and variable line-profiles likely produced by rotational modulation of its strong and likely non-spherically symmetric wind \citep[see study of][]{Rauw_2003,Naze_2021}. It is also detected as having a high proper motion with respect to its local interstellar medium pointing towards a possible dynamical ejection after a supernova event. Therefore, we conclude that intrinsic variability is the most probable explanation for the spectroscopic features which could be interpreted as signatures of a secondary companion and, hence, also modify the spectroscopic binary status of this star to LS.

The same decision was taken for HD\,175876 after checking a more extended spectroscopic data\,set (not included here) obtained in the final phases of development of this paper, and taking into account the fact that this star was also independently classified as presumably single by \cite{Cazorla_2017a}, using a different set of spectra. This decision is also supported by the study presented in \citet{2021_Gregor}, where the authors indicate that the peculiar behavior of the spectral line profiles with strong photometric variability (as the case of this two stars) could be just a consequence of the occurrence of non-radial pulsations.

In contrast, we decided to keep the other two targets (HD\,124314 and HD\,91651) as 'LPV/SB2?'. On the one hand, HD\,124314 was found by \citet[][see also Sect.~\ref{visual_components} and Table~\ref{table:info_imaging}]{Sana_2014} to be composed by two objects\footnote{In addition, there is a third companion located at $\sim$2.7~arcsec.} separated by less than 2~mas and having a difference in magnitude of $\sim$1.4 mag. Also, \citet{Sota_2014} marked this star as a likely SB2 from a preliminary inspection of some available high resolution spectra, and \citet{De_Becker_2013} identified this system as a colliding-wind binary. On the other hand, we refer the reader to Sect.~\ref{tess}, were a detailed discussion about the detected variability in the {\it TESS} light curve of HD\,91651 is presented, indicating that the frequency peak located at $\sim$\,7.4~d$^{-1}$ could be produced by a lower mass $\beta$~Cep companion instead of the star itself -- although this statement cannot be definitely confirmed without a proper asteroseismic modelling of the stars and a much larger spectroscopic data~set.

\section{Supplementary tables}
In this appendix, we present the number of spectra we got for each target, by listing the  spectrographs we used as well (Table \ref{table:sample}). In Table \ref{table:phys_parameters}, we present the fundamental physical parameters of fast rotators based on \citet{Holgado_phd} and \citet{Holgado_2020}. Table \ref{table:info_imaging} presents the detailed information about visible components (if any) in a programme sample we found by analyzing {\it Gaia} data and WDSC. In Table \ref{tab-app-eb-ev}, we list some information of interest for the SB1 stars in the slow-rotating domain.

\begin{table*}[!t]
\centering                 
\caption{Number of available spectra and total time-span of the compiled spectroscopic observations for the main working sample of 50 fast-rotating Galactic O-type stars ordered by increasing the \vsini~estimates. For completeness, the four SB2 systems identified as having at least one of the components with a \vsini\ larger than 200~\kms\ are quoted at the bottom of the table. Spectral classifications (SpC) from \citet{Sota_2011, Sota_2014} and \citet{MaizApellaniz_2016}.}              
\label{table:sample}      
\begin{tabular}{l l c c c c  c r} 
\hline
\hline
Name & SpC & \# sp.  & \# sp. & \# sp. & \# sp. &  \# Total sp.  & Time-span \\
     &     &  FEROS  & FIES   & HERMES & STELLA &  ${\rm SNR}_{\rm L}$>5         & [days] \\
\hline
BD+36$^{\circ}$4145   & O8.5\,V(n)        & .   & 3   & 5  & .   & 8    & 1358.21 \\ 
HD\,216532            & O8.5\,V(n)        & .   & 4   & 6  & 8   & 18   & 3401.82 \\ 
HD\,163892            & O9.5\,IV(n)       & 10  & 2   & 4  & 9   & 25   & 4788.68 \\
HD\,210839            & O6.5\,I(n)fp      & .   & 26  & 77 & 9   & 112  & 4400.05 \\ 
HD\,308813            & O9.7\,IV(n)       & 5   & .   & .  & .   & 5    &  352.94  \\
HD\,36879             & O7\,V(n)((f))     & .   & 5   & 2  & 7   & 14   & 4423.11 \\
HD\,37737             & O9.5\,II-III(n)   & .   & 14  & 5  & 6   & 25   & 3392.01 \\
HD\,152200            & O9.7\,IV(n)       & 4   & .   & .  & .   & 4    &    3.89 \\ 
HD\,97434             & O7.5\,III(n)((f)) & 3   & .   & .  & .   & 3    & 2867.09 \\
HD\,24912             & O7.5\,III(n)((f)) & .   & 27  & 81 & 8   & 116  & 4425.13 \\ 
BD+60$^{\circ}$2522   & O6.5\,(n)fp       & .   & 3   & 5  & 3   & 11   & 3390.83 \\ 
HD\,89137             & ON9.7\,II(n)      & 3   & .   & .  & .   & 3    &  745.01 \\ 
BD+60$^{\circ}$134    & O5.5\,V(n)((f))   & .   & 4   & .  & .   & 4    &   47.93 \\ 
HD\,172175            & O6.5\,I(n)fp      & 1   & 1   & 2  & .   & 4    & 3798.48 \\ 
HD\,14442             & O5\,n(f)p         & .   & .   & 3  & .   & 3    & 1636.25 \\ 
HD\,165246            & O8\,V(n)          & 12  & 2   & 4  & .   & 18   & 5424.00 \\ 
HD\,5689              & O7\,Vn((f))       & .   & 2   & 4  & .   & 6    & 1699.06 \\ 
HD\,124314            & O6\,IV(n)((f))    & 15  & .   & .  & .   & 15   & 3150.13 \\
HD\,192281            & O4.5\,IV(n)(f)    & .   & 4   & 5  & 8   & 17   & 3785.60 \\ 
HD\,76556             & O6\,IV(n)((f))p   & 4   & .   & .  & .   & 4    & 1394.12 \\ 
HD\,41997             & O7.5\,Vn((f))     & .   & 3   & 2  & 6   & 11   & 2284.06 \\ 
HD\,124979            & O7.5\,IV(n)((f))  & 6   & .   & .  & .   & 6    & 1419.00 \\
HD\,155913            & O4.5\,Vn((f))     & 8   & .   & .  & .   & 8    & 2103.19 \\
HD\,175876            & O6.5\,III(n)(f)   & 6   & 4   & 5  & 10  & 25   & 4763.66 \\
HD\,15137             & O9.5\,II-IIIn     & .   & 4   & 4  & 7   & 15   & 4028.98 \\
HD\,28446A            & O9.7\,IIn         & .   & 6   & 2  & 6   & 14   & 2929.98 \\
HD\,15642             & O9.5\,II-IIIn     & .   & 4   & 5  & 1   & 10   & 3392.00 \\
HD\,90087             & O9.2\,III(n)      & 4   & .   & .  & .   & 4    & 1142.99 \\
HD\,165174            & O9.7\,IIn         & .   & 5   & 5  & 9   & 19   & 3577.01 \\
HD\,52266             & O9.5\,IIIn        & 3   & 4   & 3  & 7   & 17   & 4248.22 \\
HD\,91651             & ON9.5\,IIIn       & 9   & .   & .  & .   & 9    & 2159.92 \\
HD\,228841            & O6.5\,Vn((f))     & .   & 6   & 4  & .   & 10   & 3574.18 \\
HD\,52533             & O8.5\,IVn         & 1   & 5   & 2  & .   & 8    & 4251.28 \\
BD+60$^{\circ}$513    & O7\,Vn            & .   & 2   & 3  & .   & 5    & 1731.24 \\
HD\,229232            & O4\,Vn((f))       & .   & 5   & 1  & .   & 6    & 1696.23 \\
HD\,13268             & ON\,8.5IIIn       & .   & 4   & 5  & 3   & 12   & 3630.24 \\
HD\,41161             & O8\,Vn            & .   & 5   & 3  & 8   & 16   & 4425.14 \\
HD\,149452   & O9\,IVn           & 5  & .   & .  & .   & 5    &   30.90 \\
HD\,203064            & O7.5\,IIIn((f))   & .   & 6   & 26 & 9   & 41   & 4422.97 \\
HD\,326331            & O8\,IVn((f))      & 24  & .   & .  & .   & 24   & 4070.94 \\
HD\,46485             & O7\,V((f))nvar?   & .   & 3   & 1  & 1   & 5    & 3383.98 \\
HD\,46056A            & O8\,Vn            & 1   & 3   & 3  & 0   & 7    & 3563.16 \\
HD\,117490            & ON9.5\,IIInn      & 10  & .   & .  & .   & 10   & 2518.04 \\
HD\,102415   & ON9\,IV:nn        & 6  & .   & .  & .   & 6  & 1100.03 \\
HD\,93521             & O9.5\,IIInn       & .   & 22  & 11 & 7   & 40   & 4610.61 \\
HD\,217086            & O7\,Vnn((f))z     & .   & 5   & 4  & 2   & 11   & 3770.74 \\
HD\,14434             & O5.5\,IVnn(f)p    & .   & 4   & 1  & .   & 5    & 3388.88 \\
HD\,191423            & ON9\,II-IIInn     & .   & 3   & 6  & 5   & 14   & 3399.75 \\
HD\,149757            & O9.2\,IVnn        & 3   & 177 & 20 & 10  & 210  & 4433.73 \\ 
ALS\,12370            & O6.5\,Vnn((f))    & .   & 4   & .  & .   & 4    &   48.11 \\ 
\hline
HD\,175514            & O7\,V(n)((f))z\,+\,B  & 2   & 1   & 4  & .   & 7    & .   \\
HD\,165921            & O7\,V(n)\,z\,+B0:\,V: & 1   & 7   & 1  & .   & 9    & .   \\
HD\,100213            & O8\,V(n)z\,+\,B0V(n)  & 8   & .   & .  & .   & 8    & .   \\
HDE\,228854           & O6\,IVn\,+\,O5\,Vn    & .   & 2   & .  & .   & 2    & .   \\
\hline
\end{tabular}
\end{table*}


\begin{table*}[!t]
{\tiny
\centering                 
\caption{Basic information about fundamental physical parameters of our sample of fast-rotating O-type stars ordered by increasing the \vsini~estimates. Extracted from \citet{Holgado_phd} and \citet{Holgado_2020}. The last two columns include information about the RUWE parameter from {\it Gaia}-EDR3 and the computed distances by \cite{Bailer_Jones_2021}. Uncertainties in brackets corresponds to the uncertainty in distance while the ones without brackets are the result of the {\sc iacob-gbat} analysis.}  
\label{table:phys_parameters}     
\begin{tabular}{l c c c c c c c c} 
\hline
\hline
Name  &  \Teff   &  log($\mathcal{L/L_{\odot}}$) & $R$ & log($L$/$L_{\odot}$)  &  $v_{\rm crit}$  & \vsini/$v_{\rm crit}$ & RUWE & distBJ \\
      &    [kK]  &  [dex]                        & [$R_{\odot}$] & [dex]         &   [\kms]           &      &  [kpc] \\
\hline
BD+36$^{\circ}$4145   & 35.8$\pm$0.9 & 3.75$\pm$0.16 &     9.3$\pm$0.2  ($\pm$0.2) &   5.1$\pm$0.03  ($\pm$0.02) &  673 & 0.3 & 1.12 & 1.4 \\
HD\,216532            & 35.3$\pm$0.6 & 3.54$\pm$0.08 &    6.8$\pm$0.1  ($\pm$0.15) &   4.8$\pm$0.02  ($\pm$0.02) &  724 & 0.3 & 1.57 & 0.7 \\
HD\,163892            & 32.8$\pm$0.5 & 3.68$\pm$0.06 &    9.3$\pm$0.1  ($\pm$0.35) &  4.95$\pm$0.02  ($\pm$0.04) &  620 & 0.3 & 0.69 & 1.3 \\
HD\,210839            & 35.8$\pm$0.5 & 4.14$\pm$0.05 &   19.2$\pm$0.2  ($\pm$1.35) &  5.74$\pm$0.01  ($\pm$0.06) &  530 & 0.4 & 0.97 & 0.8 \\
HD\,308813            & 31.8$\pm$0.5 & 3.52$\pm$0.06 &    6.9$\pm$0.1  ($\pm$0.25) &  4.65$\pm$0.02  ($\pm$0.03) &  641 & 0.3 & 0.78 & 2.4 \\
HD\,36879             & 36.9$\pm$0.5 & 3.89$\pm$0.05 &    11.9$\pm$0.1  ($\pm$0.6) &  5.37$\pm$0.02  ($\pm$0.04) &  651 & 0.3 & 1.01 & 1.7 \\
HD\,37737             & 30.0$\pm$0.5 & 3.80$\pm$0.04 &    10.3$\pm$0.1  ($\pm$0.9) &  4.88$\pm$0.02  ($\pm$0.08) &  524 & 0.4 & 2.53 & 1.4 \\
HD\,152200            & 30.4$\pm$0.7 & 3.63$\pm$0.12 &     6.8$\pm$0.1  ($\pm$0.2) &  4.54$\pm$0.02  ($\pm$0.03) &  579 & 0.4 & 0.86 & 1.4 \\
HD\,97434             & 34.8$\pm$0.5 & 3.99$\pm$0.07 &    13.0$\pm$0.2  ($\pm$0.6) &  5.35$\pm$0.01  ($\pm$0.04) &  568 & 0.4 & 0.91 & 2.3 \\
HD\,24912             & 35.9$\pm$0.5 & 3.94$\pm$0.05 &   10.9$\pm$0.1  ($\pm$1.25) &   5.25$\pm$0.01  ($\pm$0.1) &  609 & 0.4 & 2.24 & 0.4 \\
BD+60$^{\circ}$2522   & 36.2$\pm$1.1 & 4.08$\pm$0.15 &   18.5$\pm$0.4  ($\pm$0.75) &  5.72$\pm$0.03  ($\pm$0.03) &  560 & 0.4 & 1.04 & 2.8 \\
HD\,89137             & 29.1$\pm$0.5 & 3.88$\pm$0.04 &    11.4$\pm$0.2  ($\pm$0.8) &  4.92$\pm$0.02  ($\pm$0.06) &  487 & 0.5 & 1.01 & 2.4 \\
BD+60$^{\circ}$134    & 40.7$\pm$1.8 & 3.88$\pm$0.21 &    7.7$\pm$0.2  ($\pm$0.25) &  5.16$\pm$0.05  ($\pm$0.03) &  754 & 0.3 & 0.93 & 2.7 \\
HD\,172175            & 36.2$\pm$0.5 & 4.05$\pm$0.06 &   13.5$\pm$0.1  ($\pm$0.45) &  5.45$\pm$0.01  ($\pm$0.03) &  579 & 0.4 & 0.78 & 2.6 \\
HD\,14442             & 39.1$\pm$1.3 & 4.14$\pm$0.14 &   10.8$\pm$0.2  ($\pm$0.45) &  5.38$\pm$0.04  ($\pm$0.04) &  596 & 0.4 & 0.97 & 2.6 \\
HD\,165246            & 35.9$\pm$0.7 & 3.62$\pm$0.08 &     7.8$\pm$0.1  ($\pm$0.3) &  4.96$\pm$0.02  ($\pm$0.04) &  715 & 0.4 & 0.94 & 1.2 \\
HD\,5689              & 36.8$\pm$1.0 & 3.93$\pm$0.11 &    11.1$\pm$0.2  ($\pm$0.5) &  5.31$\pm$0.03  ($\pm$0.04) &  638 & 0.4 & 0.98 & 2.8 \\
HD\,124314            & 37.0$\pm$0.5 & 3.95$\pm$0.05 &   17.9$\pm$0.2  ($\pm$1.15) &  5.73$\pm$0.02  ($\pm$0.06) &  670 & 0.4 & 0.89 & 1.6 \\
HD\,192281            & 40.8$\pm$1.1 & 4.01$\pm$0.09 &     9.7$\pm$0.2  ($\pm$0.2) &  5.37$\pm$0.04  ($\pm$0.01) &  699 & 0.4 & 0.96 & 1.2 \\
HD\,76556             & 37.9$\pm$0.5 & 3.81$\pm$0.08 &    13.1$\pm$0.1  ($\pm$0.6) &   5.5$\pm$0.01  ($\pm$0.04) &  702 & 0.4 & 0.97 & 1.8 \\
HD\,41997             & 35.8$\pm$0.5 & 3.86$\pm$0.06 &    10.8$\pm$0.1  ($\pm$0.5) &  5.23$\pm$0.02  ($\pm$0.04) &  638 & 0.4 & 1.59 & 1.7 \\
HD\,124979            & 34.9$\pm$0.9 & 3.98$\pm$0.07 &    14.1$\pm$0.3  ($\pm$1.4) &  5.42$\pm$0.03  ($\pm$0.09) &  571 & 0.5 & 1.06 & 3.6 \\
HD\,155913            & 42.5$\pm$1.5 & 3.88$\pm$0.13 &    8.9$\pm$0.1  ($\pm$0.25) &  5.36$\pm$0.04  ($\pm$0.02) &  795 & 0.4 & 0.96 & 1.2 \\
HD\,175876            & 36.1$\pm$0.6 & 4.03$\pm$0.05 &   14.3$\pm$0.2  ($\pm$1.45) &   5.5$\pm$0.02  ($\pm$0.09) &  580 & 0.5 & 0.97 & 2.3 \\
HD\,15137             & 30.3$\pm$0.5 & 3.84$\pm$0.04 &    10.7$\pm$0.2  ($\pm$0.8) &  4.94$\pm$0.02  ($\pm$0.06) &  518 & 0.6 & 0.89 & 2.0 \\ 
HD\,28446A            & 29.8$\pm$0.5 & 3.64$\pm$0.06 &   13.7$\pm$1.5  ($\pm$0.55) &  5.04$\pm$0.02  ($\pm$0.04) &  562 & 0.5 & 0.94 & 0.8 \\
HD\,15642             & 29.9$\pm$0.8 & 3.77$\pm$0.06 &    9.6$\pm$0.2  ($\pm$0.45) &  4.81$\pm$0.03  ($\pm$0.04) &  531 & 0.6 & 1.22 & 2.3 \\
HD\,90087             & 31.6$\pm$0.6 & 3.89$\pm$0.08 &   11.8$\pm$0.2  ($\pm$0.65) &   5.1$\pm$0.02  ($\pm$0.05) &  535 & 0.6 & 0.84 & 2.2 \\
HD\,165174            & 30.2$\pm$0.8 & 3.86$\pm$0.09 &    10.7$\pm$0.2  ($\pm$0.5) &  4.93$\pm$0.03  ($\pm$0.04) &  514 & 0.6 & 0.86 & 1.0 \\
HD\,52266             & 32.2$\pm$0.8 & 3.79$\pm$0.10 &     8.9$\pm$0.1  ($\pm$0.4) &  4.89$\pm$0.02  ($\pm$0.04) &  576 & 0.5 & 0.87 & 1.4 \\
HD\,91651             & 31.8$\pm$0.8 & 3.84$\pm$0.07 &    6.1$\pm$0.1  ($\pm$0.25) &  4.53$\pm$0.02  ($\pm$0.04) &  557 & 0.6 & 0.95 & 1.8 \\
HD\,228841            & 37.7$\pm$1.4 & 3.82$\pm$0.15 &    9.0$\pm$0.2  ($\pm$0.15) &  5.16$\pm$0.04  ($\pm$0.02) &  701 & 0.4 & 1.01 & 1.7 \\
HD\,52533             & 35.2$\pm$0.5 & 3.60$\pm$0.06 &     7.9$\pm$0.1  ($\pm$0.6) &  4.92$\pm$0.02  ($\pm$0.06) &  704 & 0.4 & 0.81 & 1.7 \\
BD+60$^{\circ}$513    & 35.8$\pm$1.0 & 3.75$\pm$0.11 &     9.4$\pm$0.2  ($\pm$0.3) &  5.11$\pm$0.04  ($\pm$0.03) &  674 & 0.5 & 0.96 & 2.0 \\
HD\,229232            & 42.9$\pm$2.2 & 4.11$\pm$0.13 &    9.8$\pm$0.3  ($\pm$0.15) &  5.46$\pm$0.06  ($\pm$0.01) &  693 & 0.5 & 0.90 & 1.6 \\
HD\,13268             & 34.2$\pm$0.5 & 3.92$\pm$0.05 &     8.7$\pm$0.1  ($\pm$0.4) &  4.96$\pm$0.02  ($\pm$0.04) &  584 & 0.5 & 0.98 & 1.8 \\
HD\,41161             & 35.2$\pm$0.6 & 3.74$\pm$0.06 &   10.1$\pm$0.1  ($\pm$0.75) &  5.15$\pm$0.02  ($\pm$0.06) &  665 & 0.5 & 1.00 & 1.4 \\
HD\,149452            & 33.7$\pm$0.8 & 3.77$\pm$0.08 &    9.8$\pm$0.2  ($\pm$0.25) &  5.04$\pm$0.02  ($\pm$0.02) &  616 & 0.5 & 1.22 & 1.4 \\
HD\,203064            & 35.3$\pm$0.5 & 3.89$\pm$0.04 &    11.8$\pm$0.1  ($\pm$1.0) &  5.29$\pm$0.01  ($\pm$0.08) &  613 & 0.5 & 0.87 & 0.7 \\
HD\,326331            & 34.9$\pm$0.5 & 3.82$\pm$0.05 &    10.8$\pm$0.1  ($\pm$0.3) &  5.19$\pm$0.02  ($\pm$0.02) &  628 & 0.5 & 1.62 & 1.4 \\
HD\,46485             & 36.1$\pm$0.7 & 3.74$\pm$0.05 &     7.5$\pm$0.1  ($\pm$0.2) &  4.93$\pm$0.03  ($\pm$0.02) &  683 & 0.5 & 0.74 & 1.2 \\
HD\,46056             & 35.5$\pm$0.8 & 3.58$\pm$0.10 &    7.1$\pm$0.1  ($\pm$0.25) &  4.85$\pm$0.03  ($\pm$0.03) &  718 & 0.5 & 0.90 & 1.4 \\
HD\,117490            & 31.6$\pm$0.7 & 3.65$\pm$0.07 &    6.9$\pm$0.1  ($\pm$0.35) &  4.63$\pm$0.03  ($\pm$0.04) &  605 & 0.6 & 0.78 & 2.1 \\
HD\,102415            & 33.1$\pm$1.1 & 3.55$\pm$0.12 &    7.2$\pm$0.1  ($\pm$0.25) &  4.74$\pm$0.04  ($\pm$0.03) &  669 & 0.6 & 0.95 & 2.1 \\
HD\,93521             & 31.7$\pm$0.8 & 3.61$\pm$0.09 &     6.5$\pm$0.1  ($\pm$0.6) &  4.59$\pm$0.03  ($\pm$0.08) &  615 & 0.6 & 1.11 & 1.2 \\
HD\,217086            & 37.7$\pm$0.8 & 3.70$\pm$0.08 &   10.0$\pm$0.1  ($\pm$0.15) &  5.26$\pm$0.03  ($\pm$0.01) &  736 & 0.5 & 0.81 & 0.8 \\
HD\,14434             & 38.6$\pm$1.1 & 3.78$\pm$0.12 &     9.3$\pm$0.2  ($\pm$0.4) &  5.24$\pm$0.04  ($\pm$0.04) &  734 & 0.5 & 1.01 & 2.2 \\
HD\,191423            & 32.3$\pm$0.8 & 3.72$\pm$0.07 &     9.2$\pm$0.1  ($\pm$0.4) &  4.92$\pm$0.03  ($\pm$0.04) &  602 & 0.7 & 1.00 & 1.7 \\
HD\,149757            & 32.0$\pm$0.5 & 3.75$\pm$0.05 &     7.9$\pm$0.1  ($\pm$0.9) &   4.77$\pm$0.02  ($\pm$0.1) &  583 & 0.7 & 4.49 & 0.1 \\
ALS\,12370            & 39.0$\pm$1.8 & 3.63$\pm$0.22 &    10.1$\pm$0.2  ($\pm$0.9) &  5.32$\pm$0.05  ($\pm$0.08) &  792 & 0.6 & 0.86 & 4.8 \\
\hline
\end{tabular}
}
\end{table*}


\begin{table*}[!t]
\centering                 
\caption{Information about visible components for our sample of fast rotators. Only those cases with detected companions within 2 arcmin and having a difference in magnitude smaller than 3 mag are quoted in this table. Companions found within 1 arcmin, and between 1 and 2 arcmin are presented separately. A couple of companions with a difference in magnitude larger than 3 mag but a separation smaller than 2 arcsec are also included. Companions marked with an asterisk were detected by \citet{Sana_2014}.}
\label{table:info_imaging}      
\begin{tabular}{l l l c c c c c c c} 
\hline
\hline
Name  &  SB? &  EB?  & RW?  & \multicolumn{3}{c}{Info compon. < 1~arcmin} & \multicolumn{3}{c}{Info compon.: 1\,--\,2 arcmin} \\
      &      &       &      & \# comp. & $\Delta$V & ang.~dist. & \# comp. & $\Delta$V & ang.~dist. \\
      &      &       &      &          & (mag)     & (arcsec)   &          & mag       & (arcsec) \\
\hline
HD\,216532          & LS       &  n    & n  & 0    & ...  & ...              & 1    & -2.8 & 83  \\
HD\,308813          & SB1      &  n    & n  & 0    & ...  & ...              & 1    & +2.1 & 71  \\
HD\,152200          & SB1      &  n   & n  & 0    & ...  & ...              & 1    & -2.9 & 63  \\
HD\,97434           & LS       &  n    & n   & 0    & ...  & ...              & 2    & -1.5 & 67  \\
                    & ...      &  ...  & ... & ...  & ...  & ...              & ...  & -2.5 & 87  \\
HD\,165246          & SB1      &  EB   & n   & 1    & -2.8 & 0.0$^{*}$        & 0    & ...  & ... \\
HD\,124314          & LPV/SB2? &  ...  & ... & 2    & -1.4 & 0.0$^{*}$        & 0    &  ...  & ... \\
                    & ...      &  ...  & ... & ...  & -1.9 & 2.7$^{*}$        & ...  & ...   & ... \\
HD\,76556           & LS      &  n    & n   & 2    & -3.0 & 0.0$^{*}$        & 1    & -0.6 & 97  \\
HD\,28446A          & LS       &  n    & n   & 1    & -1.1 & 12               & 0    & ...  & ... \\
HD\,15642           & LS       &  n    & n   & 0    & ...  & ...              & 1    & -2.8 & 75  \\
HD\,52533           & SB1      &  EB   & n   & 3    & -3.8 & 0.6              & 0    & ...  & ...  \\
                    & ...      &  ...  & ... & ...  & -1.1 & 23               & ...  & ...  & ...  \\
                    & ...      &  ...  & ... & ...  & -3.0 & 54               & ...  & ...  & ...   \\
BD+60$^{\circ}$513  & LS       &  n    & n   & 1    & -0.8  & 38              & 0    & ...  & ...   \\
HD\,229232          & LS       &  n    & y   & 1    & -2.3 & 37               & 0    & ...  & ...  \\
HD\,14442           & LS       &  n    & n   & 0    & ...  & ...              & 1    & -2.0 & 97  \\
HD\,149452          & LS       &  n    & y   & 1    & -1.6 & 50               & 0    & ...  & ... \\
HD\,326331          & LS       &  n    & n   & 2    & -6.0 & 1.1$^{*}$        & 2    & +0.9 & 82 \\
                    & ...      &  ...  & ... & ...  & -2.5 & 7                & ...  & ...  & ... \\
HD\,46056A          & LS       &  n    & n   & 1    & -2.7 & 10               & 0    & ...  & ... \\
HD\,117490          & LS       &  n    & y   & 0    & ...  & ...              & 2    & -2.0 & 67 \\
                    & ...      &  ...  & ... & ...  & ...  & ...              & ...  & -3.0 & 75 \\
HD\,102415          & LS       &  n    & n   & 0    & ...  & ...              & 1    & -2.6 & 67 \\
\hline
\end{tabular}
\end{table*}

\begin{table}[!t]
\centering                 
\caption{Some information of interest for the SB1 stars in the low \vsini\ sample for which we have detected signatures of eclipses or ellipsoidal/reflection modulation in their {\em TESS} light curves.}              
\label{tab-app-eb-ev}     
\begin{tabular}{l r r r r r} 
\hline
\hline
ID & SpC & \vsini & \RVpp & P & Phot. \\
   &     & [\kms] & [\kms] & [d]  & var. tag\\
\hline
HD\,36486   & O9.5\,II\,N\,wk & 100 & 188 &  & EB \\  
HD\,152590  & O7.5\,Vz & 48 & 158 &  & EB \\ 
HD\,226868  & O9.7\,Iab\,p var & 95 & 143 & 5.59 & EV \\
HD\,12323   & ON9.2\,V & 121 & 58 & 1.92 & EV \\
HD\,94024   & O8\,IV & 162 & 54 & 2.46 & EV \\
HD\,53975   & O7.5\,Vz & 179 & 47 &  & EV/RM \\
BD+60$^{\circ}$498   & O9.7\,II-III & 114 & 43 &  & EB \\
\hline
\hline
\end{tabular}
\end{table}

\section{Supplementary figures}
Figure \ref{figure_p2p_err} presents the quality of \RVpp~ measurements as a function of \vsini~ and local S/N around He{\sc i}~$\lambda$5875 line. 
In Fig. \ref{dist_fast}, we present the distance distribution for all O-type stars in the slow- and fast-rotating domain. 
In Fig. \ref{figure_sb1_illustration}, we illustrate the possible type of second components based on the BPASS model simulations. 

\begin{figure*}[]
\centering
\includegraphics[width=0.49\textwidth]{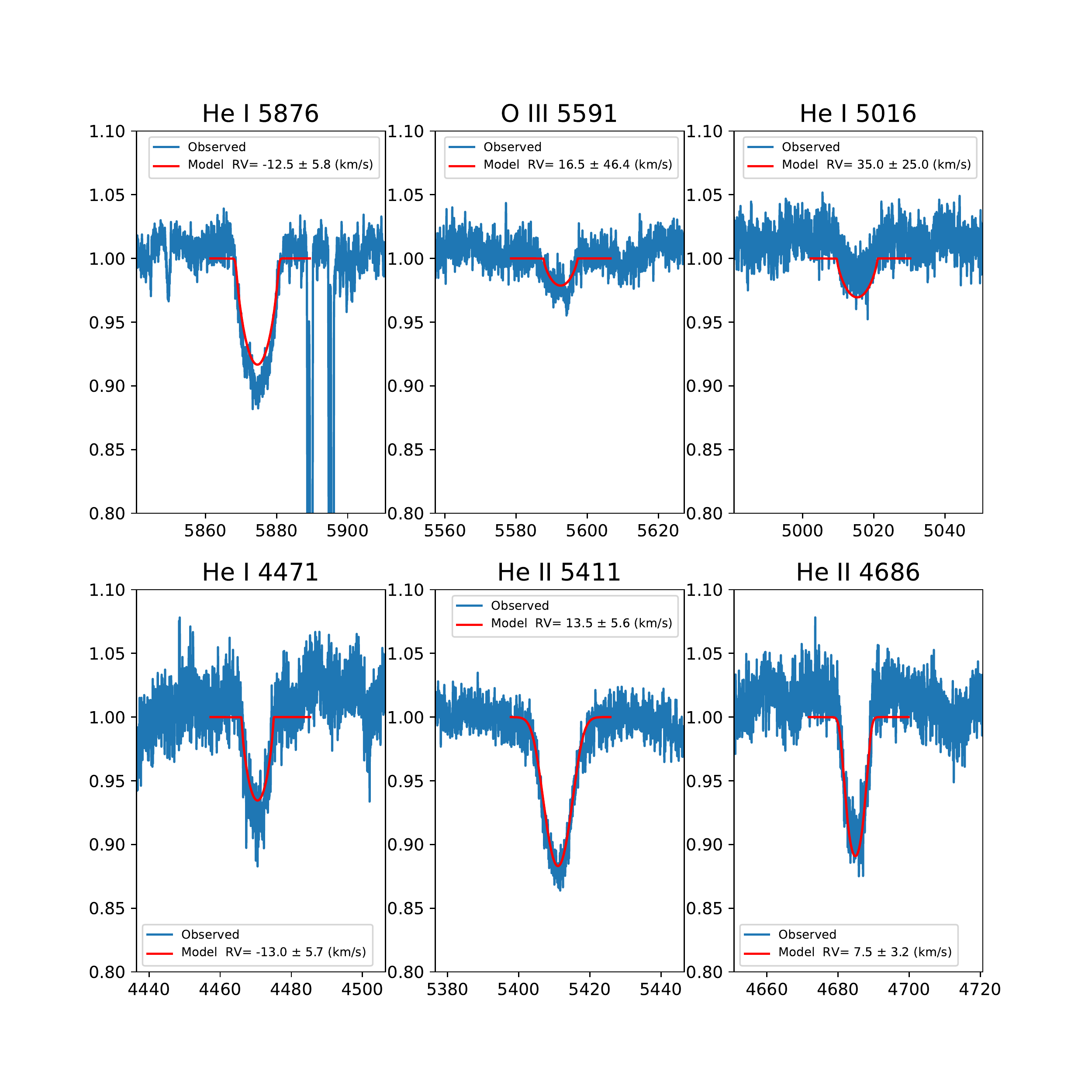}
\includegraphics[width=0.49\textwidth]{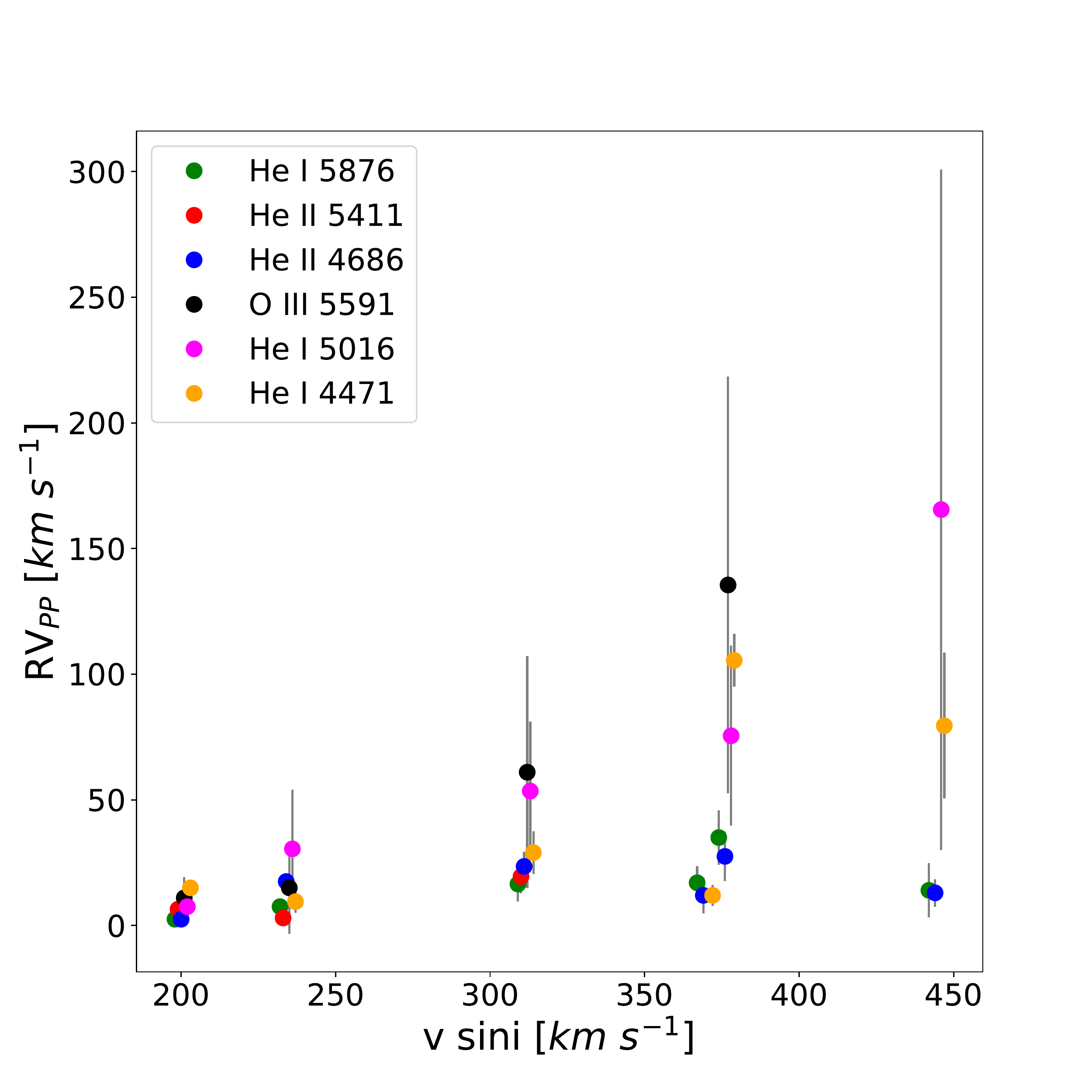}
\caption{ Example of line profile fitting of HD\,228841 for six diagnostic lines of O-type stars (left panel). \RVpp\ estimates based on all presented lines for six stars (sorted by \vsini\ values) selected as illustrative cases: BD+36$^{\circ}$4145 [O8.5\,V(n)],  BD+60$^{\circ}$134 [O5.5\,V(n)((f))], HD\,228841 [O6.5\,Vn((f))], HD\,117490 [ON9.5\,IIInn], HD\,102415 [ON9\,IV:nn], and ALS12370 [O6.5\,Vnn((f))] (right panel). Note how, for those lines that are too weak, the cross-correlation technique provides individual $RV$ measurements with a larger uncertainty (e.g. the \ion{O}{iii}\,$\lambda$5591 and \ion{He}{i}\,$\lambda$5015 lines in HD\,228841, see left panel), or no results at all. While the \ion{He}{i}\,$\lambda$5875 line is always available and provides quite accurate \RVpp\ measurements, this is not the case for the other five diagnostic lines, for which the possibility to use them critically depends on the spectral type and \vsini\ combination.} 
\label{figure_lines_example}
\end{figure*}

\begin{figure}[!t]
\centering
\includegraphics[width=0.50\textwidth]{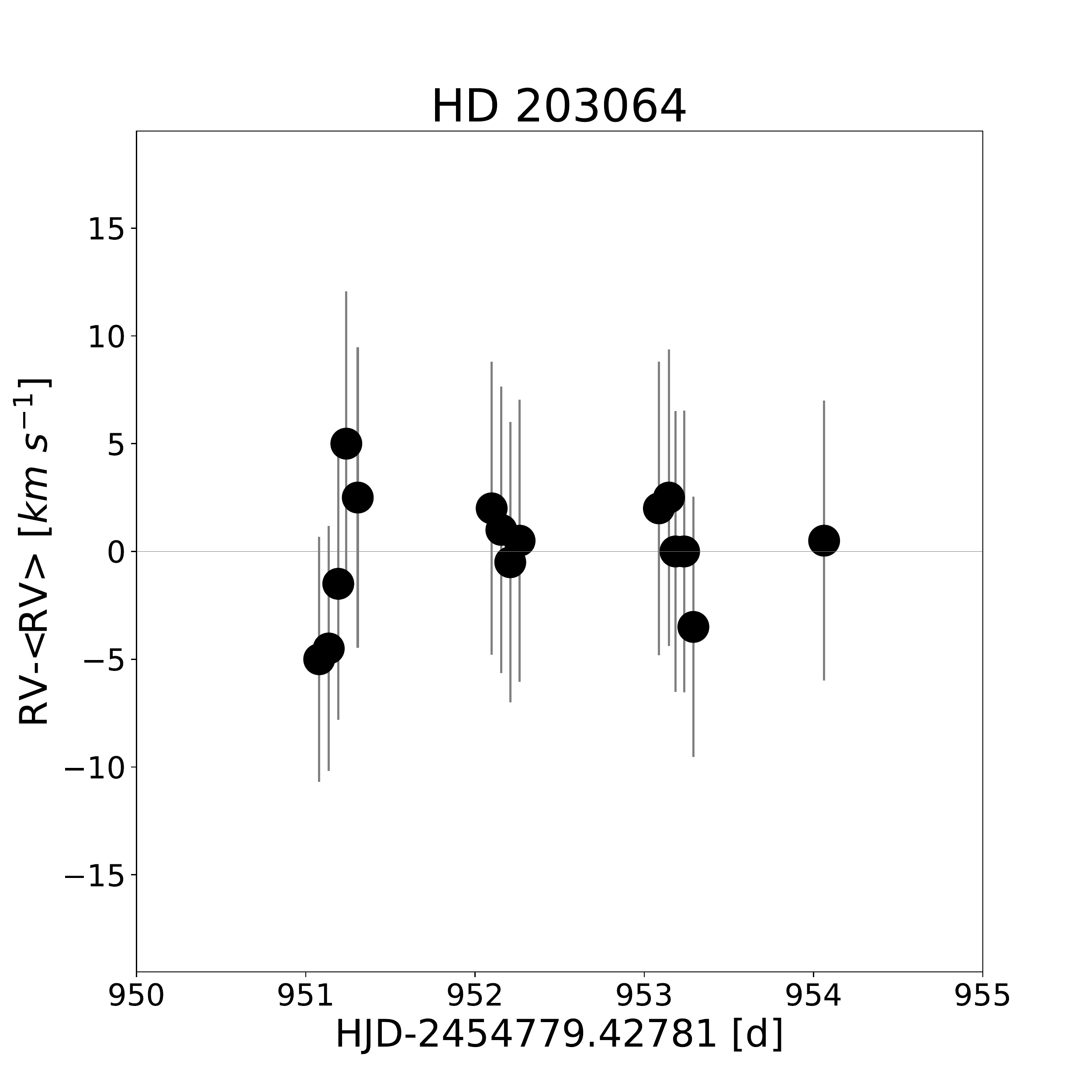}
\caption{Small portion of the radial velocity curve of HD\,203064 illustrating that the detected variability is not compatible with a 5.1~d orbit, as previously proposed, but more likely is associated with line-profile variability produced by stellar oscillations, wind variability or spots on the surface.}
\label{hd_203064_rvs}
\end{figure}


\begin{figure*}[]
\centering
\includegraphics[width=0.49\textwidth]{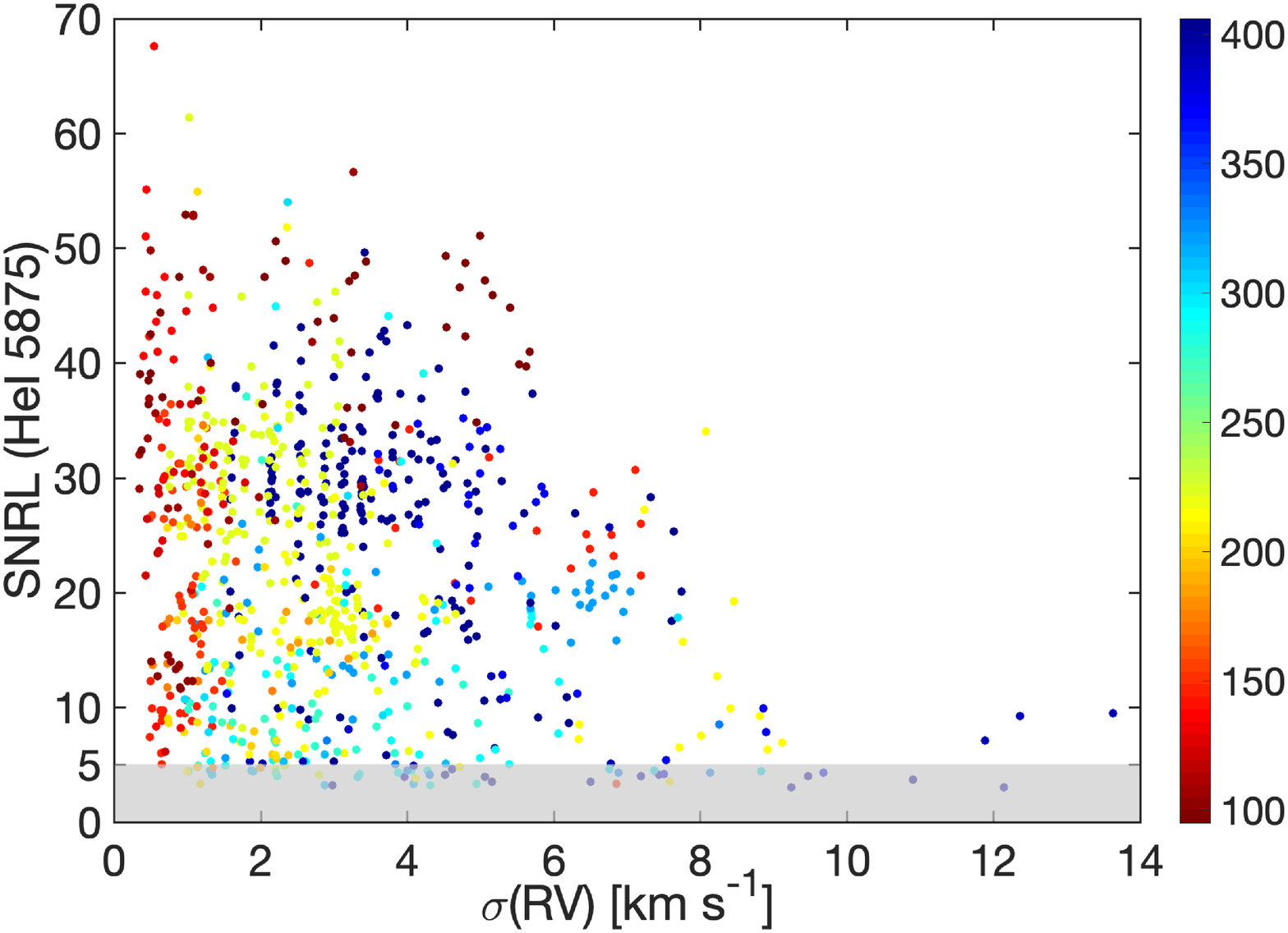}
\includegraphics[width=0.49\textwidth]{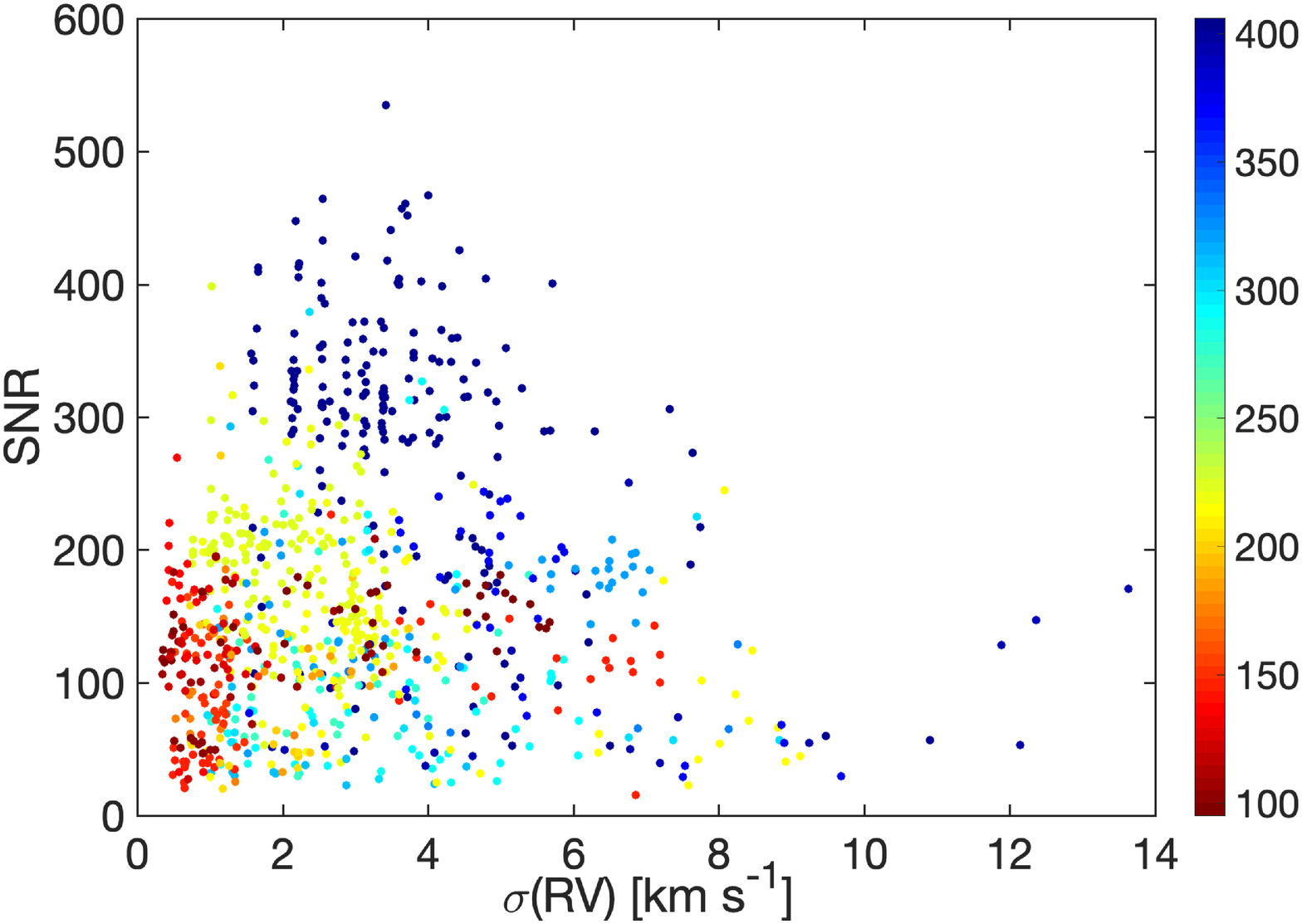}
\caption{Distribution of individual radial velocity errors vs. signal-to-noise of the He{\sc i}~$\lambda$5875 line (S/NL, left panel) and overall signal-to-noise of the spectrum (S/N, right panel) respect to the \vsini~ (the colour bars on the right) for all analyzed spectra from the working sample. The gray region represents the limit of  ${\rm S/NL}$ <5.} 
\label{figure_p2p_err}
\end{figure*}

\begin{figure*}[!t]
\centering
\includegraphics[width=0.40\textwidth]{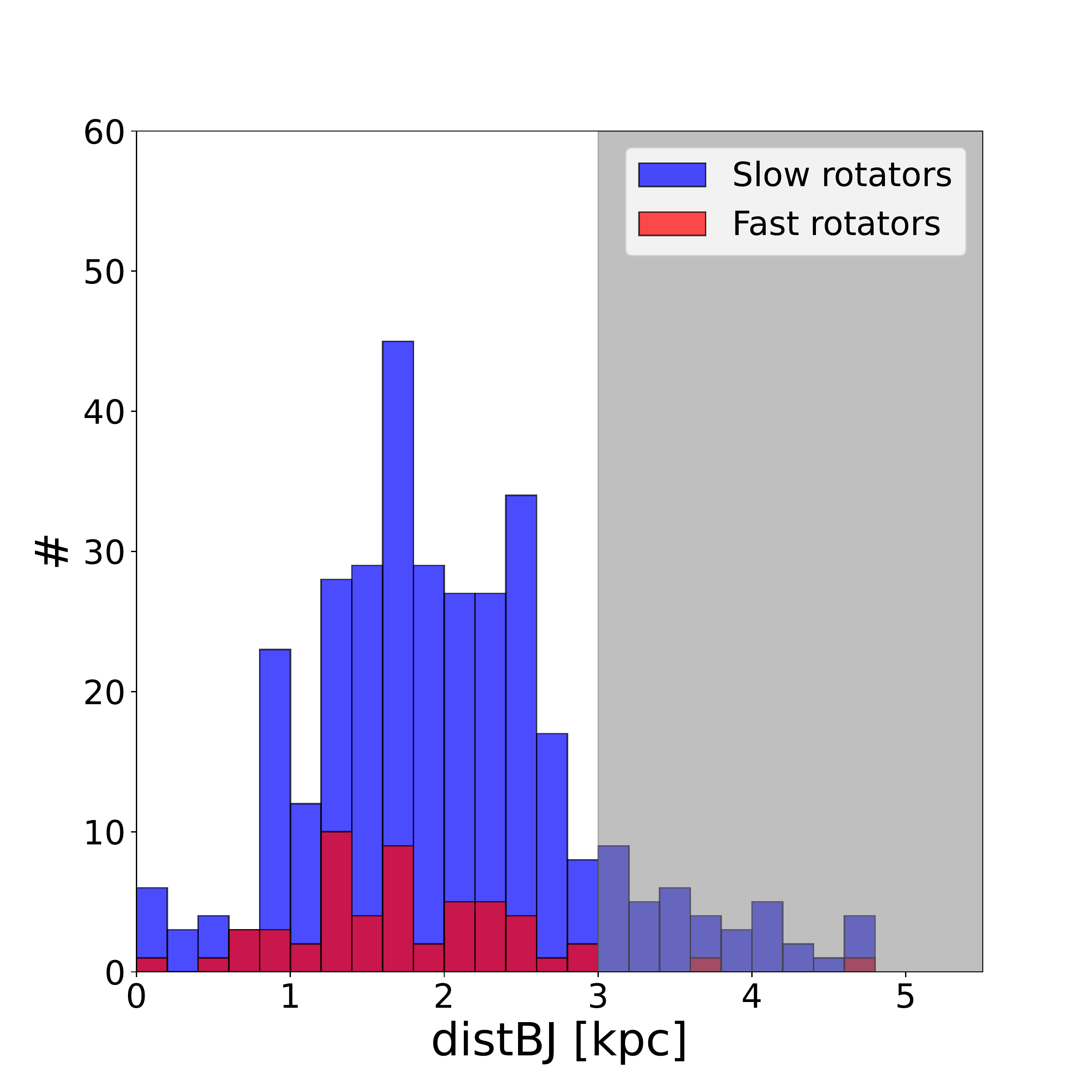}
\includegraphics[width=0.40\textwidth]{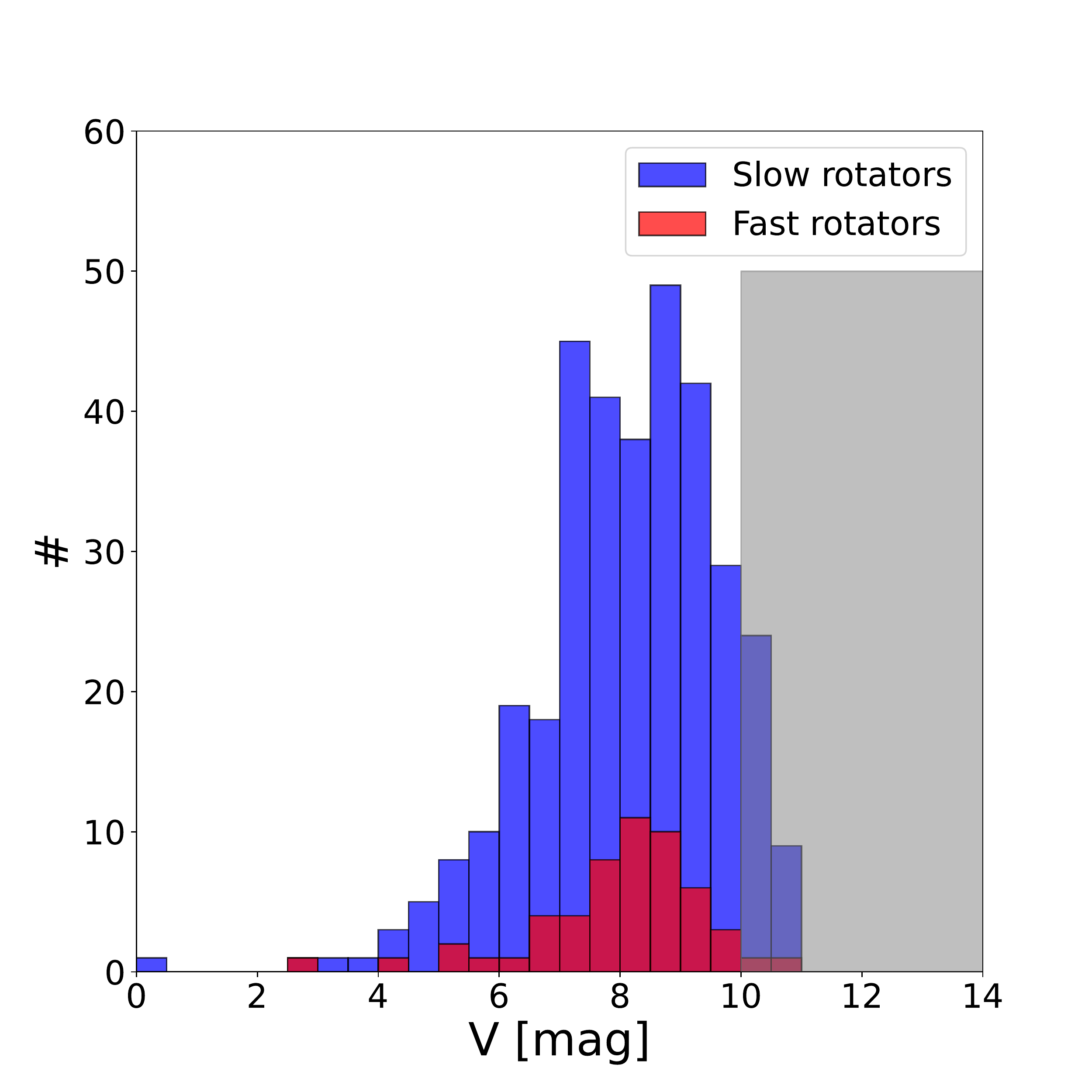}
\caption{Distance and V magnitude distribution of the 285 LS+SB1 stars and 113 SB2 systems comprising the sample of Galactic O-type stars investigated by \cite{Holgado_2020, Holgado_2022} Distance estimates from \citet{Bailer_Jones_2021}. We highlight in grey the ranges in distance and V magnitude of the sample of stars which has been excluded for the discussion presented in Section~\ref{discuss}.}
\label{dist_fast}
\end{figure*}

\begin{figure*}[!ht]
\centering
\includegraphics[width=0.40\textwidth]{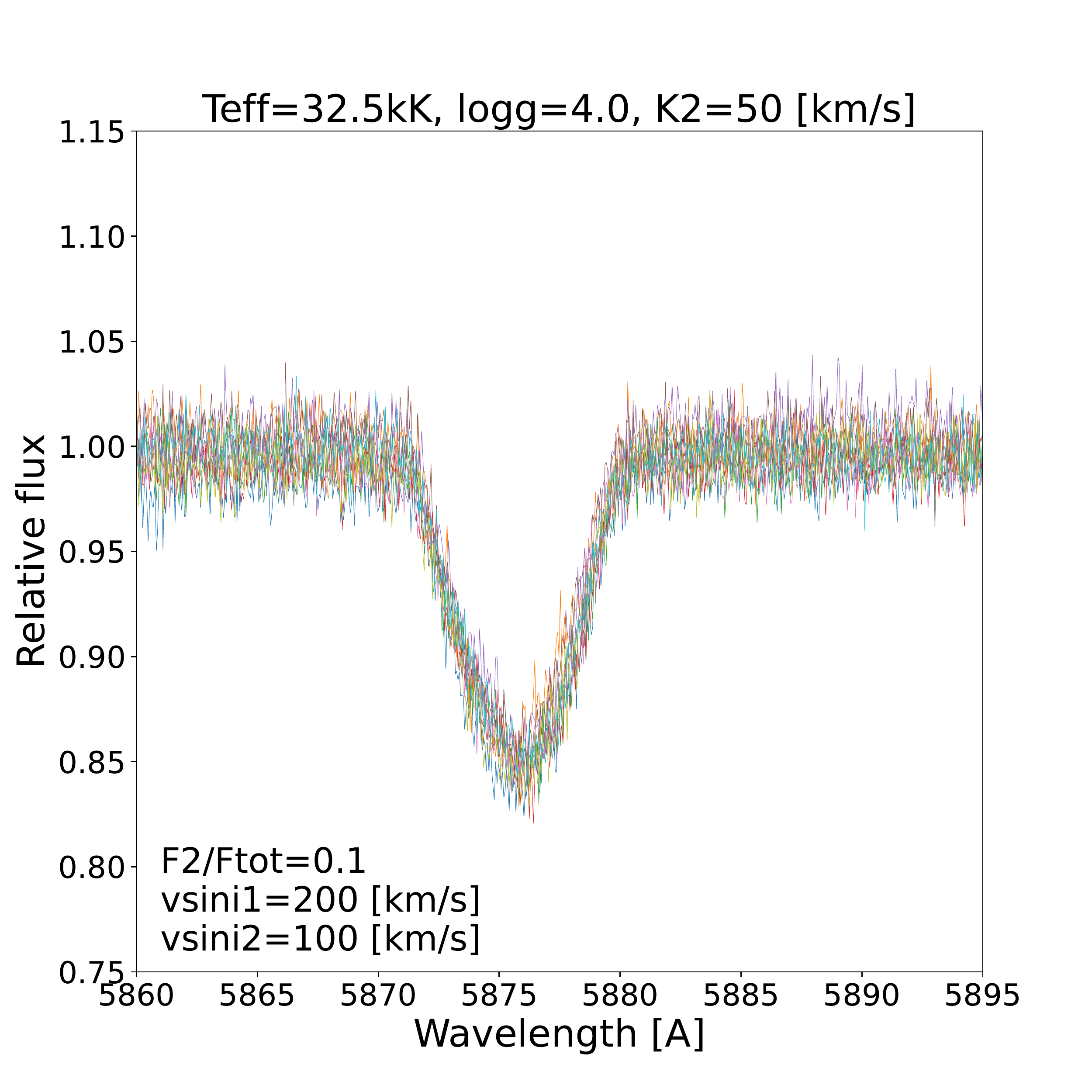}
\includegraphics[width=0.40\textwidth]{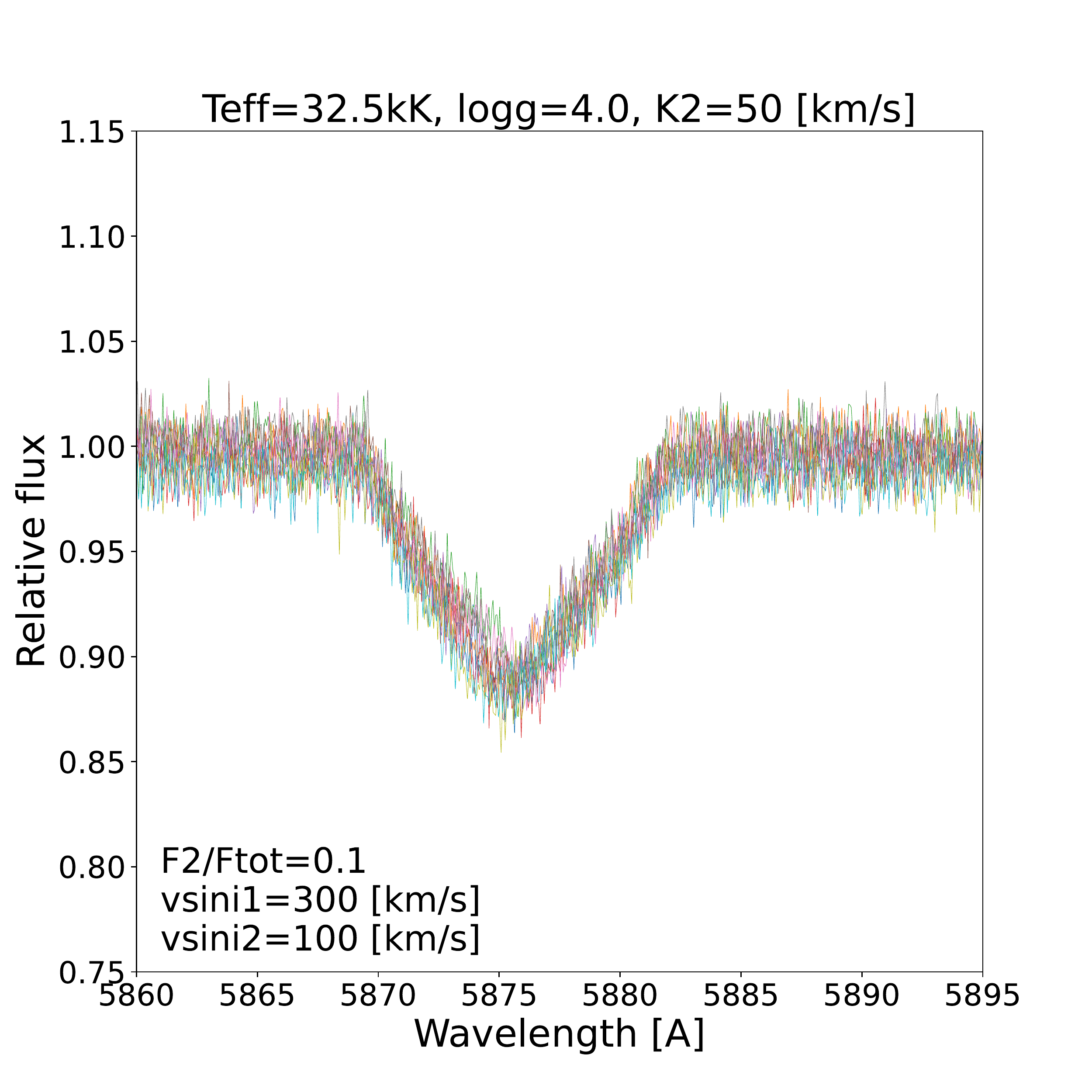}
\includegraphics[width=0.40\textwidth]{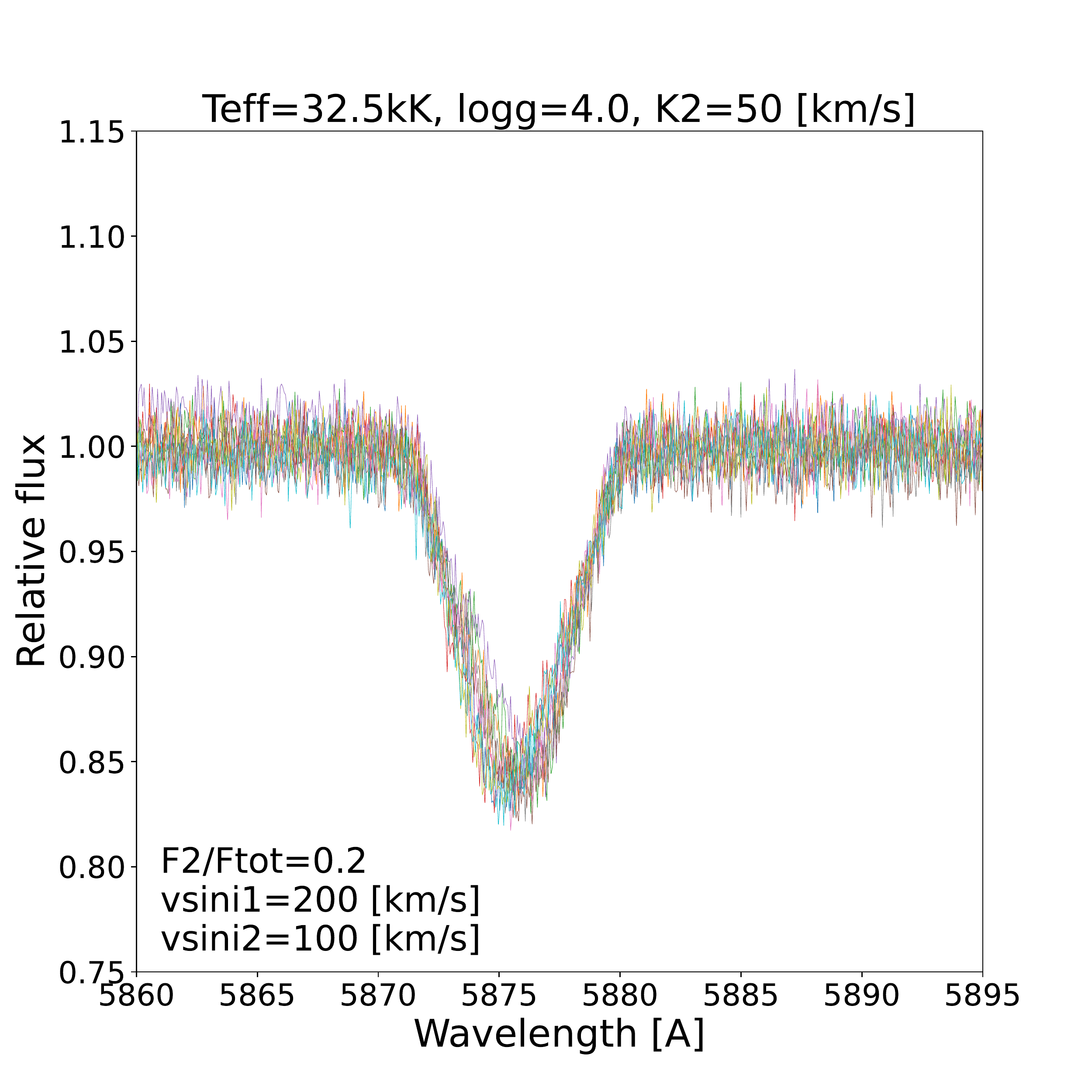}
\includegraphics[width=0.40\textwidth]{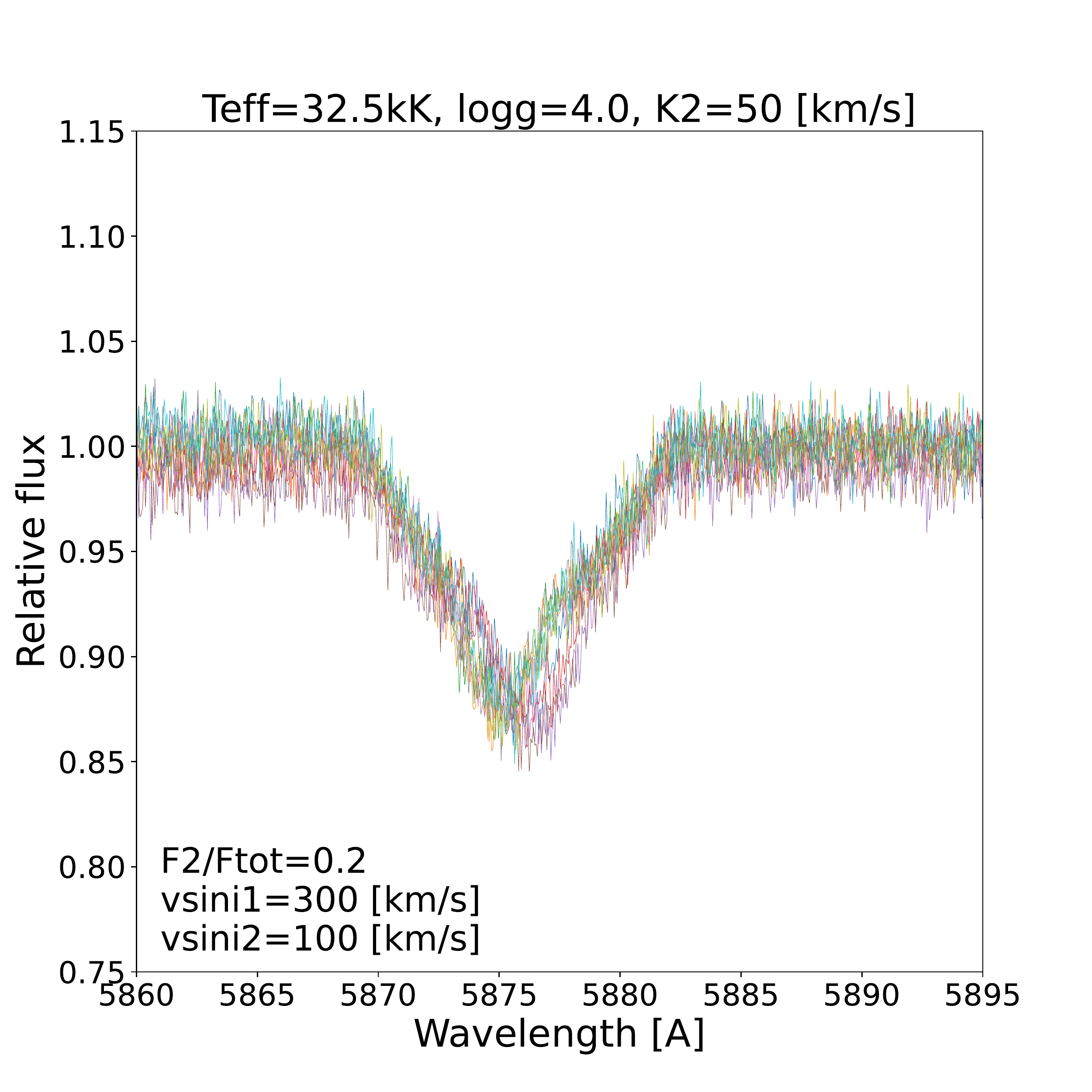}
\includegraphics[width=0.40\textwidth]{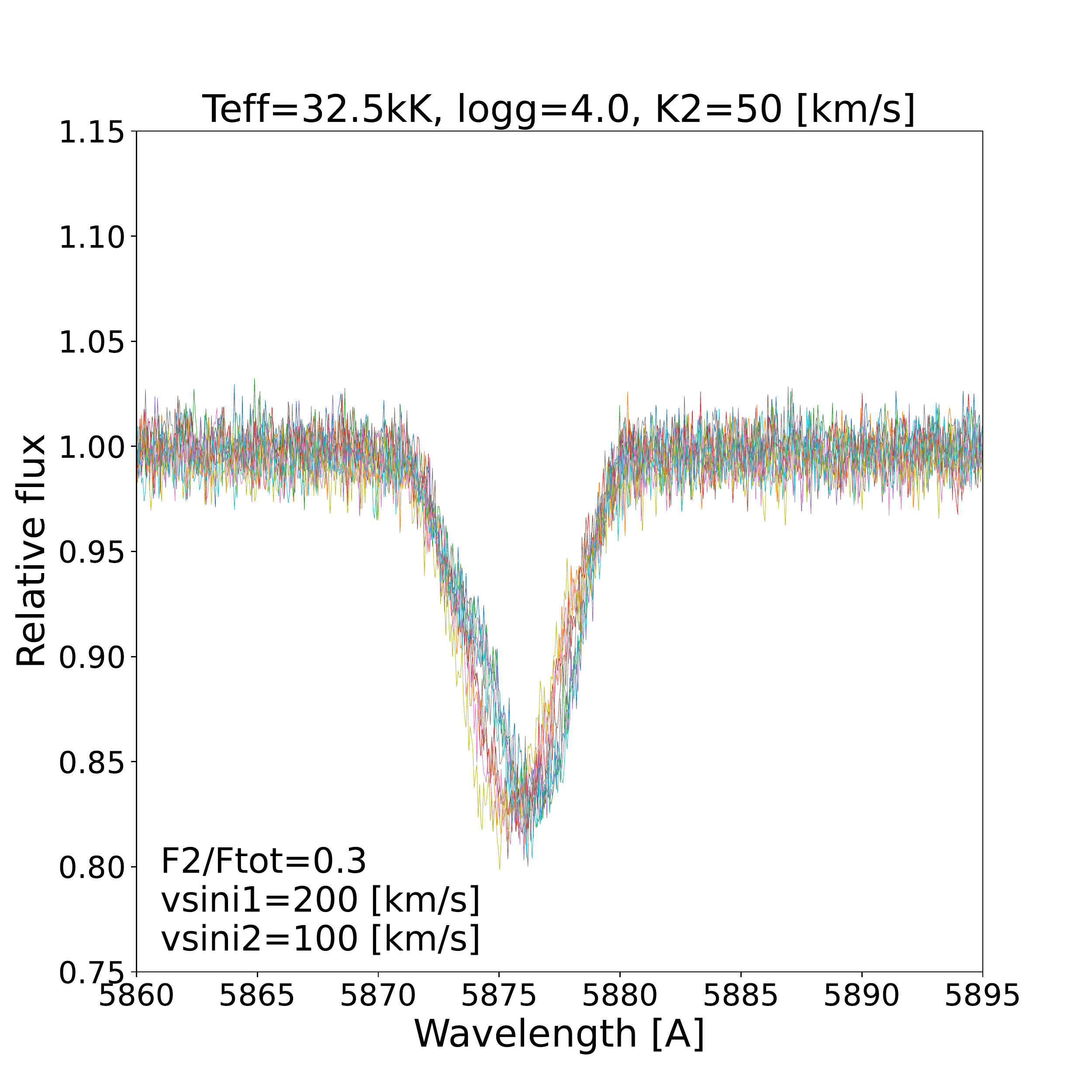}
\includegraphics[width=0.40\textwidth]{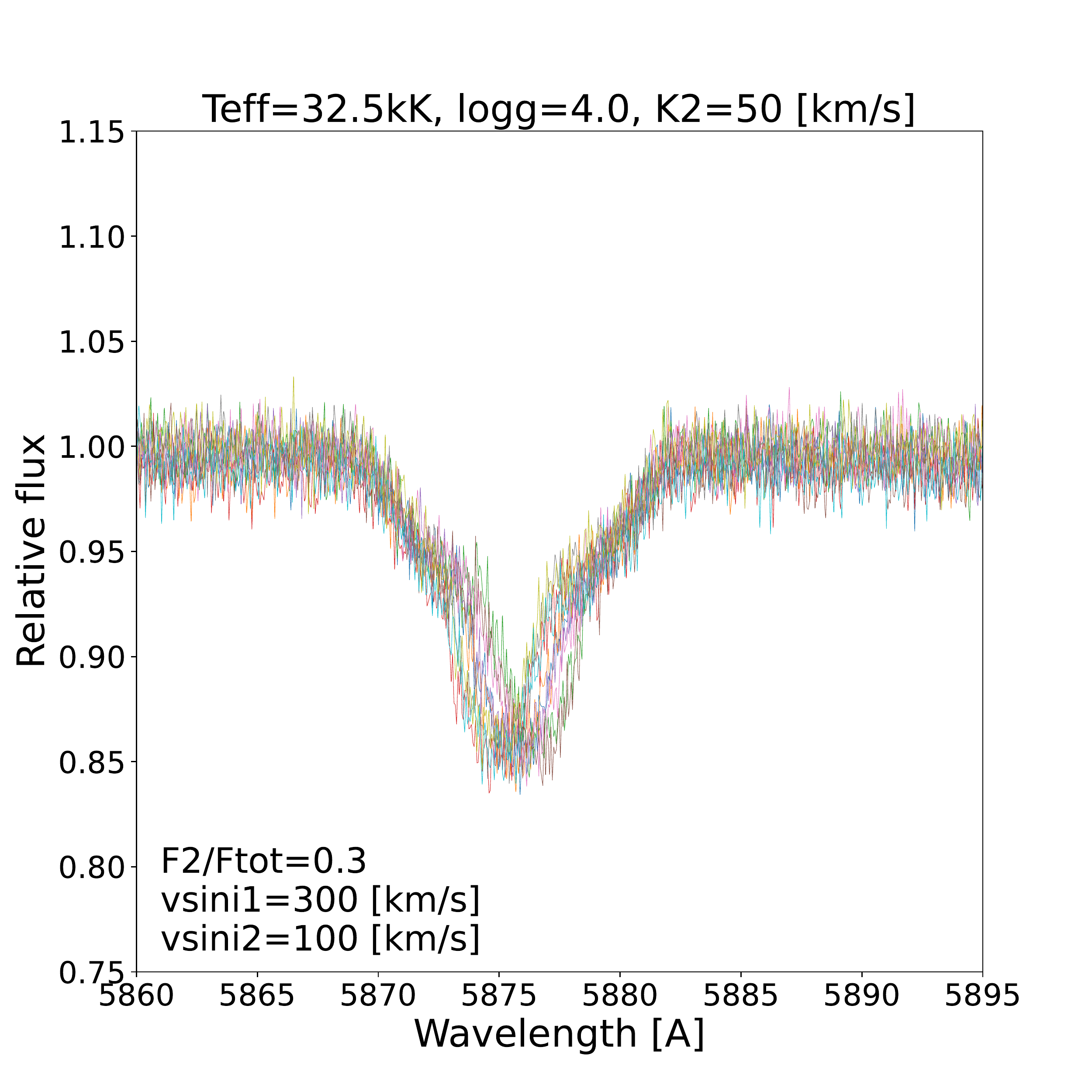}
\caption{Simulations of He{\sc i}~$\lambda$5875 line profile variation based on the TLUSTY \citep{tlusty} stellar atmosphere models. We assume that we have two stars with \Teff = 32.5 [kK] and logg = 4.0, for the primary we set up two regimes of \vsini, namely 200 and 300~\kms, without orbital velocity K1=0~\kms (left and right panel, respectively). In the case of the secondary component, we assume it has ten random K2 values in the range from 0 to 50~\kms and constant \vsini~of 100~\kms. Then we modelled the composite spectrum with the different contributions of the secondary component by varying the flux ratio $F2/F_{total}$ from 0.1 to 0.3.}
\label{spectra_simulations}
\end{figure*}

\begin{figure*}[!t]
\centering
\includegraphics[width=0.99\textwidth]{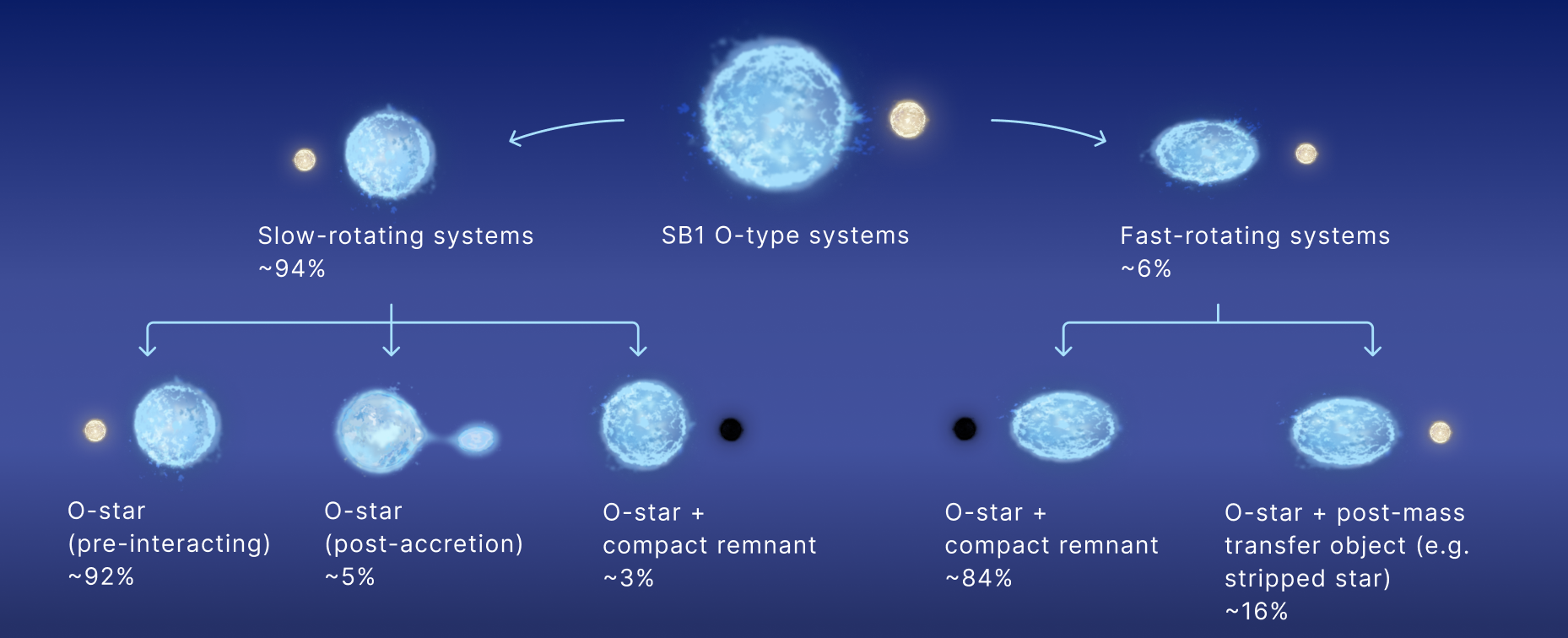}
\caption{Schematic illustration about possible nature of second components of O-type SB1 systems according to the BPASS model simulations. See Sect.~\ref{theory} for details.}
\label{figure_sb1_illustration}
\end{figure*}

\end{appendix}
\end{document}